\documentclass{article}
\usepackage[utf8]{inputenc}
\usepackage{macros}

\usepackage{graphicx}
\graphicspath{ {./images/} }

\newcommand{\OL}{\mathsf{OL}}

\DeclareMathOperator{\pdet}{pdet}
\DeclareMathOperator{\diam}{diam}

\DeclareMathOperator{\ReLU}{ReLU}

\title{Online Lewis Weight Sampling\footnote{An earlier version of this work appears in SODA 2023, which includes an error in the result about adversarially robust $\ell_p$ subspace embeddings. The current version removes this result. The other results in this paper are unaffected.}}
\author{
David P. Woodruff \\ Carnegie Mellon University \\ \texttt{dwoodruf@cs.cmu.edu} \and Taisuke Yasuda \\ Carnegie Mellon University \\ \texttt{taisukey@cs.cmu.edu}
}
\date{}

\begin{document}

\maketitle

\thispagestyle{empty}
\begin{abstract}
The seminal work of Cohen and Peng \cite{CP2015} (STOC 2015) introduced \emph{Lewis weight sampling} to the theoretical computer science community, which yields fast row sampling algorithms for approximating $d$-dimensional subspaces of $\ell_p$ up to $(1+\eps)$ relative error. Prior works have extended this important primitive to other settings, such as the online coreset and sliding window models \cite{BDMMUWZ2020} (FOCS 2020). However, these results are only for $p\in\{1,2\}$, and results for $p=1$ require a suboptimal $\tilde O(d^2/\eps^2)$ samples.

In this work, we design the first nearly optimal $\ell_p$ subspace embeddings for all $p\in(0,\infty)$ in the online coreset and sliding window models. In both models, our algorithms store $\tilde O(d/\eps^2)$ rows for $p\in(0,2)$ and $\tilde O(d^{p/2}/\eps^2)$ rows for $p\in(2,\infty)$. This answers a substantial generalization of the main open question of \cite{BDMMUWZ2020}, and gives the first results for all $p\notin\{1,2\}$ and achieves nearly optimal sample complexities for all $p$.

Towards our result, we give the first analysis of ``one-shot'' Lewis weight sampling of sampling rows proportionally to their Lewis weights, which gives a sample complexity of $\tilde O(d^{p/2}/\eps^2)$ rows for $p>2$. Previously, such a sampling scheme was only known to have a sample complexity of $\tilde O(d^{p/2}/\eps^5)$ \cite{CP2015}, whereas a bound of $\tilde O(d^{p/2}/\eps^2)$ is known if a more sophisticated recursive sampling algorithm is used \cite{MMWY2021,LT1991}. Note that the recursive sampling strategy cannot be implemented in an online setting, thus necessitating an analysis of one-shot Lewis weight sampling. Perhaps surprisingly, our analysis crucially uses a novel connection to online numerical linear algebra, \emph{even for offline Lewis weight sampling}.

As an application, we obtain the first one-pass streaming coreset algorithms for $(1+\eps)$ approximation of important generalized linear models, such as logistic regression and $p$-probit regression. Our upper bounds are parameterized by a complexity parameter $\mu$ introduced by \cite{MSSW2018}, and we also provide the first lower bounds showing that a linear dependence on $\mu$ is necessary.

\end{abstract}

\clearpage
\setcounter{page}{1}

\newpage

\section{Introduction}

We consider the problem of computing \emph{$\ell_p$ subspace embeddings} in the setting of big data analysis. In this problem, we are given a large input matrix $\bfA\in\mathbb R^{n\times d}$ where $n\gg d$, and we seek an approximation $\bfS\bfA\in\mathbb R^{r\times d}$ with $d\leq r \ll n$ such that
\begin{equation}\label{eq:subspace-embedding}
    \mbox{for all $\bfx\in\mathbb R^d$,}\qquad \norm*{\bfS\bfA\bfx}_p = (1\pm\eps)\norm*{\bfA\bfx}_p.
\end{equation}
That is, we seek a small summary of $\bfA$ which approximates every vector in its column space in the $\ell_p$ norm, where the summary takes the form $\bfS\bfA$ for some $r\times n$ matrix $\bfS\in\mathbb R^{r\times n}$. Such a dimensionality reduction result is of fundamental importance to machine learning and theoretical computer science, and the utility of such a result has been proven through a long line of work on this problem \cite{Sar2006, DDHKM2009, SW2011, WZ2013, MM2013, CP2015, WW2019, LWW2021, WY2022}. While $\bfS$ can in principle be any linear map, including a dense matrix (see, e.g., the dense Cauchy sketches of \cite{SW2011}), the best known results for $\ell_p$ subspace embeddings proceed by a sampling approach \cite{DDHKM2009, CP2015, MMWY2021}, in which $\bfS$ only has one nonzero entry per row. We focus on such an approach in this work, and refer to the number $r$ of rows of $\bfS$ as the \emph{sample complexity}.

This problem is also central to the functional analysis literature, and nearly optimal upper bounds have been known since the works of \cite{Lew1978, BLM1989, Tal1990, LT1991, Tal1995, SZ2001, Sch2011}, which are complemented by recently discovered lower bounds due to \cite{LWW2021}. These results were then turned algorithmic by a result of \cite{CP2015}, which showed that it was possible to compute a sketch $\bfS\bfA$ satisfying \eqref{eq:subspace-embedding} in nearly input sparsity time. These results gave $r = \tilde O(d/\eps^2)$ rows for $p\in[1,2]$ and $r = \tilde O(d^{p/2}/\eps^5)$ for $p>2$. Furthermore, the algorithm of \cite{CP2015} had a simple two-step procedure consisting of (1) approximating Lewis weights and then (2) sampling each row proportionally to these weights, making it an attractive algorithm in various settings. It was recently shown how to extend the results to $r = \tilde O(d/\eps^2)$ for $p\in(0,2]$ and $r = \tilde O(d^{p/2}/\eps^2)$ by \cite{MMWY2021}, through the use of a more sophisticated recursive sampling strategy due to \cite{Tal1990,LT1991}.

Despite this recent progress in the algorithmic theory of $\ell_p$ subspace embeddings, there are important problems that remain. Driven by increasingly challenging practical problems associated with the analysis of modern data sets, recent work in algorithmic data science has focused on more and more restrictive requirements and models of computation. These include:
\begin{itemize}
    \item the \emph{streaming model}, in which the data set can only be accessed through one pass through a stream of the rows of the data set
    \item the \emph{online coreset model}, which further restricts the streaming model by only allowing for a small number of rows to be irrevocably stored (i.e., cannot be thrown away to save space)
    \item the \emph{sliding window model}, in which only the $W$ most recent rows in a stream are considered as the input at any time
\end{itemize}
These models of computation address important practical requirements for applications, and we refer to a rich line of previous work, and references therein, on the motivations for studying the online \cite{CMP2020, BDMMUWZ2020} and sliding window \cite{BDMMUWZ2020, UU2021, EMMZ2022} models.

The streaming model can be addressed quite straightforwardly by employing a standard merge-and-reduce procedure to convert the offline $\ell_p$ subspace embedding algorithms of \cite{CP2015, MMWY2021} into streaming algorithms. However, the latter two extensions of the streaming model described above are more challenging to handle. For $p = 2$, \cite{CMP2020} gave nearly optimal results for the online coreset model, showing that one can maintain approximately $r = \tilde O(d/\eps^2)$ rows in an online manner, while essentially only losing a $\log\kappa^\OL$ factor, where $\kappa^\OL = \kappa^\OL(\bfA)$ is a natural quantity known as the \emph{online condition number of $\bfA$} (Definition \ref{def:online-condition-num}). The same work also shows that such a condition number dependence is required \cite[Theorem 5.1]{CMP2020}. The work of \cite{BDMMUWZ2020} then gave an elegant reduction from sliding windows to online coreset algorithms, thus achieving similar guarantees in this model. Thus, for $p = 2$, all of these questions are nearly settled.

On the other hand, for $p \neq 2$, the landscape is far worse, even for $p = 1$. The work of \cite{BDMMUWZ2020} obtained a bound of $r = \tilde O(d^2/\eps^2)$ rows in both the online coreset and sliding window models, which is loose by a factor of $d$ compared to the optimal sample complexity in the offline model. They also give a deterministic sliding window algorithm for $p=1$ sampling $\tilde O(d/\eps^2)$ rows, but this algorithm runs in exponential time. They leave the following as their main open question:

\begin{Question}\label{q:online-l1}
Can the sample complexity of $\ell_1$ subspace embeddings in the online coreset model be improved from $\tilde O(d^2/\eps^2)$ to $\tilde O(d/\eps^2)$?
\end{Question}

For all other $p\in(0,\infty)\setminus\{1,2\}$ are absent in the sliding window model. While it seems possible to extend the techniques of \cite{BDMMUWZ2020} to $p\in(0,2)$, the problems of suboptimal sample complexity or exponential running time would still remain. Additionally, for $p>2$, the behavior of Lewis weights changes substantially from that of $p<2$ (e.g., lack of monotonicity), which breaks many parts of existing approaches.

Furthermore, the consideration of the online model and its variants brings up a natural question on Lewis weight sampling, even in the offline setting: 

\begin{Question}\label{q:lewis-sampling}
Is it possible to obtain a sample complexity bound of $\tilde O(d^{p/2}/\eps^2)$ rows for $p > 2$ using the simple strategy of sampling proportionally to $\ell_p$ Lewis weights?
\end{Question}

Aside from being a more aesthetically pleasing result than the recursive sampling strategy of \cite{MMWY2021}, in certain situations such as the online settings which we consider, the application of Lewis weight sampling would only work if the algorithm is a simple scheme of sampling proportionally to weights; the recursive sampling strategy cannot work without knowledge of all of the rows of the matrix. Thus, Question \ref{q:lewis-sampling} is a central unresolved question in the study of Lewis weight sampling. We also note that Lewis weights and their sampling guarantees have been central to many recent advances in machine learning and theoretical computer science \cite{CD2021, PPP2021, MMWY2021} even beyond $\ell_p$ losses \cite{CWW2019, LWYZ2020, MRM2021, MMWY2021}, making it even more important to gain an improved understanding of Lewis weight sampling.

\subsection{Our Contributions}

\subsubsection{Online \texorpdfstring{$\ell_p$}{lp} Lewis Weight Sampling}

As our first contribution, we answer Question \ref{q:online-l1} affirmatively, achieving an online coreset for $\ell_1$ subspace embeddings with
\[
    O\parens*{\frac{d}{\eps^2}(\log n)\log(n\kappa^\OL)}
\]
rows\footnote{Note that one can compose this algorithm with itself, in an online fashion, so that $n$ here can be replaced by $O\parens*{\frac{d}{\eps^2}(\log n)\log(n\kappa^\OL)}$. For simplicity of presentation, we state our results without this optimization.}. In fact, we show much more than this, by obtaining the first online coresets for $\ell_p$ subspace embeddings for all $p\in(0,\infty)\setminus\{1,2\}$ with $(1+\eps)$ error. Our dependence on the dimension $d$ is optimal for all $p\in(0,\infty)$ up to polylogarithmic factors due to known lower bounds for $\ell_p$ subspace embeddings \cite{LWW2021}, and our dependence on $\eps$ is quadratic for $p\in(0,\infty)$, which is also optimal \cite{LWW2021}. Thus, we in fact answer a substantial generalization of Question \ref{q:online-l1}. Our results are summarized in Table \ref{tab:results}.

\begin{table}[ht]
\centering
\begin{tabular}{ c c c c c c c }
\toprule
& Sample Size & \\
\hline
$p = 1$ & $d^2/\eps^2$ & \cite[Theorem 4.1]{BDMMUWZ2020} \\
$p = 2$ & $d/\eps^2$ & \cite{CMP2020,BDMMUWZ2020} \\
\rowcolor{blue!15}$0 < p < 2$ & $d/\eps^2$ & Theorem \ref{thm:online-lewis-p>2} \\
\rowcolor{blue!15}$2 < p < \infty$ & $d^{p/2}/\eps^2$ & Theorem \ref{thm:online-lewis-p>2} \\
\bottomrule
\end{tabular}
\caption{Our results for online Lewis weight sampling. We suppress polylogarithmic factors in $n$, $\kappa^\OL$, $\eps^{-1}$.}
\label{tab:results}
\end{table}

In order to obtain our result, we make a key change over prior approaches towards online coresets for $\ell_p$ subspace embeddings \cite{CMP2020, BDMMUWZ2020}: we decouple the problem of approximating the importance of a row and approximating the matrix itself. That is, we maintain two sketches, one for approximating an online generalization of Lewis weights which we call \emph{online Lewis weights} (see Section \ref{sec:online-lewis-weights} for definitions and properties), and one which uses the online Lewis weights as importance scores in order to obtain an $\ell_p$ subspace embedding. This has two advantages: (1) we can build on prior work for spectral approximation to approximate the Lewis quadratic, and (2) by conditioning on the success of the approximation of Lewis weights, we can simply treat the online sampling process for the $\ell_p$ subspace embedding exactly as an offline sampling process, which significantly simplifies the analysis. In particular, we avoid complex sequential chaining arguments such as those considered in, e.g., \cite{RST2010, BDR2021}. This decoupled approach may be of independent interest for future work on matrix approximation and importance sampling, especially in online models and other restricted models of computation. 

\subsubsection{Offline \texorpdfstring{$\ell_p$}{lp} Lewis Weight Sampling}

En route to obtaining the result of Table \ref{tab:results}, we answer Question \ref{q:lewis-sampling} in the affirmative, thereby closing a long-standing gap in the study of $\ell_p$ Lewis weight sampling since \cite{CP2015}. We refer to Table \ref{tab:results-one-shot} for a summary of this result alongside prior results.

\begin{table}[ht]
\centering
\begin{tabular}{ c c c c c c c }
\toprule
Sample Size & Sampling Algorithm & \\
\hline
$d^{p/2}/\eps^5$ & One-Shot & \cite{CP2015,BLM1989} \\
$d^{p/2}/\eps^2$ & Recursive & \cite{MMWY2021,LT1991} \\
\rowcolor{blue!15}$d^{p/2}/\eps^2$ & One-Shot & Theorems \ref{thm:offline-lewis-weight-sampling-main}, \ref{thm:one-shot-lewis-weight-sampling} \\
\bottomrule
\end{tabular}
\caption{Our results for offline Lewis weight sampling. We suppress polylogarithmic factors in $n$ and $\eps^{-1}$.}
\label{tab:results-one-shot}
\end{table}

Perhaps surprisingly, our result crucially relies on a novel connection to online leverage scores \cite{CMP2020}, which allows us to circumvent the problem of non-monotonicity of $\ell_p$ Lewis weights for $p>2$, which was the major barrier towards achieving this result. We also give the first results for Lewis weight sampling which simultaneously achieve an optimal dependence on $d$ and $\eps$ along with a polylogarithmic dependence on the failure rate $\delta$ for any $p \neq 2$, which is crucial for certain applications such as $\ell_p$ subspace embeddings in sliding windows. Prior results had at least one problem of only achieving constant probability of success \cite{CP2015}, suboptimal dependence on $\eps$ \cite{BLM1989}, or suboptimal dependence on $d$ \cite{Sch1987}. Our analysis also allows for the use of $\ell_p$ Lewis weight approximations which satisfy weaker guarantees than the requirement of upper bounding the true $\ell_p$ Lewis weights, which can be computed in $\tilde O(\nnz(\bfA) + d^\omega)$ time \cite{Lee2016, JLS2021}, rather than $\tilde O(\nnz(\bfA) + d^{O(p)})$ time \cite{CP2015}. Thus, our Lewis weight sampling result gives ``the best of all worlds'' in terms of running time and dependencies on $d$, $\eps$, and $\delta$, up to logarithmic factors. Altogether, our results show that there is no need to sacrifice sample complexity when using $\ell_p$ Lewis weight sampling, other than logarithmic factors.

\begin{restatable}{Theorem}{LewisWeightSampling}[``Best of All Worlds'' $\ell_p$ Lewis Weight Sampling]\label{thm:offline-lewis-weight-sampling-main}
Let $p>2$ and let $\bfA\in\mathbb R^{n\times d}$. Let $\delta\in(0,1)$ be a failure rate parameter and let $\eps\in(0,1)$ be an accuracy parameter. Let $\bfw\in\mathbb R^n$ be one-sided $\ell_p$ Lewis weights (Definition \ref{def:one-sided-lewis}) with $\norm*{\bfw}_1 \leq O(d)$, which can be computed in
\[
    \tilde O(\nnz(\bfA) + d^\omega)
\]
time \cite[Theorem 5.3.1]{Lee2016}, \cite[Lemma 2.5]{JLS2021}. Let
\[
    \alpha = O\parens*{\frac{d^{p/2-1}}{\eps^2}\parens*{(\log d)^2(\log n) + \log\frac1\delta}}
\]
be an oversampling parameter. Suppose that weights $\bfs\in\mathbb R^n$ are sampled by independently setting $\bfs_i = 1/\bfp_i^{1/p}$ with probability $\bfp_i \coloneqq \min\{\alpha\bfw_i, 1\}$ and $\bfs_i = 0$ otherwise. Let $\bfS = \diag(\bfs)$. Then, with probability at least $1-\delta$,
\[
    \mbox{for all $\bfx\in\mathbb R^d$, }\norm*{\bfS\bfA}_p = (1\pm\eps)\norm*{\bfA\bfx}_p
\]
and the sample complexity of $\bfS$ is at most
\[
    r = O\parens*{\frac{d^{p/2}}{\eps^2}\parens*{(\log d)^2(\log n) + \log\frac1\delta}}.
\]
\end{restatable}

We also give similar high probability Lewis weight sampling results for $0<p<2$ in Appendix \ref{sec:high-prob-lws}, which we need for our high probability online coresets for $0<p<2$, as well as applications to sliding windows. We give a discussion of our techniques in Section \ref{sec:tech:lewis-weight-sampling}, and our proofs are contained in Section \ref{sec:one-shot}.

\subsubsection{Applications: Sliding Windows}

As an application of our results for online coresets for $\ell_p$ subspace embeddings, we obtain significantly improved results $\ell_p$ subspace embeddings in the \emph{sliding window model}. 

In the sliding window model of $\ell_p$ subspace embeddings, we are given a stream of rows $\{\bfa_i\}_{i=1}^n$ as well as a parameter $W\in\mathbb N$, which specifies the size of a window. Then, at each time $i\in[n]$, we consider the matrix $\bfA_i^W$ which denotes the $W\times d$ matrix formed by rows $i, i-1, i-2, \dots, i-W+1$, that is, the $W$ most recent rows at time $i$. Our goal is to output an $\ell_p$ subspace embedding for $\bfA_i^W$ at time $i$, for each $i\in[n]$. 

A simple observation from the work of \cite{BDMMUWZ2020} shows how to convert algorithms for online coresets for $\ell_p$ subspace embeddings into algorithms for sliding windows, by running the online coreset algorithm ``in reverse''. That is, suppose that we have maintained an online coreset $\bfS_i^W\bfA_i^W$ for $\bfA_i^W$ such that for $j\in [W]$, the last $j$ rows of $\bfS_i^W\bfA_i^W$ are a subspace embedding for the last $j$ rows of $\bfA_i^W$. Then, we can update this sketch by throwing away the first row of $\bfS_i^W\bfA_i^W$ (which could be a zero row that is maintained implicitly), appending the new row, and recomputing an online coreset for this new matrix if necessary to save space. Although we lose a factor of $(1+\eps)$ in the distortion each time we recompute the online coreset, by carrying out this process in a binary tree fashion, we only compose this approximation at most $\log n$ times per each row and thus we can set $\eps$ to be $\eps/\log n$ instead so that the total distortion is only $(1+\eps)$. Although this reduction is stated for deterministic algorithms in \cite{BDMMUWZ2020}, because we have a logarithmic dependence on $\delta$ for our coresets, we can afford to union bound over all blocks of the merge-and-reduce tree to obtain the following:

\begin{Theorem}[$\ell_p$ Lewis Weight Sampling in Sliding Windows]\label{thm:sliding-window-lewis}
Let $\bfA\in\mathbb R^{n\times d}$ and $p\in(0,\infty)$. Let $\delta\in(0,1)$ be a failure rate parameter and let $\eps\in(0,1)$ be an accuracy parameter. Let $W\in\mathbb N$ be a window size parameter. Let $\kappa$ be the stream condition number of $\bfA$, that is, any submatrix of consecutive rows of $\bfA$ has condition number at most $\kappa$. Then, there is a sliding window coreset algorithm $\mathcal A$ such that, with probability at least $1-\delta$, $\mathcal A$ outputs a weighted subset of $m$ rows with sampling matrix $\bfS$ such that
\[
    \norm*{\bfS_i^W\bfA_i^W\bfx}_p^p = (1\pm\eps)\norm*{\bfA_i^W\bfx}_p^p
\]
for all $\bfx\in\mathbb R^d$ and every $i\in[n]$, and
\[
    m = \begin{dcases}
        O\parens*{\frac{d^{p/2}}{\eps^2}\cdot(\log(n\kappa))^{p/2+1}(\log n)^2\bracks*{(\log d)^2(\log n) + \log\frac1\delta}} & p \in (2,\infty) \\
        O\parens*{\frac{d}{\eps^2}\cdot\log(n\kappa)(\log n)^2\bracks*{(\log d)^2\log n + \log\frac1\delta}} & p \in (1,2) \\
        O\parens*{\frac{d}{\eps^2}\cdot\log(n\kappa)(\log n)^2\log\frac{n}{\delta}} & p = 1 \\
        O\parens*{\frac{d}{\eps^2}\cdot\log(n\kappa)(\log n)^2\bracks*{(\log d)^3 + \log\frac1\delta}} & p\in(0,1)
    \end{dcases}
\]
\end{Theorem}
\begin{proof}
Our merge-and-reduce algorithm is exactly as described in Section 5 of \cite{BDMMUWZ2020}, except that we use a randomized algorithm with failure rate $\delta$ set to $O(\delta/n)$ rather than a deterministic algorithm as the ``reduce'' algorithm. This allows us to union bound over all $O(n)$ reduce operations used in the merge-and-reduce algorithm. The result then follows from applying Theorems \ref{thm:online-lewis-p<2} and \ref{thm:online-lewis-p>2} to get the guarantees for the ``reduce'' algorithm. 
\end{proof}

\subsubsection{Applications: Online Coresets for Generalized Linear Models}

As applications of our online approximation of Lewis weights, we obtain the first online coresets, and in fact the first one-pass streaming algorithms, for a variety of generalized linear models. 

The work of \cite{MSSW2018} gave one of the first investigations of coresets for unregularized logistic regression in the worst case, and showed that in general, logistic regression does not admit coresets with sublinear in $n$ memory. To get around this problem, they defined a natural complexity parameter $\mu(\bfA)$ which characterizes the complexity of the dataset, and gave sensitivity sampling algorithms for the logistic regression problem with sample complexity $\poly(\mu(\bfA),d,\log n,\eps^{-1})$, but required multiple passes through the data. This work was later extended to an oblivious one-pass streaming algorithm by \cite{MOW2021}, which achieved an $O(1)$-approximation using $\poly(\mu(\bfA),d,\log n)$ bits of space. The results of \cite{MRM2021}
also gave further results in this direction, which generalized the results to handle a broad class of ``nice hinge functions'' which includes the logistic, $\ReLU$, and hinge losses, and further improved the polynomial dependencies in the coreset size using $\ell_1$ Lewis weights. The recent work of \cite{MOP2022} showed that similar ideas can handle the $p$-generalized probit regression model, a generalization of the probit regression model which they introduce using the $p$-generalized Gaussian distribution \cite{DBPS2018}. They introduce a generalization of the $\mu(\bfA)$ parameter to the degree $p$ analogue $\mu_p(\bfA)$\footnote{The parameter $\mu_p(\bfA)$ \cite{MOP2022} for $p=1$ corresponds to $\mu(\bfA)$ of \cite{MSSW2018, MOW2021, MRM2021}. See Definition \ref{def:mu-p}.} and give a two-pass streaming coreset algorithm for $p\neq 2$ and a one-pass streaming coreset algorithm for $p=2$.

In the important one-pass streaming setting, the only results that apply from the above line of work are the oblivious $O(1)$-approximation algorithm of \cite{MOW2021} for logistic regression, as well as the $(1+\eps)$-approximation for the probit regression model. We remedy this situation by providing analogues of \cite{MRM2021, MOP2022} in the one-pass streaming setting via our online coresets for Lewis weights.

\paragraph{Online Coresets for Nice Hinge Functions.}

Recall first the definitions of $\mu_p$ complexity \cite{MOP2022} and nice hinge functions \cite{MRM2021}:

\begin{Definition}[$\mu_p$ Complexity (Definition 2.2 \cite{MOP2022}, Definition 2 \cite{MSSW2018})]\label{def:mu-p}
Let $\bfA\in\mathbb R^{n\times d}$ and $p\in(0,\infty)$. Then,
\[
    \mu_p(\bfA) \coloneqq \sup_{\bfA\bfx\neq 0} \frac{\sum_{\bfa_i^\top \bfx > 0} \abs*{\bfa_i^\top\bfx}^p}{\sum_{\bfa_i^\top \bfx < 0} \abs*{\bfa_i^\top\bfx}^p}.
\]
We say $\bfA$ is $\mu_p$-complex if  $\mu_p(\bfA) \leq \mu_p < \infty$. If $p = 1$, we drop the subscript and simply say that $\bfA$ is $\mu$-complex. 
\end{Definition}

\begin{Definition}[Nice Hinge Functions (Definition 7, \cite{MRM2021})]\label{def:nice-hinge}
We say that $\varphi:\mathbb R \to \mathbb R^+$ is an \emph{$(L,a_1,a_2)$-nice hinge function} if there exist universal constants $L>0$, $a_1\geq 0$, and $a_2\geq 0$ such that
\begin{itemize}
    \item $x\mapsto\varphi(x)$ is $L$-Lipschitz.
    \item $\abs*{\ReLU(x) - \varphi(x)} \leq a_1$ for all $x\in\mathbb R$.
    \item $\varphi(x) \geq a_2$ for all $x\geq 0$.
\end{itemize}
\end{Definition}

Note that nice hinge functions include the $\ReLU$, hinge loss $\max\{0,1+x\}$, and the logistic loss. The work of \cite{MRM2021} shows that oversampling from Lewis weights by a factor of $\mu(\bfA)^2/\eps^2$ yields relative error coresets for any nice hinge function. By using our $\ell_1$ Lewis weight overestimates of either Theorem \ref{thm:online-lewis-p>2} or Theorem \ref{thm:online-lewis-p<2}, we immediately extend the results of \cite{MRM2021} to the online setting:

\begin{Theorem}
Let $\varphi$ be a nice hinge function (Definition \ref{def:nice-hinge}) with $a_2 > 0$ or the $\ReLU$ loss. Let $\tilde\bfw\in\mathbb R^n$ be online $\ell_1$ Lewis weight estimates obtained by either Theorem \ref{thm:online-lewis-p>2} or Theorem \ref{thm:online-lewis-p<2} and let $T = \norm*{\tilde\bfw}_1$. Let $\bfp_i \geq C(\mu(\bfA)/\eps)^2 \max\{\tilde\bfw_i, 1/n\}$ for some $C = O(\log\frac{nT}{\eps\delta})$. For each $i\in[n]$, let
\[
    \bfs_i = \begin{cases}
        1/{\bfp_i} & \text{with probability $\bfp_i$} \\
        0 & \text{otherwise}.
    \end{cases}
\]
Then with probability at least $1-\delta$, for all $\bfx\in\mathbb R^d$,
\[
    \abs*{\sum_{i=1}^n \bfs_i\varphi([\bfA\bfx](i)) - \varphi([\bfA\bfx](i))} \leq \eps \sum_{i=1}^n \varphi([\bfA\bfx](i)).
\]
Furthermore, with probability at least $1-\delta$, there are at most
\[
    O\parens*{\frac{d\mu(\bfA)^2}{\eps^2}\log(n\kappa^\OL)\log\frac{nT}{\eps\delta}}
\]
nonzero entries $\bfs_i$.
\end{Theorem}
\begin{proof}
This is essentially a direct application of \cite[Theorem 5, Corollary 6, Theorem 8, Corollary 9]{MRM2021}. There are slight differences -- the sampling method in \cite{MRM2021} draws without replacement from the Lewis weight sampling distribution, while we sample each row independently. These differences are minor, and can be handled as we do in the proofs of Theorem \ref{thm:online-lewis-p>2} or Theorem \ref{thm:online-lewis-p<2}.
\end{proof}

\paragraph{Streaming Coresets for $p$-Probit Regression.}

The probit model is a popular generalized linear model which uses the Gaussian cdf as the link function to model binary data. This model was recently generalized to the $p$-generalized probit model (or $p$-probit model for short) in \cite{MOP2022}, which models binary data $Y_i\in\{0,1\}$ for $i\in[n]$ as
\[
    \Pr\braces*{Y_i = 1} = \E[Y_i] = \Phi_p(\bfz_i^\top \bfx),
\]
where
\[
    \Phi_p(r) = \frac{p^{1-1/p}}{2\Gamma(1/p)}\int_{-\infty}^r \exp(-\abs*{t}^p/p)~dt
\]
is the cdf of the $p$-generalized normal distribution, $\bfz_i\in\mathbb R^d$ is the feature vector for the label $y_i$, and $\bfx\in\mathbb R^d$ is a parameter vector. We then define
\[
    \psi_p(x) = -\log(\Phi_p(-x))
\]
to be the $p$-probit loss, and the negative log likelihood of the dataset under the parameter vector $\bfx\in\mathbb R^d$ can be written as
\[
    \mathcal L_p(\bfx) \coloneqq \sum_{i=1}^n \psi_p([\bfA\bfx](i))
\]
where the $i$th row of $\bfA$ is $\bfa_i = -(2y_i - 1)\bfz_i$ \cite{MOP2022}. We give the following streaming coreset theorem. Note that due to the reservoir sampler, this result is not an online coreset, in the sense that the rows are not selected irrevocably.

\begin{Theorem}[One-Pass Streaming Coresets for $p$-Probit Regression]\label{thm:p-probit-coresets}
Let $\tilde\bfw\in\mathbb R^n$ be online $\ell_1$ Lewis weight estimates obtained by either Theorem \ref{thm:online-lewis-p>2} or Theorem \ref{thm:online-lewis-p<2} and let $T = \norm*{\tilde\bfw}_1$.
Then, there is a one-pass streaming algorithm which computes a coreset $\bfs\in\mathbb R^n$ such that
with probability at least $1-\delta$, for all $\bfx\in\mathbb R^d$,
\[
    \abs*{\sum_{i=1}^n \bfs_i\psi_p([\bfA\bfx](i)) - \psi_p([\bfA\bfx](i))} \leq \eps \sum_{i=1}^n \psi_p([\bfA\bfx](i)).
\]
Furthermore, with probability at least $1-\delta$, there are at most
\[
    O\parens*{\frac{Sd}{\eps^2}(\log S)\log\frac{\mu_p(\bfA)}{\eps}}
\]
nonzero entries $\bfs_i$, for
\[
    S = O(\mu_p(\bfA))(d\log(n\kappa^\OL))^{1\lor(p/2)}.
\]
\end{Theorem}
\begin{proof}
It is shown in Lemma 2.10 that the $\psi_p$-sensitivities are bounded as
\[
    \sup_{\bfA\bfx\neq 0}\frac{\psi_p([\bfA\bfx](i))}{\mathcal L_p(\bfx)} \leq O(\mu_p(\bfA))\parens*{\frac1n + \bfs_i^p(\bfA)},
\]
where $\bfs_i^p(\bfA)$ are the $\ell_p$ sensitivities (Equation \eqref{eq:lp-sensitivity}). We show that our approximate online Lewis weights bound the $\ell_p$ sensitivities, which allow us to sample from them. For $0<p<2$, the approximate online Lewis weights $\tilde\bfw_i$ upper bound the true Lewis weights by Lemma \ref{lem:online-lewis-bound}, which means that they upper bound the $\ell_p$ sensitivities (note that we do not use our sampling-based Lewis weight estimation algorithm, which requires splitting of rows which can affect the $p$-probit loss). For $p>2$, we use Corollary \ref{cor:online-sensitivities} to argue that $\norm*{\tilde\bfw}_1^{p/2-1}\tilde\bfw_i$ is an upper bound on the $\ell_p$ sensitivities. Finally, combining the result with the VC dimension bound in Lemma 2.9 of \cite{MOP2022} and the sensitivity framework results of \cite{FSS2020} (see also Appendix B of \cite{MOP2022} for a discussion of the sensitivity framework in this context), as well as the sampling scheme using weighted reservoir sampling using these sensitivities as in \cite{MOP2022} yields the desired result.
\end{proof}

We note that our coreset size also substantially improves the polynomial factors of those in \cite{MOP2022}, who gave bounds of
\[
    S = \begin{cases}
        O(\mu_p(\bfA) d) & p = 2 \\
        O(\mu_p(\bfA) d^p (d\log d)^2 & p \in [1, 2) \\
        O(\mu_p(\bfA) d^{2p} (d\log d)^2 & p\in(2,\infty)
    \end{cases}
\]
for $S$ as used in Theorem \ref{thm:p-probit-coresets}.

\paragraph{New Lower Bounds.}

To complement our improved algorithmic results, we provide the first lower bounds on the size of mergeable\footnote{By mergeable, we mean that a coreset for $\bfA$ and a coreset for $\bfB$ can be combined into a coreset for $\bfA\circ\bfB$.} coresets for $p$-probit regression and logistic regression, for instances with bounded $\mu_p$-complexity. Despite a line of work on guarantees for generalized linear models parameterized by $\mu$-complexity, lower bounds depending on $\mu$ were previously unknown \cite{MSSW2018, MOW2021, MRM2021, MOP2022}. Our work shows that the linear dependence on $\mu$ in Theorem \ref{thm:p-probit-coresets} as well as the results of \cite{MOP2022} are tight. 

As done in previous work \cite{MSSW2018}, we consider the following communication game: Alice has a dataset $\bfA\in\mathbb R^{n_1\times d}$ and Bob has a dataset $\bfB\in\mathbb R^{n_2\times d}$, and Alice must send a single message $M$ to Bob such that Bob can output an approximate solution to the logistic regression problem for the concatenated dataset $\bfA\circ \bfB$ using just $M$ and $\bfB$. We then show lower bounds on the number of bits required by $M$. This in particular includes mergeable coreset algorithms, since $M$ here can be taken to be the coreset, which can approximately solve logistic regression on $\bfA\circ\bfB$ in combination with $\bfB$. Because our hard instances are only in polylogarithmically many dimensions, if the weights are specified using polylogarithmically many bits, then $M$ is a lower bound on the size of the coreset, up to polylogarithmic factors.

\begin{Theorem}[Informal Version of Theorems \ref{thm:mu-p-probit-lb}, \ref{thm:mu-logreg-lb}]
There exists $\bfA\in\mathbb R^{m\times d}$ with $d = O(\log^2 m)$ and $\mu_p$-complexity at most $O(m)$ for any $1 \leq p < \infty$ such that for any $1 \leq \Delta \leq O(m^{1/3})$, a mergeable coreset which approximates the optimal $p$-probit cost or the logistic regression cost up to a $\Delta$ relative error must have size at least $\tilde\Omega(m/\Delta)$. In particular, a constant factor approximation to the optimal $p$-probit regression cost or logistic regression cost for a $\mu_p$-complex dataset requires $\tilde\Omega(\mu_p)$ points in the coreset.
\end{Theorem}

\subsection{Our Techniques}

\subsubsection{Online Coresets for \texorpdfstring{$\ell_p$}{lp} Subspace Embeddings}

We first discuss previous approaches towards online coresets for $\ell_p$ subspace embeddings as well as their shortcomings, and then discuss our ideas which allow us to overcome them.

For $\ell_2$ \cite{CMP2020}, a spectral approximation to $\bfA$ simultaneously yields both an $\ell_2$ subspace embedding as well as estimates to the online leverage scores of $\bfA$, and yields nearly optimal sample complexities. A na\"ive generalization of this to $\ell_p$ leads to the idea of using an $\ell_p$ subspace embedding in order to estimate sampling probabilities. This strategy naturally lends itself to the technique of \emph{sensitivity sampling}, in which rows are selected with probability proportional to the sensitivity score defined by
\begin{equation}\label{eq:lp-sensitivity}
    \bfs_i^p(\bfA) \coloneqq \sup_{\bfx\in\mathbb R^d}\frac{\abs*{\angle*{\bfa_i,\bfx}}^p}{\norm*{\bfA\bfx}_p^p},
\end{equation}
or its online variants. Indeed, given an $\ell_p$ subspace embedding for $\bfA$, one can approximate the denominator of the above term, as well as the numerator, given access to that row, and this is how \cite{BDMMUWZ2020} proceeds to obtain their online $\ell_1$ subspace embedding result. However, the problem is that sensitivity sampling is not known to admit efficient chaining arguments, and leads to a loss of a factor of $d$ in the sample complexity compared to Lewis weight sampling. Although one would ideally have estimates to Lewis weights (or an appropriate online generalization), the challenge faced by \cite{BDMMUWZ2020} is that they cannot estimate these Lewis weights given an $\ell_p$ subspace embedding of $\bfA$.

One of the main observations necessary towards achieving our results is the \emph{decoupling of the Lewis weight approximation step and the sampling step}. That is, we maintain two different sketches of $\bfA$, one for approximating online Lewis weights, and one for getting an $\ell_p$ subspace embedding. By taking this approach, we also get the additional benefit of a much simpler analysis: by using fresh randomness for the subspace embedding sampling, we are able to avoid using a martingale argument for the subspace embedding, and only use such arguments for approximating the Lewis weights. With Lewis weight estimates in hand, we simply appeal to standard offline analyses of Lewis weight sampling. That is, \emph{the only ``online'' aspect of online $\ell_p$ subspace embedding is in the Lewis weight estimation}. We view the simplicity of the analysis as one of our main strengths of this work.

With this idea in hand, we show how to \emph{deterministically} estimate online $\ell_p$ Lewis weights, by maintaining an online variation on the Lewis quadratic. For $p<2$, we show that these weights in fact bound the true $\ell_p$ Lewis weights, and also have a small sum. This is enough to use these weights as sampling weights, using existing results on $\ell_p$ Lewis weight sampling. For $p>2$, while these approximated weights do not necessarily bound the true $\ell_p$ Lewis weights, we show that they satisfy a one-sided generalization of the property of Lewis weights, which were recently shown to be sufficient to make the Lewis weight sampling arguments work \cite{WY2022}. Our result follows.

For $p\in(0,2)$, we also analyze a sampling-based Lewis weight estimation algorithm analogous to the online leverage score estimation algorithm of \cite{CMP2020}, in which randomly sampled rows are used to estimate the $\ell_p$ Lewis weights. While quite similar to the analysis of \cite{CMP2020}, a straight adaptation does not work due to complications that arise due to introducing the Lewis weights, and we use a variation on the idea of splitting rows in order to control the matrix martingale. This alternate algorithm may be simpler to implement, and may be more attractive in certain cases. This algorithm also has the benefit that it keeps no ``side information'', apart from row samples that are stored online.

Our deterministic Lewis weight approximation algorithm is discussed in Section \ref{sec:online-lewis-weights}, while our sampling-based Lewis weight approximation algorithm for $0<p<2$ is discussed in Section \ref{sec:sample-lw-est}. In Section \ref{sec:batch}, we describe a variant of our algorithm to work in batches of rows, which allows the algorithm to run in input sparsity time. 

\subsubsection{Improved Analysis of Lewis Weight Sampling}\label{sec:tech:lewis-weight-sampling}

Our basic framework for the analysis of one-shot Lewis weight sampling is based around a reduction idea of \cite{CP2015}, as well as an adaptation of this technique by \cite{CD2021}. This reduction shows that for $p\in(0,2)$, in order to bound the expected error of the one-shot Lewis sampling procedure, it suffices to bound the expected error of a sampling procedure which samples each row with probability $1/2$, under the condition that the input matrix has uniformly bounded Lewis weights. 

\paragraph{The \cite{CP2015} reduction.}

The idea of \cite{CP2015} is roughly as follows. The expected error of the one-shot Lewis sampling procedure can be written as
\[
    \E_\bfs \bracks*{\sup_{\norm*{\bfA\bfx}_p = 1}\abs*{\norm*{\bfS\bfA\bfx}_p^p - 1}}.
\]
We wish to show that this quantity is bounded by $\eps$. Now one can note that the quantity inside the absolute is a zero mean random variable. This fact, combined with a standard symmetrization argument, bounds this quantity by
\begin{equation}\label{eq:expect-intro}
    \E_\bfs \E_{\bfsigma}\bracks*{\sup_{\norm*{\bfA\bfx}_p = 1}\abs*{\sum_{i=1}^n \bfsigma_i \abs*{[\bfS\bfA\bfx](i)}^p}},
\end{equation}
where $\bfsigma_i$ are independent Rademacher variables. This expression is, in fact, exactly an expression that is bounded by previous works on Lewis weight sampling from the functional analysis literature \cite{Tal1990, LT1991, Tal1995, SZ2001}, which bounds this expectation by roughly $\eps$, whenever the $\ell_p$ Lewis weights of the matrix are at most $\eps^2$. The idea now is that, heuristically, we would expect the Lewis weights of the matrix $\bfS\bfA$ to be at most $\eps^2$ if $\bfS$ samples each row $i$ with probability roughly $\bfp_i = \bfw_i^p(\bfA) / \eps^2$ and scale the result by $1/\bfp_i^{1/p}$. Indeed, if two rows of $\bfA$ are the same up to a scaling, then the Lewis weights would simply differ by that scaling factor, raised to the $p$th power. Thus, the Lewis weight of $\bfa_i / \bfp_i^{1/p}$ would be expected to be $1 / \bfp_i$ times larger than that of $\bfa_i$, which would just be $\eps^2$. The problem here is that $\bfS\bfA$ and $\bfA$ do not belong to the same matrix. To address this problem, \cite{CP2015} make the observation that, using the monotonicity of Lewis weights for $p\in(0,2)$ \cite[Lemma 5.5]{CP2015}, one can bound \eqref{eq:expect-intro} by the corresponding quantity using the matrix $\bfA''$, which concatenates $\bfS\bfA$ with $\bfA$. This allows one to argue that the rows of $\bfA''$ corresponding to $\bfS\bfA$ indeed do have Lewis weights bounded by $\eps^2$. Furthermore, one can replace $\bfA$ in $\bfA''$ by a version $\bfA'$ of $\bfA$ which splits rows with large Lewis weight into multiple copies, so that every row of $\bfA''$ in fact has Lewis weight bounded by $\eps^2$. The result then follows by applying existing bounds from \cite{Tal1990, LT1991, Tal1995, SZ2001}.

\paragraph{Circumventing non-monotonicity via online Lewis weights.}

The part of the above argument which breaks for $p>2$ is precisely in the lack of monotonicity for $p>2$. That is, if we add a row to a matrix $\bfA$, then the Lewis weights of the existing rows may in fact increase. What we would really like in order for the above proof to go through is to define a monotonic version of $\ell_p$ Lewis weights which behaves well with respect to row additions. This motivates a connection to ideas from \emph{online numerical linear algebra} \cite{CMP2020, BDMMUWZ2020, WY2022}. 

The change we make to the above argument is the following: rather than considering the $\ell_p$ Lewis weights for the matrix $\bfA''$, we define a version of \emph{online} Lewis weights for $\bfA''$ as follows. We first use the existing weights for the split up version $\bfA'$ of $\bfA$. Then, we define weights for $\bfS\bfA$ by treating them as a batch of rows which arrive after the rows $\bfA'$. We show that such weights exist and thus satisfy a batch monotonicity, which is sufficient to show that the weights are bounded by $\eps^2$. Furthermore, we can show that these weights also have a small sum. We also show that the weights defined in this way satisfy a one-sided Lewis weight property, which we show is sufficient to make the result of \cite{LT1991} still go through. Our result follows.

Our full proof of this result can be found in Section \ref{sec:one-shot}.

\subsection{Independent and Concurrent Work}

The independent and concurrent work of \cite{CLS2022} solves the problem of online active $\ell_p$ regression, and obtains similar results on online $\ell_p$ subspace embeddings. Their result proves a bound of roughly $\tilde O(d/\eps^2)$ for $p\in[1, 2]$, and proceeds by storing an $\ell_p$ subspace embedding formed using a merge-and-reduce strategy \cite{BDMMUWZ2020}, which they then use to estimate their version of online Lewis weights, which have a slightly different definition than what we use. Our approach is similar to theirs in the sense that both approaches rely on using separate, decoupled sketches as ``side information'' to estimate the Lewis weights and the online $\ell_p$ subspace embedding. Our use of using simpler side information is inspired by \cite{CLS2022}; in a previous version of this draft, the side information we used was a deterministic online coreset algorithm for $\ell_2$ subspace embeddings by \cite{BDMMUWZ2020, CMP2020}. Upon seeing \cite{CLS2022}, we simplified the side information we use to just exactly and deterministically maintaining the online Lewis quadratic itself.

Our algorithm is arguably simpler, as we only need to maintain a Lewis quadratic for which the update is just rank one updates to a quadratic, whereas the algorithm of \cite{CLS2022} requires a merge-and-reduce procedure, and loses $\log n$ factors in the side information memory. We also offer an online coreset algorithm for $p < 2$ which does not use side information at all, so that the only storage needed are reweighted input rows that are stored in an online manner. We also answer their open question on handling $p > 2$; in particular, we circumvent the lack of monotonicity for $p > 2$ by defining online $\ell_p$ Lewis weights with a one-sided Lewis weight property, which we show is sufficient for sampling an $\ell_p$ subspace embedding.

\subsection{Open Directions}

We suggest some future directions concerning Lewis weight sampling and $\ell_p$ subspace embeddings. 

Perhaps the most important question in this area, which has been studied extensively since the works of \cite{BLM1989, Tal1990, LT1991, Tal1995}, is on obtaining optimal bounds for the $\ell_p$ subspace embedding problem: what is the smallest number of dimensions required for a $(1+\eps)$-approximate $\ell_p$ subspace embedding? For $p = 2$, \cite{BSS2012} obtained a bound of $O(\eps^{-2}d)$, which is optimal up to constant factors. On the other hand, for $p\neq 2$, all known upper bounds suffer polylogarithmic losses of the form $\eps^{-2}d^{1\lor(p/2)}\poly\log d$. 

For $p>2$, lower bounds are still lacking, even when we restrict to subset selection-based approaches. That is, if $\bfS$ is a diagonal matrix with $m$ nonzero entries such that $\norm*{\bfS\bfA\bfx}_p = (1\pm\eps)\norm*{\bfA\bfx}_p$ for every $\bfx\in\mathbb R^d$, then must $m$ be $\Omega(d^{p/2}/\eps^2)$? The current best known bound due to \cite{LWW2021} is $\Omega(d^{p/2} + 1/\eps^2)$. 

Finally, we ask whether sensitivity sampling yields $\ell_p$ subspace embeddings with $\eps^{-2}d^{1\lor(p/2)}\poly\log d$ rows, or even $\poly(\eps^{-1})d^{1\lor(p/2)}\poly\log d$ rows. Currently, the only known way to get this bound is to sample proportionally to Lewis weights. Is it possible that sampling proportionally to sensitivity scores \eqref{eq:lp-sensitivity} also gives such a bound, or is there a lower bound ruling out this sampling algorithm?

\section{Preliminaries}

\subsection{Lewis Weights}

Lewis weights were initially discovered in the functional analysis community by \cite{Lew1978}, who used them to obtain optimal bounds on distances between subspaces of $\ell_2$ and $\ell_p$, in the sense of Banach--Mazur distance. The use of Lewis weights as sampling probabilities for approximating $d$-dimensional subspaces of $L_p$ was first introduced by \cite{Sch1987}, whose results were subsequently refined and extended by \cite{BLM1989, Tal1990, LT1991, Tal1995, SZ2001}. This technique was then brought to the algorithms community \cite{CP2015}, whose definition of Lewis weights we give below:

\begin{Definition}[$\ell_p$ Lewis weights \cite{CP2015}]\label{def:lewis-weights}
Let $\bfA\in\mathbb R^{n\times d}$ and $p\in(0,\infty)$. Then, the $\ell_p$ Lewis weights are the unique weights $\bfw\in\mathbb R^n$ which satisfy
\[
    \bfw_i = \bracks*{\bfa_i^\top(\bfA^\top\bfW^{1-2/p}\bfA)^- \bfa_i}^{p/2},
\]
where $\bfW = \diag(\bfw)$. We denote these weights as $\bfw_i^p(\bfA)$, or $\bfW_i^p(\bfA)$ for its diagonal matrix.
\end{Definition}

While this definition is recursive since $\bfw$ appears on both sides of the equation, the existence of such weights is nonetheless proven by \cite{Lew1978, SZ2001, CP2015}. Furthermore, \cite{CP2015} give efficient, and in fact input sparsity time, algorithms for approximating these weights. Algorithms for approximating Lewis weights have subsequently been improved by works such as \cite{Lee2016, FLPS2022, JLS2021}. 

It has since been shown that relaxations of Definition \ref{def:lewis-weights} can be more useful, which admit more efficient approximation algorithms while still leading to similar guarantees for applications \cite{JLS2021, WY2022}. We refer to this as the $\gamma$-one-sided $\ell_p$ Lewis property:

\begin{Definition}[One-sided $\ell_p$ Lewis weights and bases \cite{WY2022}]\label{def:one-sided-lewis}
Let $\bfA\in\mathbb R^{n\times d}$ and $p\in(0,\infty)$. Let $\gamma\in(0,1]$. Then, weights $\bfw\in\mathbb R^n$ are \emph{$\gamma$-one-sided $\ell_p$ Lewis weights} if
\[
    \bfw_i \geq \gamma \cdot \bftau_i(\bfW^{1/2-1/p}\bfA),
\]
where $\bfW\coloneqq\diag(\bfw)$, or equivalently,
\[
    \bfw_i \geq \gamma^{p/2} \bracks*{\bfa_i^\top(\bfA^\top\bfW^{1-2/p}\bfA)\bfa_i}^{p/2}.
\]
If $\gamma = 1$, we just say that $\bfw$ are \emph{one-sided $\ell_p$ Lewis weights} Let $\bfR\in\mathbb R^{d\times d}$ be a change of basis matrix such that $\bfW^{1/2-1/p}\bfA\bfR$ has orthonormal columns. Then, $\bfA\bfR$ is a \emph{one-sided $\ell_p$ Lewis basis}.
\end{Definition}

We note that \cite{JLS2021, WY2022} used this definition only with $\gamma = 1$. The flexibility of allowing for $\gamma = \Theta(1) < 1$ will, as we show, not affect any of the bounds for sampling, while allowing for more convenient approximation when handling rows in a batched online setting Section \ref{sec:batch}.

We will also frequently refer to the associated quadratic form, which is $\bfA^\top\bfW^{1-2/p}\bfA$. Several other properties similar to those of Lewis weights carry over to one-sided Lewis weights, which will be useful to us. The following properties of $\gamma$-one-sided $\ell_p$ Lewis weights are proven in \cite{WY2022} for $\gamma = 1$. The extension to $\gamma\in(0,1]$ is straightforward from their proof.

\begin{Lemma}[Lemmas 2.8 and 2.10 of \cite{WY2022}]\label{lem:oslw-sensitivity}
Let $\bfA\in\mathbb R^{n\times d}$ and $p\in(0,\infty)$. Let $\bfw$ be $\gamma$-one-sided $\ell_p$ Lewis weights for $\bfA$ and let $\bfR$ be a one-sided $\ell_p$ Lewis basis. Then,
\begin{itemize}
    \item for every $i\in[n]$,
    \[
        \frac{\bfw_i}{\gamma^{p/2}} \geq \norm*{\bfe_i^\top\bfA\bfR}_2^p
    \]
    \item for all $\bfx\in\mathbb R^d$,
    \[
        \norm*{\bfW^{1/2-1/p}\bfA\bfx}_2 \leq \begin{cases} \norm*{\bfw}_1^{1/2-1/p}\norm*{\bfA\bfx}_p & \text{if $p\geq 2$} \\
        \frac1{\gamma^{1/p-1/2}}\norm*{\bfA\bfx}_p & \text{if $p < 2$}
        \end{cases}
    \]
    \item for every $i\in[n]$, 
    \[
        \sup_{\bfx\in\rowspan(\bfA)\setminus\{0\}} \frac{\abs*{\angle*{\bfa_i,\bfx}}^p}{\norm*{\bfA\bfx}_p^p} \leq 
        \begin{cases} \norm*{\bfw}_1^{p/2-1}\bfw_i & \text{if $p\geq 2$} \\
        \frac1{\gamma^{1-p/2}}\bfw_i & \text{if $p < 2$}
        \end{cases}
    \]
\end{itemize}
\end{Lemma}

Similarly, we have the following simple modification of a result from \cite{JLS2021}:

\begin{Lemma}[Lemma 2.6, \cite{JLS2021}]\label{lem:lp-overestimate}
Let $p>2$. Let $\bfw\in\mathbb R^n$ be $\gamma$-one-sided $\ell_p$ Lewis weights for $\bfA\in\mathbb R^{n\times d}$. Then, for all $\bfx\in\mathbb R^d$,
\[
    \norm*{\bfA\bfx}_p \leq \frac1{\gamma^{1/p-1/2}}\norm*{\bfW^{1/2-1/p}\bfA\bfx}_2.
\]
\end{Lemma}

\subsubsection{Lemmas from Linear Algebra}

We record a couple of linear algebraic lemmas that we will use repeatedly.

\begin{Lemma}\label{lem:pinv-psd}
Let $\bfR = \bfV\tilde\bfR\bfV^\top \in\mathbb R^{d\times d}$ where $\tilde\bfR\in\mathbb R^{r\times r}$ is a symmetric positive definite matrix and $\bfV\in\mathbb R^{d\times r}$ has orthonormal columns. Then,
\[
    \bfR^- = \bfV\tilde\bfR^{-1}\bfV^\top.
\]
\end{Lemma}
\begin{proof}
Note that $\bfV\tilde\bfR^{-1}\bfV^\top$ is an inverse for the column space of $\bfR$, i.e.,
\[
    \bfR(\bfV\tilde\bfR^{-1}\bfV^\top)\bfR = \bfV\tilde\bfR\tilde\bfR^{-1}\tilde\bfR\bfV^\top = \bfR
\]
and a weak inverse, i.e.,
\[
    (\bfV\tilde\bfR^{-1}\bfV^\top)\bfR(\bfV\tilde\bfR^{-1}\bfV^\top) = \bfV\tilde\bfR^{-1}\bfV^\top.
\]
One can also easily check that both $\bfR(\bfV\tilde\bfR^{-1}\bfV^\top)$ and $(\bfV\tilde\bfR^{-1}\bfV^\top)\bfR$ are Hermitian. Thus, $\bfV\tilde\bfR^{-1}\bfV^\top$ is uniquely determined to be the pseudoinverse of $\bfR$.
\end{proof}

\begin{Lemma}\label{lem:psd-flip}
Let $0\preceq\bfR\preceq\bfS \in \mathbb R^{d\times d}$ by symmetric positive semidefinite matrices. Let $\bfa\in\rowspan(\bfR)$. Then,
\[
    \bfa^\top\bfR^-\bfa \geq \bfa^\top\bfS^-\bfa.
\]
\end{Lemma}
\begin{proof}
Let $\bfV\in\mathbb R^{d\times r}$ be an orthonormal basis for $V\coloneqq\rowspan(\bfR)$, where $r = \dim(V)$. Let $\bfP = \bfV\bfV^\top$ be the projection matrix onto $V$. Write $\bfa = \bfV\bfb$ for $\bfb\in\mathbb R^r$ and $\bfR = \bfV\tilde\bfR\bfV^\top$, $\bfP\bfS\bfP = \bfV\tilde\bfS\bfV^\top$ for $\tilde\bfR,\tilde\bfS\in\mathbb R^{r\times r}$. Then, we have that
\[
    \bfa^\top\bfR^-\bfa = \bfb^\top\bfV^\top(\bfV\tilde\bfR\bfV^\top)^-\bfV\bfb = \bfb^\top \tilde\bfR^{-1}\bfb
\]
and
\[
    \bfa^\top\bfS^-\bfa = \bfb^\top\bfV^\top(\bfV\tilde\bfS\bfV^\top)^-\bfV\bfb = \bfb^\top\tilde\bfS^{-1}\bfb. 
\]
Furthermore, for all $\bfx\in\mathbb R^r$, we have that
\[
    \bfx^\top\bfV^\top\bfR\bfV\bfx \leq \bfx^\top\bfV^\top\bfS\bfV\bfx = \bfx^\top\bfV^\top\bfP\bfS\bfP\bfV\bfx
\]
so $\tilde\bfR \preceq \tilde\bfS$, meaning that $\tilde\bfR^{-1} \succeq \tilde\bfS^{-1}$. Thus,
\[
    \bfa^\top\bfR^-\bfa = \bfb^\top \tilde\bfR^{-1}\bfb \geq \bfb^\top \tilde\bfS^{-1}\bfb = \bfa^\top\bfS^-\bfa.\qedhere
\]
\end{proof}
\section{Online Lewis Weights}\label{sec:online-lewis-weights}

In this section, we introduce both known and new results in online numerical linear algebra, especially pertaining to online Lewis weights.

For a matrix $\bfA\in\mathbb R^{n\times d}$, $\bfA_j\in\mathbb R^{j\times d}$ denotes the submatrix of $\bfA$ formed by the first $j$ rows. The following notion of online leverage scores was introduced by \cite{CMP2020, BDMMUWZ2020}:

\begin{Definition}[Online Leverage Scores]\label{def:ols}
Let $\bfA\in\mathbb R^{n\times d}$. Then, for each $i\in[n]$, the $i$th online leverage score is defined as
\[
    \bftau_i^{\OL}(\bfA) \coloneqq \begin{cases}
        \min\braces*{\bfa_i^\top(\bfA_{i-1}^\top\bfA_{i-1})^{-}\bfa_i, 1} & \text{if $\bfa_i\in\rowspan(\bfA_{i-1})$} \\
        1 & \text{otherwise}
    \end{cases}
\]
\end{Definition}

It is not hard to see that the online leverage scores are at least the standard leverage scores. It can also be shown that the sum of online leverage scores is not much more than the sum of the standard leverage scores.

\begin{Definition}\label{def:online-condition-num}
Let $\bfA\in\mathbb R^{n\times d}$. Define the \emph{online condition number} of $\bfA$ to be
\[
    \kappa^\OL = \kappa^\OL(\bfA) \coloneqq \norm*{\bfA}_2\max_{i=1}^n \norm*{\bfA_i^-}_2.
\]
\end{Definition}

\begin{Lemma}[Sum of Online Leverage Scores \cite{CMP2020}]\label{lem:sum-of-ols}
Let $\bfA\in\mathbb R^{n\times d}$. Then,
\[
    \sum_{i=1}^n \bftau_i^\OL(\bfA) \leq O(d\log \kappa^{\OL})
\]
\end{Lemma}
\begin{proof}
This follows from setting $\lambda = (\max_{i=1}^n \norm*{\bfA_i^-}_2)^{-1}$, i.e., the minimum singular value for any $\bfA_i$ for $i\in[n]$, in Theorem 2.2 of \cite{CMP2020}. The result follows by noticing that with this choice of $\lambda$, the online ridge leverage scores used in \cite{CMP2020} are the same as the online leverage scores, up to constant factors.
\end{proof}

We now give the definition for the online $\ell_p$ Lewis weights, which are defined analogously to online leverage scores (Definition \ref{def:ols}). Similar definitions have appeared in \cite{BDMMUWZ2020, WY2022}.

\begin{Definition}[Online $\ell_p$ Lewis Weights]
Let $\bfA\in\mathbb R^{n\times d}$ and $0 < p < \infty$. Then, for each $i\in[n]$, the $i$th online $\ell_p$ Lewis weight is defined as
\[
    \bfw_i^{p,\OL}(\bfA) \coloneqq \begin{cases}
        \min\braces*{\bracks*{\bfa_i^\top(\bfA_{i-1}^\top\bfW^{p,\OL}(\bfA)_{i-1}^{1-2/p}\bfA_{i-1})^{-}\bfa_i}^{p/2}, 1} & \text{if $\bfa_i\in\rowspan(\bfA_{i-1})$} \\
        1 & \text{otherwise}
    \end{cases}
\]
where $\bfW^{p,\OL}(\bfA)_{j}$ is the $j\times j$ diagonal matrix with $\bfW^{p,\OL}(\bfA)_j(i,i) = \bfw_i^{p,\OL}(\bfA)$. 
\end{Definition}

Note that by maintaining the online Lewis quadratic $\bfA_{i-1}^\top\bfW^{p,\OL}(\bfA)_{i-1}^{1-2/p}\bfA_{i-1}$, we can access $\bfw_i^{p,\OL}(\bfA)$ upon the arrival of row $\bfa_i$ by using only $O(d^2)$ floating point numbers of ``side information'', i.e., memory that is not stored in the form of reweighted rows of $\bfA$. 

We first show that for $0 < p < 2$, the online Lewis weights upper bound Lewis weights.

\begin{Lemma}\label{lem:online-lewis-bound}
Let $\bfA\in\mathbb R^{n\times d}$ and $0 < p < 2$. Then, for each $i\in[n]$,
\[
    \bfw_i^p(\bfA) \leq \bfw_i^{p,\OL}(\bfA)
\]
\end{Lemma}
\begin{proof}
We proceed by induction. It suffices to consider the case when $\bfw_i^{p,\OL}(\bfA) < 1$, since $\bfw_i^p(\bfA)\leq 1$ for every $i\in[n]$. In particular, $\bfa_i\in\rowspan(\bfA_{i-1})$ and
\[
    \bfw_i^{p,\OL}(\bfA) = \bracks*{\bfa_i^\top(\bfA_{i-1}^\top\bfW^{p,\OL}(\bfA)_{i-1}^{1-2/p}\bfA_{i-1})^{-}\bfa_i}^{p/2}.
\]
Then, since $1 - \frac2p < 0$, we have that
\begin{align*}
    \bfW^{p,\OL}(\bfA)_{i-1} \succeq \bfW^p(\bfA)_{i-1} \succ 0 &\implies \bfW^{p,\OL}(\bfA)_{i-1}^{1-2/p} \preceq \bfW^p(\bfA)_{i-1}^{1-2/p} \\
    &\implies \bfA_{i-1}^\top(\bfW^{p,\OL}(\bfA)_{i-1}^{1-2/p} - \bfW^p(\bfA)_{i-1}^{1-2/p})\bfA_{i-1} \preceq 0 \\
    &\implies \bfA_{i-1}^\top\bfW^{p,\OL}(\bfA)_{i-1}^{1-2/p}\bfA_{i-1}  \preceq  \bfA_{i-1}^\top\bfW^p(\bfA)_{i-1}^{1-2/p}\bfA_{i-1}.
\end{align*}
By Lemma \ref{lem:psd-flip}, it follows that for every $\bfa\in\rowspan(\bfA_{i-1})$,
\[
    \bfa^\top(\bfA_{i-1}^\top\bfW^{p,\OL}(\bfA)_{i-1}^{1-2/p}\bfA_{i-1})^{-}\bfa \geq \bfa^\top(\bfA_{i-1}^\top\bfW^p(\bfA)_{i-1}^{1-2/p}\bfA_{i-1})^- \bfa.
\]
Similarly, we have that
\begin{align*}
    \bfa^\top(\bfA_{i-1}^\top\bfW^p(\bfA)_{i-1}^{1-2/p}\bfA_{i-1})^- \bfa &\geq \bfa^\top(\bfA^\top\bfW^p(\bfA)^{1-2/p}\bfA)^- \bfa
\end{align*}
for every $\bfa\in\rowspan(\bfA_{i-1})$. The result follows by taking $p/2$-th roots on the chain of inequalities. 
\end{proof}

Note that for $p > 2$, the above proof fails since $1 - \frac{2}{p} > 0$, which causes the inequalities to go the wrong way. Nevertheless, we show that these weights satisfy the \emph{one-sided Lewis property}, which is in fact sufficient to make the chaining argument go through.

\begin{Lemma}[One-Sided Lewis Property of Online Lewis Weights]\label{lem:one-sided-online-lw}
Let $\bfA\in\mathbb R^{n\times d}$ and $0 < p < \infty$. Then, for each $i\in[n]$,
\[
    \bfw_i^{p,\OL}(\bfA) \geq \bftau_i(\bfW^{p,\OL}(\bfA)^{1/2-1/p}\bfA).
\]
\end{Lemma}
\begin{proof}
We already have the result when $\bfa_i\notin\rowspan(\bfA_{i-1})$, so we assume $\bfa_i\in\rowspan(\bfA_{i-1})$. Similarly, we can assume that $\bfw_i^{p,\OL}(\bfA) < 1$. In this case,
\begin{align*}
    \bfw_i^{p,\OL}(\bfA) &= \bracks*{\bfa_i^\top(\bfA_{i-1}^\top\bfW^{p,\OL}(\bfA)_{i-1}^{1-2/p}\bfA_{i-1})^-\bfa_i}^{p/2}
\end{align*}
which rearranges to
\[
    \bfw_i^{p,\OL}(\bfA) = (\bfw_i^{p,\OL}(\bfA)^{1/2-1/p}\bfa_i)^\top(\bfA_{i-1}^\top\bfW^{p,\OL}(\bfA)_{i-1}^{1-2/p}\bfA_{i-1})^-(\bfw_i^{p,\OL}(\bfA)^{1/2-1/p}\bfa_i).
\]
By Lemma \ref{lem:psd-flip}, this is bounded below by
\[
    (\bfw_i^{p,\OL}(\bfA)^{1/2-1/p}\bfa_i)^\top(\bfA^\top\bfW^{p,\OL}(\bfA)^{1-2/p}\bfA)^-(\bfw_i^{p,\OL}(\bfA)^{1/2-1/p}\bfa_i) = \bftau_i(\bfW^{p,\OL}(\bfA)^{1/2-1/p}\bfA),
\]
which is the claimed result.
\end{proof}

\subsection{The Sum of Online Lewis Weights}

Finally, we bound the sum of online Lewis weights, using bounds on the sum of online leverage scores. Our proof substantially simplifies the proofs of \cite[Lemma 4.7, Lemma 5.15]{BDMMUWZ2020}, which relied on an elaborate argument involving recursive applications of a ``whack-a-mole'' lemma of \cite{CLMMPS2015}, and also slightly improves the bound by logarithmic factors.

\begin{Lemma}[Sum of Online Lewis Weights]\label{lem:online-lewis-sum-bound}
Let $\bfA\in\mathbb R^{n\times d}$ and $0 < p < \infty$. Then,
\[
    \sum_{i=1}^n \bfw_i^{p,\OL}(\bfA) \leq O(d)\log(n\kappa^\OL(\bfA)).
\]
\end{Lemma}
\begin{proof}
Our analysis is similar to those given by \cite{CMP2020} and \cite{WY2022}. For $\bfw_i^{p,\OL}(\bfA) < 1$, we have that
\[
    \bfw_i^{p,\OL}(\bfA) = \bracks*{\bfa_i^\top(\bfA_{i-1}^\top\bfW^{p,\OL}(\bfA)_{i-1}^{1-2/p}\bfA_{i-1})^{-}\bfa_i}^{p/2}.
\]
This rearranges to
\[
    \bfw_i^{p,\OL}(\bfA) = (\bfw_i^{p,\OL}(\bfA)^{1/2-1/p}\bfa_i)^\top(\bfA_{i-1}^\top\bfW^{p,\OL}(\bfA)_{i-1}^{1-2/p}\bfA_{i-1})^{-}(\bfw_i^{p,\OL}(\bfA)^{1/2-1/p}\bfa_i),
\]
which is exactly the $i$th online leverage score of $\bfW^{p,\OL}(\bfA)^{1/2-1/p}\bfA$. Similar reasoning for $\bfw_i^{p,\OL}(\bfA) = 1$ shows that $\bfw_i^{p,\OL}(\bfA) = \bftau_i^\OL(\bfW^{p,\OL}(\bfA)^{1/2-1/p}\bfA)$. Thus,
\[
    \sum_{i=1}^n \bfw_i^{p,\OL}(\bfA) = \sum_{i=1}^n \bftau_i^{\OL}(\bfW^{p,\OL}(\bfA)^{1/2-1/p}\bfA) \leq O(d\log\kappa^\OL(\bfW^{p,\OL}(\bfA)^{1/2-1/p}\bfA))
\]
by Lemma \ref{lem:sum-of-ols}. If $p < 2$, then we have for any $\bfx\in\mathbb R^d$ and $i\in[n]$ that
\[
    \norm*{\bfA_i\bfx}_2 \leq \norm*{\bfW^{p,\OL}(\bfA_i)^{1/2-1/p}\bfA_i\bfx}_2 \leq \norm*{\bfW^{p}(\bfA_i)^{1/2-1/p}\bfA_i\bfx}_2 \leq d^{\abs*{1/2-1/p}}\norm*{\bfA_i\bfx}_p \leq (nd)^{\abs*{1/2-1/p}}\norm*{\bfA_i\bfx}_2,
\]
so $\kappa^\OL(\bfA) = \poly(n)\kappa^\OL(\bfW^{p,\OL}(\bfA)^{1/2-1/p}\bfA)$. If $p > 2$, then by Lemma \ref{lem:lp-overestimate},
\[
    \norm*{\bfA_i\bfx}_p \leq \norm*{\bfW^{p,\OL}(\bfA_i)^{1/2-1/p}\bfA_i\bfx}_2 \leq \norm*{\bfA_i\bfx}_2.
\]
Thus,
\[
    \sum_{i=1}^n \bfw_i^{p,\OL}(\bfA) \leq O\parens*{d}\log(n\kappa^\OL(\bfA)).\qedhere
\]
\end{proof}

As a result, of our online Lewis weights as well as our new one-shot Lewis weight sampling result of Theorem \ref{thm:one-shot-lewis-weight-sampling} for $p>2$ or Theorem \ref{thm:lewis-weight-sampling-0<p<2} for $p<2$, we obtain our main online sampling result:

\begin{Theorem}\label{thm:online-lewis-p>2}
    Let $\bfA\in\mathbb R^{n\times d}$ and $p\in(0,\infty)$. Let $\delta\in(0,1)$ be a failure rate parameter and let $\eps\in(0,1)$ be an accuracy parameter. Then there is an online coreset algorithm $\mathcal A$ such that, with probability at least $1-\delta$, $\mathcal A$ outputs a weighted subset of $m$ rows with sampling matrix $\bfS$ such that
    \[
        \norm*{\bfS_i\bfA_i\bfx}_p^p = (1\pm\eps)\norm*{\bfA_i\bfx}_p^p
    \]
    for all $\bfx\in\mathbb R^d$, for every $i\in[n]$, and
    \[
        m = \begin{dcases}
            O\parens*{\frac{d^{p/2}}{\eps^2}}(\log(n\kappa^\OL))^{p/2+1}\bracks*{(\log d)^2(\log n) + \log\frac1\delta} & p\in(2,\infty) \\
            O\parens*{\frac{d}{\eps^2}}\log(n\kappa^\OL)\bracks*{(\log d)^2\log n + \log\frac1\delta} & p \in (1,2) \\
            O\parens*{\frac{d}{\eps^2}}\log(n\kappa^\OL)\log\frac{n}{\delta} & p = 1 \\
            O\parens*{\frac{d}{\eps^2}}\log(n\kappa^\OL)\bracks*{(\log d)^3 + \log\frac1\delta} & p\in(0,1)
        \end{dcases}
    \]
\end{Theorem}

By using the fact that one-sided $\ell_p$ Lewis weights bound the $\ell_p$ sensitivities (Lemma \ref{lem:oslw-sensitivity}) as well as the above bound on the online $\ell_p$ Lewis weights, we obtain a bound on the sum of $\ell_p$ online sensitivities. This significantly generalizes Lemma 4.7 of \cite{BDMMUWZ2020} and shaves a log factor.

\begin{Corollary}[Sum of Online Sensitivities]\label{cor:online-sensitivities}
Let $p\in(0,\infty)$ and $\bfA\in\mathbb R^{n\times d}$. Define the $i$th online $\ell_p$ sensitivity as
\[
    \bfs_i^{p,\OL}(\bfA) \coloneqq \begin{cases}
    \min\braces*{1, \sup_{\norm*{\bfA_{i-1}\bfx}_p = 1}\abs*{[\bfA\bfx](i)}^p} & \text{if $\bfa_i\in\rowspan(\bfA_{i-1})$} \\
    1 & \text{otherwise}
    \end{cases}
\]
Then,
\[
    \sum_{i=1}^n \bfs_i^{p,\OL}(\bfA) \leq O(d\log(n\kappa^\OL(\bfA))^{1\lor (p/2)}
\]
\end{Corollary}
\begin{proof}
We instead bound the sum of scores
\[
    \sup_{\norm*{\bfA_{i}\bfx}_p = 1}\abs*{[\bfA\bfx](i)}^p.
\]
Note that this is within a factor of $2$ of $\bfs_i^{p,\OL}$, and is just $\bfs_i^p(\bfA_i)$. Indeed, for an $\ell_p$ unit vector $\bfA_i\bfx$ with $\abs*{[\bfA\bfx](i)}^p = a$, if $a \geq 1/2$, then we are done, and otherwise,
\[
    \frac{a}{1-a} - a = \frac{a^2}{1-a} \leq \frac12 \frac{a}{1-a} \implies \frac{a}{1-a} \leq 2a.
\]
Then, by Lemma 2.10 of \cite{WY2022}, we have that
\[
    \sup_{\norm*{\bfA_{i}\bfx}_p = 1}\abs*{[\bfA\bfx](i)}^p \leq \norm*{\bfw_i^{p,\OL}(\bfA_i)}_1 ^{0\lor(p/2-1)}\bfw_i^{p,\OL}(\bfA_i)
\]
so summing over $i$ yields the desired result.
\end{proof}

The above proof applies to any set of weights $\bfw$ that satisfy an \emph{online one-sided Lewis weights} property; we simply need to replace Lemma 2.10 of \cite{WY2022} with Lemma \ref{lem:oslw-sensitivity}. This flexibility is useful when one desires an algorithmic approximation on the online sensitivities.

\begin{Definition}[Online $\gamma$-One-sided $\ell_p$ Lewis Weights]\label{def:online-oslw}
Let $\bfA\in\mathbb R^{n\times d}$ and $0 < p < \infty$. Let $\gamma\in(0,1]$. Then, $\bfw\in\mathbb R^n$ are \emph{one-sided online $\ell_p$ Lewis weights} if for each $i\in[n]$, $\bfw$ restricted to the first $i$ rows are $\gamma$-one-sided $\ell_p$ Lewis weights (Definition \ref{def:one-sided-lewis}) of $\bfA_i$. 
\end{Definition}

\begin{Corollary}
Let $\bfw$ be online $\gamma$-one-sided $\ell_p$ Lewis weights for $\bfA$. Then,
\[
    \sum_{i=1}^n \bfs_i^{p,\OL}(\bfA)\leq O\parens*{\frac{\norm*{\bfw}_1^{1\lor(p/2)}}{\gamma^{0\lor(1/p-1/2)}}}
\]
\end{Corollary}

\section{Batch Processing of Rows}\label{sec:batch}

In many practical scenarios, it may be convenient to consider variants of the online coreset model in which multiple row may be processed in batches at a time, in order to save on running time. For example, this may corresponding to data arriving in packets over the internet. Such improvements are considered in \cite{CMP2020, BDMMUWZ2020} for online leverage scores. We show how this can be done, even for online Lewis weights.

Recall that highly efficient algorithms for approximating leverage scores are known, running in time $\tilde O(\nnz(\bfA) + d^\omega)$ for $(1+\eps)$-factor approximations to the leverage scores \cite{SS2011, DMMW2012, CLMMPS2015}. We let $\textsc{ApproxLev}(\bfA,\eps)$ refer to such a routine.

\subsection{Batch Online Lewis Weights, \texorpdfstring{$0<p<4$}{0 < p < 4}}

\begin{algorithm}
	\caption{Batch Online Lewis Weights, $p\in(0,4)$}
	\textbf{input:} Previous Lewis quadratic $\bfM$, new rows $\bfA\in\mathbb R^{n\times d}$, $p\in(0,4)$. \\
	\textbf{output:} Batch online Lewis weight $\braces*{\tilde\bfw_i}_{i=1}^n$.
	\begin{algorithmic}[1] %
	    \State $\tilde\bfw_i^{(0)} \gets 1$ for all $i\in[n]$
	    \For{$t\in[T]$}
	        \State $\bfB \gets \begin{bmatrix} \diag(\tilde\bfw^{(t-1)})^{1/2-1/p}\bfA \\ \bfM^{1/2} \end{bmatrix}$
	        \State $\tilde\bfw_i^{(t)} \gets \textsc{ApproxLev}(\bfB,1/10)$ for $i\in[n]$
	    \EndFor
	    \State \Return $\tilde\bfw^{(T)}$
	\end{algorithmic}\label{alg:batch-online-lewis-p<4}
\end{algorithm}

We recall the following notation of \cite{CP2015}. For two nonnegative numbers $v, w$, we denote $v\approx_\alpha w$ to mean $v/\alpha \leq w \leq \alpha v$. We extend this to nonnegative vectors $\bfv,\bfw$, as well as to symmetric PSD matrices via the L\"owner order. 

We first adapt Lemma 3.2 of \cite{CP2015} to the batch online setting:

\begin{Lemma}\label{lem:cp2015-lem32-analogue}
Let $\bfA\in\mathbb R^{n\times d}$ and let $\bfM\in\mathbb R^{d\times d}$ be a symmetric PSD matrix. Suppose that nonnegative vectors $\bfv,\bfw\in\mathbb R^n$ satisfy $\bfv\approx_\alpha\bfw$ for $\alpha\geq1$. Then,
\[
    [\bfa_i^\top(\bfA^\top\bfV^{1-2/p}\bfA + \bfM)^{-}\bfa_i]^{p/2} \approx_{\alpha^{\abs*{p/2-1}}}[\bfa_i^\top(\bfA^\top\bfW^{1-2/p}\bfA + \bfM)^{-}\bfa_i]^{p/2}.
\]
\end{Lemma}
\begin{proof}
For $\bfv\approx_\alpha\bfw$, we have $\bfV^{1-2/p}\approx_{\alpha^{\abs*{1-2/p}}}\bfW^{1-2/p}$. Then,  $\bfA^\top\bfV^{1-2/p}\bfA\approx_{\alpha^{\abs*{1-2/p}}} \bfA^\top\bfW^{1-2/p}\bfA$, so $\bfA^\top\bfV^{1-2/p}\bfA + \bfM \approx_{\alpha^{\abs*{1-2/p}}} \bfA^\top\bfW^{1-2/p}\bfA + \bfM$. By using Lemma \ref{lem:psd-flip} to apply $\bfa_i$ to the pseudoinverse quadratic form and taking $p/2$-th powers, we find that
\[
    [\bfa_i^\top(\bfA^\top\bfV^{1-2/p}\bfA + \bfM)^{-}\bfa_i]^{p/2} \approx_{\alpha^{\abs*{p/2-1}}}[\bfa_i^\top(\bfA^\top\bfW^{1-2/p}\bfA + \bfM)^{-}\bfa_i]^{p/2}.\qedhere
\]
\end{proof}

By the Banach fixed point theorem, Lemma \ref{lem:cp2015-lem32-analogue} implies the existence of the following weights:

\begin{Corollary}[Batch Online $\ell_p$ Lewis Weights]\label{cor:batch-online-lw}
Let $\bfA\in\mathbb R^{n\times d}$, let $\bfM\in\mathbb R^{d\times d}$ be a symmetric PSD matrix, and let $0<p<4$. There exists weights $\bfw_i^p(\bfA;\bfM)$ such that
\[
    \bfw_i^p(\bfA;\bfM) = \bracks*{\bfa_i^\top(\bfM + \bfA^\top\bfW^p(\bfA;\bfM)^{1-p/2}\bfA)^{-}\bfa_i}^{p/2}.
\]
\end{Corollary}

We now show that Algorithm \ref{alg:batch-online-lewis-p<4} returns multiplicative approximations to the batch online $\ell_p$ Lewis weights of Corollary \ref{cor:batch-online-lw}. We start by showing that after one iteration, we obtain a good approximation within $\poly(n)$ factors, by adapting Lemma 3.5 of \cite{CP2015}.

\begin{Lemma}
After $t=1$ in Algorithm \ref{alg:batch-online-lewis-p<4}, we have that $\tilde\bfw_i \approx_{\beta^{p/2}n^{\abs*{p/2-1}}}\bfw_i^p(\bfA;\bfM)$ for $\beta=1+1/10$.
\end{Lemma}
\begin{proof}
By rearranging the guarantee of $\bfw_i^p(\bfA;\bfM)$ in Corollary \ref{cor:batch-online-lw}, we see that $\bfw_i^p(\bfA;\bfM)$ satisfies
\[
    \bfw_i = (\bfw_i^{1/2-1/p}\bfa_i)^\top(\bfM + \bfA^\top\bfW^p(\bfA;\bfM)^{1-p/2}\bfA)^{-}(\bfw_i^{1/2-1/p}\bfa_i).
\]
That is, $\bfw_i^p(\bfA;\bfM)$ are exactly the first $n$ leverage scores of the matrix
\[
    \bfB = \begin{bmatrix} \bfW^p(\bfA;\bfM)^{1/2-1/p}\bfA \\ \bfM^{1/2} \end{bmatrix}.
\]
Thus, there exists a change of basis $\bfR$ such that $\bfB\bfR$ has orthonormal columns. We then rename $\bfA$ to $\bfA\bfR$ and $\bfM^{1/2}$ to $\bfM^{1/2}\bfR$ and proceed by assuming that $\bfB$ has orthonormal columns. Then, $\bfB^\top\bfB = \bfI_d$. We then claim that $\bfA^\top\bfA + \bfM \approx_{n^{\abs*{1-2/p}}} \bfI_d$. 

Note that $\norm*{\bfb_i}_2^2 = \bfw_i^p(\bfA;\bfM)$. Then for any unit vector $\bfu$, we have that
\[
    1 = \bfu^\top \bfu = \bfu^\top \bfB^\top\bfB\bfu = \sum_{i=1}^n (\bfu^\top\bfb_i)^2 + \bfu^\top\bfM\bfu = \sum_{i=1}^n \bfw_i^p(\bfA;\bfM)[(\bfw_i^p(\bfA;\bfM)^{-1}\bfu^\top\bfb_i)^2] + \bfu^\top\bfM\bfu
\]
while
\begin{align*}
    \bfu^\top\bfA^\top\bfA\bfu + \bfu^\top\bfM\bfu &= \sum_{i=1}^n \bfw_i^p(\bfA;\bfM)^{2/p-1}(\bfu^\top\bfb_i)^2 + \bfu^\top\bfM\bfu \\
    &= \sum_{i=1}^n \bfw_i^p(\bfA;\bfM)^{2/p}[\bfw_i^p(\bfA;\bfM)^{-1}(\bfu^\top\bfb_i)^2] + \bfu^\top\bfM\bfu.
\end{align*}
Then just as reasoned in \cite{CP2015}, the worst case distortion is $n^{\abs*{p/2-1}}$ between these two quantities. 
\end{proof}

\begin{Corollary}
Let $T = O(\log\log n)$ and let $\tilde\bfw$ be the output of Algorithm \ref{alg:batch-online-lewis-p<4}. Then, $\tilde\bfw\approx_{O(1)} \bfw_i^p(\bfA;\bfM)$.
\end{Corollary}
\begin{proof}
The total multiplicative contribution of the blowups from $\beta$ is at most $\beta^{\frac{p/2}{1-\abs*{p/2-1}}}$ for $\beta = 1+1/10$. The contribution from the starting error is at most $n^{\abs*{p/2-1}^T}$. Thus, for $T = O(\log\log n)$, we obtain an $O(1)$-approximation to the batch online $\ell_p$ Lewis weights. 
\end{proof}

Using these, we obtain that batch processing yields weights that are still one-sided $\ell_p$ Lewis weights, and are bounded by the online leverage scores of a reweighted matrix.

\begin{Lemma}\label{lem:batch-online-lewis}
Let $p\in(0,4)$, let $\bfA\in\mathbb R^{n\times d}$, and let $\bfB\in\mathbb R^{m\times d}$. Let 
\[
    \bfC = \begin{bmatrix}\bfB \\ \bfA\end{bmatrix}
\]
Let $\tilde\bfw_i$ for $i\in[m]$ be $\gamma$-one-sided $\ell_p$ Lewis weights of $\bfB$. Let $\bfM = \bfB^\top\tilde\bfW^{1-2/p}\bfB$, and let $\hat\bfw$ satisfy $\bfw_i^p(\bfA;\bfM)\leq \hat\bfw \leq \lambda\bfw_i^p(\bfA;\bfM)$. Then, the concatenation $\bfw = \tilde\bfw \circ \hat\bfw$ are $\gamma$-one-sided $\ell_p$ Lewis weights of $\bfC$ if $p\in(0,2]$ and $\min\{\gamma, \lambda^{1-p/2}\}$-one-sided $\ell_p$ Lewis weights of $\bfC$ if $p\in(2,4)$. Furthermore, if
\[ 
    \bfw_i \leq O(1) \bftau_i^{\OL}(\bfW^{1/2-1/p}\bfC)
\]
for the first $m$ rows of $\bfC$, then this is also true for all the rows. The weights $\hat\bfw$ can be computed in time $\tilde O(\nnz(\bfA)+d^\omega)$ given $\bfM$.
\end{Lemma}
\begin{proof}
For the last $n$ rows, if $0 < p\leq 2$, we have by Lemma \ref{lem:psd-flip} that
\begin{align*}
    \hat\bfw_i \geq \bfw_i^p(\bfA;\bfM) &= \bracks*{\bfa_i^\top(\bfA^\top\bfW^p(\bfA;\bfM)^{1-2/p}\bfA + \bfM)^{-}\bfa_i}^{p/2} \\
    &\geq \bracks*{\bfa_i^\top(\bfA^\top\hat\bfW^{1-2/p}\bfA + \bfM)^{-}\bfa_i}^{p/2} \\
    &= \bracks*{\bfa_i^\top(\bfC^\top \bfW^{1-2/p}\bfC)^{-}\bfa_i}^{p/2}
\end{align*}
and if $2 < p < 4$, that
\begin{align*}
    \hat\bfw_i \geq \bfw_i^p(\bfA;\bfM) &= \lambda^{1-p/2}\bracks*{\bfa_i^\top(\bfA^\top\bfW^p(\bfA;\bfM)^{1-2/p}\bfA + \bfM)^{-}\bfa_i}^{p/2} \\
    &\geq \lambda^{1-p/2}\bracks*{\bfa_i^\top(\bfA^\top\hat\bfW^{1-2/p}\bfA + \bfM)^{-}\bfa_i}^{p/2} \\
    &= \lambda^{1-p/2}\bracks*{\bfa_i^\top(\bfC^\top \bfW^{1-2/p}\bfC)^{-}\bfa_i}^{p/2}
\end{align*}
For the first $m$ rows, we have by Lemma \ref{lem:psd-flip} that
\begin{align*}
    \tilde\bfw_i &\geq \gamma\bracks*{\bfb_i^\top(\bfM)^{-}\bfb_i}^{p/2} \\
    &\geq \gamma\bracks*{\bfb_i^\top(\bfA^\top\hat\bfW^{1-2/p}\bfA + \bfM)^{-}\bfb_i}^{p/2} \\
    &= \gamma\bracks*{\bfb_i^\top(\bfC^\top \bfW^{1-2/p}\bfC)^{-}\bfb_i}^{p/2}.
\end{align*}
These rearrange to the statement that $\bfw$ are $\gamma$-one-sided $\ell_p$ Lewis weights. 

Furthermore, we have that
\begin{align*}
    \hat\bfw_i \leq O(1)\bfw_i^p(\bfA;\bfM) &= O(1)\bracks*{\bfa_i^\top(\bfA^\top\bfW^p(\bfA;\bfM)^{1-2/p}\bfA + \bfM)^{-}\bfa_i}^{p/2} \\
    &\leq O(1)\bracks*{\bfa_i^\top(\bfA^\top\hat\bfW^{1-2/p}\bfA + \bfM)^{-}\bfa_i}^{p/2} \\
    &\leq O(1)\bracks*{\bfa_i^\top(\bfC_{i-1}^\top \bfW_{i-1}^{1-2/p}\bfC_{i-1})^{-}\bfa_i}^{p/2}
\end{align*}
which rearranges to the statement that $\bfw_i \leq O(1) \bftau_i^{\OL}(\bfW^{1/2-1/p}\bfC)$.
\end{proof}

\subsection{Tensor Trick for Batch Online Lewis Weights, \texorpdfstring{$4\leq p<\infty$}{4 ≤ p < inf}}\label{sec:tensor-trick}

For $p\geq 4$, the iterative algorithm of Algorithm \ref{alg:batch-online-lewis-p<4} does not work directly. Instead, we use a trick of \cite{MMMWZ2022} to reduce $p\geq 4$ to the case of $p < 4$ as follows. Let $k$ be an integer large enough so that $2\leq p/k < 4$ (i.e., $k = \floor{p/4}+1$). Then, we define the Khatri--Rao power $\bfA^{\otimes k}\in\mathbb R^{n\times d^{k}}$, where the $i$th row is the $k$-fold tensor product $\bfa_i^{\otimes k} = \bfa_i \otimes \bfa_i \otimes \dots \otimes \bfa_i$ of $\bfa_i$ with itself. Note then that for any $\bfx\in\mathbb R^d$,
\[
    \angle*{\bfa_i^{\otimes k}, \bfx^{\otimes k}} = \angle*{\bfa_i,\bfx}^{k},
\]
so
\[
    \norm*{\bfA^{\otimes k}\bfx^{\otimes k}}_{p/k}^{p/k} = \sum_{i=1}^n \abs*{\angle*{\bfa_i^{\otimes k}, \bfx^{\otimes k}}}^{p/k} = \sum_{i=1}^n \abs*{\angle*{\bfa_i,\bfx}}^{p} = \norm*{\bfA\bfx}_p^p.
\]
Thus, it suffices to compute sensitivity upper bounds and subspace embeddings for $\ell_{p/k}$ on the Khatri--Rao power matrix $\bfA^{\otimes k}$ in order to handle $\ell_p$ for $\bfA$. Note that the $\ell_{p/k}$ Lewis weights of $\bfA^{\otimes k}$ sum to $(d^k)^{p/2k} = d^{p/2}$, so the upper bound on the sum of sensitivities is as desired. We may then just apply Lemma \ref{lem:batch-online-lewis} on the Khatri--Rao power matrix.

\subsection{Batch Online Lewis Weights, \texorpdfstring{$2\leq p<\infty$}{2 ≤ p < inf}}

While the tensor trick of Section \ref{sec:tensor-trick} is sufficient for many cases, it is still desirable to avoid this and directly get an analogue of Corollary \ref{cor:batch-online-lw}, for example in our Section \ref{sec:one-shot}. We will obtain this by modifying the classical convex program argument for the existence of $\ell_p$ Lewis weights \cite{Lew1978, SZ2001, CP2015}. This also immediately leads to polynomial time algorithms \cite{CP2015}. 

\begin{Lemma}[Batch Online $\ell_p$ Lewis Weights, $2\leq p < \infty$]\label{lem:batch-online-lewis-convex-program}
Let $\bfA\in\mathbb R^{n\times d}$, let $\bfM = \bfL^\top\bfL\in\mathbb R^{d\times d}$ be a symmetric PSD matrix, and let $2\leq p < \infty$. There exists weights $\bfw\in\mathbb R^n$ such that for $i\in[n]$,
\[
    \bfw_i = \parens*{\frac{p}{2}}^{\frac{p/2}{1-2/p}}(\bfa_i^\top(\bfA^\top\bfW^{1-2/p}\bfA+\bfM)^{-1}\bfa_i)^{p/2}
\]
and
\[
    \sum_{i=1}^n \bfw_i \leq \parens*{\frac{p}2}^{\frac1{1-2/p}}d.
\]
\end{Lemma}
\begin{proof}
Consider the following optimization problem over symmetric PSD matrices $\bfQ$:
\begin{align*}
    \mbox{maximize} \qquad& \det(\bfQ) \\
    \mbox{subject to} \qquad& \sum_{i=1}^n (\bfa_i^\top \bfQ\bfa_i)^{p/2} + \sum_{j=1}^d \bfl_j^\top\bfQ\bfl_j \leq d \\
    & \bfQ\succeq 0
\end{align*}
where $\bfa_i$ is the $i$th row of $\bfA$ and $\bfl_j$ is the $j$th row of $\bfL$. Let $\bfQ$ be any matrix which attains this maximum. Note then that
\[
    \sum_{i=1}^n (\bfa_i^\top \bfQ\bfa_i)^{p/2} + \sum_{j=1}^d \bfl_j^\top\bfQ\bfl_j = d
\]
since otherwise scaling $\bfQ$ up can increase the objective function. Furthermore, by considering Lagrange multipliers, the gradient of the constraint is some scalar $C$ times the gradient of of the objective, so
\[
    \sum_{i=1}^n \frac{p}{2}(\bfa_i^\top \bfQ\bfa_i)^{p/2-1}\bfa_i\bfa_i^\top + \sum_{j=1}^d \bfl_j\bfl_j^\top = C\det(\bfQ)\bfQ^{-1}.
\]
We now define
\[
    \bfw_i \coloneqq \parens*{\frac{p}{2}}^{\frac1{1-2/p}} (\bfa_i^\top\bfQ\bfa_i)^{p/2}.
\]
Then, we have that
\[
    \bfA^\top\bfW^{1-2/p}\bfA + \bfM = C\det(\bfQ)\bfQ^{-1}
\]
for $\bfW = \diag(\bfw)$. Rearranging, we have that
\[
    \bfQ = C\det(\bfQ)(\bfA^\top\bfW^{1-2/p}\bfA + \bfM)^{-1}
\]
so
\[
    \bfw_i = \parens*{\frac{p}{2}}^{\frac1{1-2/p}} (\bfa_i\bfQ\bfa_i)^{p/2} = \parens*{\frac{p}{2}}^{\frac1{1-2/p}} (C\det(\bfQ))^{p/2} [\bfa_i^\top(\bfA^\top\bfW^{1-2/p}\bfA + \bfM)^{-1}\bfa_i]^{p/2}
\]
and thus
\begin{align*}
    \bfw_i &= \parens*{\frac{p}{2}}^{\frac{2/p}{1-2/p}} (C\det(\bfQ)) [(\bfw_i^{1/2-1/p}\bfa_i)^\top(\bfA^\top\bfW^{1-2/p}\bfA + \bfM)^{-1}(\bfw_i^{1/2-1/p}\bfa_i)] \\
    &= \parens*{\frac{p}{2}}^{\frac{2/p}{1-2/p}} (C\det(\bfQ)) \bftau_i(\bfB)
\end{align*}
where $\bfB$ is the vertical concatenation of $\bfW^{1/2-1/p}\bfA$ and $\bfL$. Note also that for rows $j$ corresponding to $\bfL$ in $\bfB$, we have that
\[
    (C\det(\bfQ))\bftau_j(\bfB) = (C\det(\bfQ))\bfl_j^\top(\bfA^\top\bfW^{1-2/p}\bfA + \bfM)^{-1}\bfl_j = \bfl_j^\top\bfQ\bfl_j.
\]
Now by the normalization constraint, we have that
\[
    \sum_{i=1}^n \parens*{\frac2p}^{\frac1{1-2/p}}\bfw_i + \sum_{j=1}^d \bfl_j^\top\bfQ\bfl_j = \sum_{i=1}^n (\bfa_i^\top \bfQ\bfa_i)^{p/2} + \sum_{j=1}^d \bfl_j^\top\bfQ\bfl_j = d.
\]
However,
\[
    \parens*{\frac2p}^{\frac1{1-2/p}}\bfw_i = \parens*{\frac{p}2}^{\frac{-1}{1-2/p}}\parens*{\frac{p}{2}}^{\frac{2/p}{1-2/p}} (C\det(\bfQ)) \bftau_i(\bfB) = \frac2p (C\det(\bfQ))\bftau_i(\bfB)
\]
so we must have that $p/2 = C\det(\bfQ)$. The result follows.
\end{proof}

\begin{Remark}
Note that if we set $\bfM = 0$ and redefine $\bfw_i' \coloneqq \bfw_i / (p/2)^{\frac1{1-2/p}}$, then we will retrieve the usual definition of $\ell_p$ Lewis weights.
\end{Remark}
\section{Nearly Optimal One-Shot Lewis Weight Sampling, \texorpdfstring{$2<p<\infty$}{2 < p < inf}}\label{sec:one-shot}

The following is a generalization of what is known for Lewis weights to the setting of one-sided Lewis weights, with higher moments.

\begin{restatable}{Theorem}{LedouxTalagrand}\label{thm:one-sided-lt}
    Let $\bfA\in\mathbb R^{n\times d}$ and $p\in(2,\infty)$. Let $\bfw$ be $\gamma$-one-sided $\ell_p$ Lewis weights for $\bfA$. Suppose that
    \[
        \frac{\bfw_i}{d} \leq \beta
    \]
    for all $i\in[n]$, for some $\beta>0$. Define the quantity
    \[
        \Lambda \coloneqq \sup_{\norm*{\bfA\bfx}_p = 1}\abs*{\sum_{i=1}^n \bfsigma_i \abs*{\angle*{\bfa_i,\bfx}}^p}
    \]
    Then,
    \[
        \E_{\bfsigma}[\Lambda^l] \leq \bracks*{O(1)\beta\cdot T_\bfw^{p/2}[[\gamma^{-1}(\log d)^2(\log n)]^{1+1/l}+l]}^{l/2}
    \]
    for any $l\geq 1$, where $\bfsigma = \{\bfsigma_i\}_{i=1}^n$ are independent Rademacher variables.
\end{restatable}

We work out the proof of this result in detail in Appendix \ref{sec:one-sided-lewis-moment-bound}. Using this, we obtain the following analysis for a one-shot Lewis weight sampling algorithm.

\begin{Theorem}[High probability one-shot Lewis weight sampling]\label{thm:one-shot-lewis-weight-sampling}
Let $\bfA\in\mathbb R^{n\times d}$ and $2 < p < \infty$. Let $\delta\in(0,1)$ be a failure rate parameter and let $\eps\in(0,1)$ be an accuracy parameter. Let $\bfw\in\mathbb R^n$ be $\gamma$-one-sided $\ell_p$ Lewis weights that sum to $T_\bfw$. Suppose that we set $\bfs_i = 1/\bfp_i^{1/p}$ with probability $\bfp_i$, where
\[
    \bfp_i \geq \min\braces*{\parens*{\frac{(p/2)^{\frac1{1-2/p}}}{\gamma}}^{p/2}\frac{\bfw_i}{d\beta},1},
\]
for
\[
    \beta = \frac{\eps^2}{T^{p/2}[\gamma^{-1}(\log d)^2(\log n)+\log\frac1\delta]} \in (0, 1)
\]
with
\[
    T\coloneqq T_\bfw + O(d) = O(T_\bfw).
\]
Then, with probability at least $1-\delta$,
\[
    \norm*{\bfS\bfA\bfx}_p^p = (1\pm O(\eps))\norm*{\bfA\bfx}_p^p
\]
for all $\bfx\in\mathbb R^d$, and $\bfs$ samples at most
\[
    O\parens*{\frac{T^{p/2}}{\eps^2}\frac{T_\bfw}{d}\bracks*{\gamma^{-1}(\log d)^2(\log n)+\log\frac1\delta}}
\]
rows of $\bfA$ with probability at least $1-\delta$.
\end{Theorem}
\begin{proof}
We WLOG assume that $\bfA$ is just the subset of rows of the original matrix such that $\bfp_i < 1$. In particular, we may assume that $\bfw_i < 1$ for all $i\in[n]$. 

\paragraph{Symmetrization.}

We wish to bound
\begin{equation}\label{eq:moment}
    \E_{\bfs}\bracks*{\sup_{\norm*{\bfA\bfx}_p = 1}\abs*{\parens*{\sum_{i=1}^n\abs*{\angle*{\bfs_i\bfa_i,\bfx}}^p}-1}^l},
\end{equation}
for $l = O(\log\frac1\delta)$. For convenience, we in fact take $l = \max\{O(\log\frac1\delta), \log n\}$, which does not affect the bounds. Indeed, if we can bound this by $(C\cdot\eps)^l$ for $C>0$, then by Markov's inequality,
\[
    \Pr\braces*{\sup_{\norm*{\bfA\bfx}_p = 1}\abs*{\parens*{\sum_{i=1}^n\abs*{\angle*{\bfs_i\bfa_i,\bfx}}^p}-1} \geq \frac{C\cdot \eps}{\delta^{1/l}}} = \Pr\braces*{\sup_{\norm*{\bfA\bfx}_p = 1}\abs*{\parens*{\sum_{i=1}^n\abs*{\angle*{\bfs_i\bfa_i,\bfx}}^p}-1}^l \geq \frac{(C\cdot\eps)^l}{\delta}} \leq \delta,
\]
which implies that
\[
    \Pr\braces*{\mbox{for all $\bfx\in\mathbb R^d$, }\norm*{\bfS\bfA\bfx}_p^p = (1\pm O(\eps))\norm*{\bfA\bfx}} \geq 1 - \delta.
\]

By a standard symmetrization procedure \cite[Lemma 7.4]{CP2015}, \eqref{eq:moment} is bounded by
\[
    2^l \E_{\bfs}\E_{\bfsigma}\bracks*{\sup_{\norm*{\bfA\bfx}_p = 1}\abs*{\sum_{i=1}^n\bfsigma_i\abs*{\angle*{\bfs_i\bfa_i,\bfx}}^p}^l}.
\]
We now define a new matrix $\bfA''$ along with weights $\bfw_i''$ which we will show satisfy the following:
\begin{Condition}\label{cnd:flat-lewis-weights}
    \leavevmode
\begin{itemize}
    \item $\bfw''$ are $\min\{\gamma,\Omega(1)\}$-one-sided Lewis weights for $\bfA''$
    \item ${\bfw_i''}/{d} \leq \beta$
    \item $\sum_{i}\bfw_i'' = O(T)$
\end{itemize}
\end{Condition}
These three items will allow us to apply Theorem \ref{thm:one-sided-lt} on $\bfA''$ with weights $\bfw''$. 

\paragraph{Flattening $\bfA$.}

Let $\bfA'$ be the original matrix $\bfA$ with the same weights $\bfw$, except that whenever $\bfw_i \geq d\beta$, we replace $\bfa_i$ with $k$ copies of $\bfa_i / k^{1/p}$ for $k = \ceil*{1/d\beta}$, and set $\bfw_j' = \bfw_i / k$ for each copy $j$ of the original row $i$. Note then that
\[
    \bfA^\top\bfW^{1-2/p}\bfA = \bfA'^\top\bfW'^{1-2/p}\bfA'
\]
since
\[
    \bfw_i^{1-2/p} \bfa_i\bfa_i^\top = k\cdot \parens*{\frac{\bfw_i}{k}}^{1-2/p} \parens*{\frac{\bfa_i}{k^{1/p}}}\parens*{\frac{\bfa_i}{k^{1/p}}}^\top.
\]
Then,
\begin{align*}
    \bfw_j' = \frac{\bfw_i}{k} &\geq \frac{\gamma\cdot \bftau_i(\bfW^{1/2-1/p}\bfA)}{k} \\
    &= \gamma\cdot \frac{(\bfw_i^{1/2-1/p}\bfa_i)^\top (\bfA^\top\bfW^{1-2/p}\bfA)^- (\bfw_i^{1/2-1/p}\bfa_i)}{k} \\
    &= \gamma\cdot k^{1-2/p}\cdot k^{2/p}\frac{(\bfw_j'^{1/2-1/p}\bfa_i/k^{1/p})^\top (\bfA'^\top\bfW'^{1-2/p}\bfA')^- (\bfw_j'^{1/2-1/p}\bfa_i/k^{1/p})}{k} \\
    &= \gamma\cdot \bftau_j(\bfW'^{1/2-1/p}\bfA')
\end{align*}
so $\bfw_j'$ are $\gamma$-one-sided $\ell_p$ Lewis weights for $\bfA'$. Clearly, we also have $\bfw_j' \leq d\beta$.

\paragraph{Extending $\bfw'$ via Batch Online Lewis Weights.}

We now define $\bfA''$ to be the matrix
\[
    \bfA'' = \begin{bmatrix}\bfA' \\ \bfS\bfA\end{bmatrix}
\]
where $\bfS\bfA$ is the sampled matrix. We then set $\bfw_i''$ to be $\bfw_i'$ for rows corresponding to $\bfA'$, and we set $\bfw_i''$ to be the batch online $\ell_p$ Lewis weights of $\bfS\bfA$ with respect to $\bfM = \bfA'^\top\bfW'^{1-2/p}\bfA' = \bfA\bfW^{1-2/p}\bfA$, as given by Lemma \ref{lem:batch-online-lewis-convex-program}, for rows corresponding to $\bfS\bfA$. 

We now show that $\bfw''$ satisfy Condition \ref{cnd:flat-lewis-weights}. We start with the first item. For rows $i$ corresponding to $\bfA'$, this follows from the fact that
\[
    \bfw_i'' = \bfw_i' \geq \gamma \cdot \bftau_i(\bfW'^{1/2-1/p}\bfA') \geq \gamma \cdot \bftau_i(\bfW''^{1/2-1/p}\bfA'')
\]
by the monotonicity of leverage scores under row additions. For rows $i$ corresponding to $\bfS\bfA$, this follows from the guarantees of Lemma \ref{lem:batch-online-lewis-convex-program}, which rearranges to the statement that
\[
    \bfw_i'' = \parens*{\frac{p}{2}}^{\frac{1}{1-2/p}}\bftau_i(\bfW''^{1/2-1/p}\bfA'').
\]
We now show the second item of Condition \ref{cnd:flat-lewis-weights}. This is immediate by definition of $\bfw''$ for rows corresponding to $\bfA'$. For rows corresponding to $\bfS\bfA$, letting $\hat\bfw$ denote the weights $\bfw''$ restricted to the rows of $\bfS\bfA$,
\begin{align*}
    \hat\bfw_i &= \parens*{\frac{p}{2}}^{\frac{p/2}{1-2/p}}((\bfs_i\bfa_i)^\top((\bfS\bfA)\hat\bfW^{1-2/p}(\bfS\bfA) + \bfA^\top\bfW\bfA)^{-1}(\bfs_i\bfa_i))^{p/2} \\
    &\leq \parens*{\frac{p}{2}}^{\frac{p/2}{1-2/p}}\bfs_i^p(\bfa_i^\top(\bfA^\top\bfW\bfA)^{-1}\bfa_i)^{p/2} \\
    &\leq \parens*{\frac{p}{2}}^{\frac{p/2}{1-2/p}}\frac1{\bfp_i}\frac{\bfw_i}{\gamma^{p/2}} \leq d\beta.
\end{align*}
The third item follows from the fact that the weights restricted to $\bfS\bfA$ sum to at most $O(d)$. Then, by Theorem \ref{thm:one-sided-lt},
\begin{align*}
    2^l\E_{\bfsigma}\bracks*{\sup_{\norm*{\bfA''\bfx}_p = 1}\abs*{\sum_{i=1}^n \bfsigma_i \abs*{\angle*{\bfa_i'',\bfx}}^p}^l} &\leq \bracks*{O(1)\beta\cdot T^{p/2}[[\gamma^{-1}(\log d)^2(\log n)]^{1+1/l}+l]}^{l/2} \leq O(\eps)^l.
\end{align*}

\paragraph{High probability error bounds.}
We now finish the argument as in \cite{CP2015, CD2021}. For a given fixing of $\bfs$, let
\[
    F_\bfs = \sup_{\norm*{\bfA\bfx}_p = 1}\abs*{\norm*{\bfS\bfA\bfx}_p^p - 1}.
\]
Then for the corresponding $\bfA''$, we have for all $\bfx$ that
\[
    \norm*{\bfA''\bfx}_p^p \leq (2+F_\bfs)\norm*{\bfA\bfx}_p^p,
\]
so
\[
    \sup_{\norm*{\bfA\bfx}_p = 1}\abs*{\sum_{i=1}^{n'}\bfsigma_i\abs*{\angle*{\bfa_i'',\bfx}}^p} \leq (2+F_\bfs)\sup_{\norm*{\bfA''\bfx}_p = 1}\abs*{\sum_{i=1}^{n'}\bfsigma_i\abs*{\angle*{\bfa_i'',\bfx}}^p} \leq O(2+F_\bfs)\eps.
\]
Altogether, we have that
\[
    \E_{\bfs}[F_\bfs^l] \leq \E_\bfs[O\parens*{2+F_\bfs}^l\eps^l] \leq (O(1)^l +  \E[F_\bfs^l])O(\eps)^l
\]
which means that $\E_\bfs[F_\bfs] \leq O(\eps)^l$.
\end{proof}

Theorem \ref{thm:offline-lewis-weight-sampling-main} is a simple corollary of this result by specializing to one-sided $\ell_p$ Lewis weights as given in \cite{Lee2016, JLS2021}.
\section{Sampling-Based Online Lewis Weight Estimation, \texorpdfstring{$0<p<2$}{0<p<2}}\label{sec:sample-lw-est}

\subsection{Flattening Online Lewis Weights}

We first show a reduction to the case where all online Lewis weights that do not increase the rank of the rows are uniformly bounded, using the idea of splitting rows. This will be useful for the sampling-based method, by controlling the worst-case behavior of the matrix martingale for use in a matrix Freedman's inequality. Note that we only need to split rows which $\bfw_i^{p,\OL}(\bfA) < 1$, since otherwise, these rows are sampled with probability $1$ and thus do not affect the matrix Freedman. Note also that such a splitting is isometric in $\ell_p$ (i.e., $\norm*{\bfA'\bfx}_p = \norm*{\bfA\bfx}_p$ for all $\bfx\in\mathbb R^d$), so the pseudo condition number can only change by at most $\poly(n)$ factors, and thus the bound on the sum of online $\ell_p$ Lewis weights via Lemma \ref{lem:online-lewis-sum-bound} is unaffected.

\begin{Lemma}[Online Lewis Weight Flattening]\label{lem:flatten}
Let $\bfA\in\mathbb R^{n\times d}$ and $0 < p < 2$. Let $\beta\in(0,1)$ be a cutoff parameter. Let $\bfA'$ be the matrix formed by replacing $\bfa_i$ by $k\coloneqq \ceil*{1/\beta}$ copies of $\bfa_i / k^{1/p}$ whenever $\bfw_i^{p,\OL}(\bfA) < 1$. Then, for every row $j$ for the new matrix $\bfA'$ which comes from such a row,
\[
    \bfw_j^{p,\OL}(\bfA') \leq \beta.
\]
\end{Lemma}
\begin{proof}
For the rows $i$ considered in this lemma, we have that
\[
    \bfw_i^{p,\OL}(\bfA)^{2/p} = \bracks*{\bfa_i^\top(\bfA_{i-1}^\top\bfW^{p,\OL}(\bfA)_{i-1}^{1-2/p}\bfA_{i-1})^{-}\bfa_i}^{p/2}
\]

We will inductively show the desired result, along with the invariant that
\begin{equation}\label{eq:invariant}
    \bfA_{i}^\top\bfW^{p,\OL}(\bfA)_{i}^{1-2/p}\bfA_{i} \preceq \bfA_{j}'^\top\bfW^{p,\OL}(\bfA')_{j}^{1-2/p}\bfA_{j}'
\end{equation}
whenever $j$ is the last row formed as one of the $k$ copies of row $i\in[n]$ in the original matrix.

Let $j$ be a row formed as one of the $k$ copies of row $i\in[n]$ in the original matrix $\bfA$. Then,
\[
    \bfA_{i-1}^\top\bfW^{p,\OL}(\bfA)_{i-1}^{1-2/p}\bfA_{i-1} \preceq \bfA_{j-1}'^\top\bfW^{p,\OL}(\bfA')_{j-1}^{1-2/p}\bfA_{j-1}'
\]
since we only add rows over the first such index $j$, so by Lemma \ref{lem:pinv-psd}, we have that
\[
    \bfw_i^{p,\OL}(\bfA)^{2/p} = \bfa_i^\top(\bfA_{i-1}^\top\bfW^{p,\OL}(\bfA)_{i-1}^{1-2/p}\bfA_{i-1})^-\bfa_i \geq \bfa_i^\top(\bfA_{j-1}'^\top\bfW^{p,\OL}(\bfA')_{j-1}^{1-2/p}\bfA_{j-1}')\bfa_i.
\]
Then dividing both sides by $k^{2/p}$, we have that
\[
    \parens*{\frac{\bfw_i^{p,\OL}(\bfA)}{k}}^{2/p} \geq \parens*{\frac{\bfa_i}{k^{1/p}}}^\top(\bfA_{j-1}'^\top\bfW^{p,\OL}(\bfA')_{j-1}^{1-2/p}\bfA_{j-1}')\parens*{\frac{\bfa_i}{k^{1/p}}} = \bfw_j^{p,\OL}(\bfA')^{2/p},
\]
that is, $\frac1k\bfw_i^{p,\OL}(\bfA) \geq \bfw_j^{p,\OL}(\bfA')$. It follows that $\bfw_j^{p,\OL}(\bfA') \leq 1/k \leq \beta$. Finally, if $S$ is the set of all $k$ rows formed from row $i\in[n]$ in the original matrix, then
\[
    \sum_{j\in S} \bfw_j^{p,\OL}(\bfA')^{1-2/p}\parens*{\frac{\bfa_i}{k^{1/p}}}\parens*{\frac{\bfa_i}{k^{1/p}}}^\top \succeq \sum_{j\in S} \parens*{\frac{\bfw_i^{p,\OL}(\bfA)}{k}}^{1-2/p}\parens*{\frac{\bfa_i}{k^{1/p}}}\parens*{\frac{\bfa_i}{k^{1/p}}}^\top = \bfw_i^{p,\OL}(\bfA)\bfa_i\bfa_i^\top
\]
where we have used that $1 - 2/p < 0$. This establishes the invariant \eqref{eq:invariant}.
\end{proof}

\subsection{Online Lewis Weight Estimation}

We now show how to obtain Lewis weight overestimates $\tilde\bfw_i$ that have a small sum.

\begin{algorithm}
	\caption{Sampling-Based Online Lewis Weight Estimation}
	\textbf{input:} $\bfA\in\mathbb R^{n\times d}$, $p\in(0,2)$, oversampling parameter $\alpha\in(0,1)$. \\
	\textbf{output:} Online Lewis weight estimates $\braces*{\tilde\bfw_i}_{i=1}^n$.
	\begin{algorithmic}[1] %
        \State $\tilde\bfA_0 \gets \varnothing$ \Comment{Matrix estimate for sketching Lewis quadratic}
        \For{$i\in[n]$}
            \State $\tilde\bfw_i \gets \bracks*{ \bfa_i^\top(\tilde\bfA_{i-1}\tilde\bfW_{i-1}^{1-2/p}\tilde\bfA_{i-1})^-\bfa_i}^{p/2}$
            \State $\bfp_i \gets \min\braces*{\frac1\alpha\tilde\bfw_i, 1}$
            \State $\tilde\bfA_i \coloneqq \begin{cases}\begin{bmatrix}\tilde\bfA_{i-1} \\ \bfa_i / \sqrt\bfp_i\end{bmatrix} & \text{with probability $\bfp_i$} \\ \tilde\bfA_{i-1} & \text{otherwise} \end{cases}$
        \EndFor
        \State \Return $\braces*{\tilde\bfw_i}_{i=1}^n$
	\end{algorithmic}\label{alg:online-lewis-weight-estimate}
\end{algorithm}

\subsubsection{Lower Bound on Online Lewis Weight Estimates}

We first show that the approximate online Lewis weights are bounded below by the true Lewis weights, analogously to Lemma 3.3 of \cite{CMP2020}, which shows how to approximate online leverage scores. We will need the matrix Freedman's inequality \cite{Tro2011}:

\begin{Theorem}[Matrix Freedman's Inequality \cite{Tro2011}]\label{thm:matrix-freedman}
Let $\bfY_0, \bfY_1, \dots, \bfY_n$ be a matrix whose values are self-adjoint matrices with dimension $d$, and let $\bfX_1, \ldots, \bfX_n$ be the difference sequence. Assume that the difference sequence is uniformly bounded in the sense that
    \begin{align*}
        \norm{\bfX_k}_2 &\leq R\mbox{ almost surely, for } k = 1, \ldots, n.
    \end{align*}
    Define the predictable quadratic variation process of the martingale:
    \begin{align*}
        \bfW_k := \sum_{j=1}^k \E_{j - 1}\bracks*{\bfX_j^2}\mbox{, for }k = 1, \ldots, n.
    \end{align*}
    Then, for all $\eps > 0$ and $\sigma^2 > 0$,
    \begin{align*}
        \Pr\braces{\norm*{\bfY_n}_2 \geq \eps\mbox{ and } \norm{\bfW_n}_2 \leq \sigma^2} &\leq d\cdot\exp\left(-\frac{-\eps^2/2}{\sigma^2+R\eps/3}\right).
    \end{align*}
\end{Theorem}

\begin{Theorem}[Online Lewis weight estimates bound Lewis weights]\label{thm:estimate-lb}
Let $\bfA\in\mathbb R^{n\times d}$ and $0 < p < 2$. Furthermore, suppose that all the online Lewis weights of $\bfA$ are bounded by $\beta$ whenever $i$ does not increase the rank, that is, $\bfw_i^{p,\OL}(\bfA) \leq \beta$ for each $i\in[n]$ with $\bfa_i\in\rowspan(\bfA_{i-1})$. Furthermore, suppose that
\[
    \alpha + \beta \leq \frac{\eps^2}{\log\frac{d}{\delta}}
\]
for $\delta\in(0,1)$. Then, with probability at least $1-\delta$, we have for all $i\in[n]$ that
\[
    \tilde\bfw_i^{p} \geq \frac1{1+\eps}\bfw_i^p(\bfA),
\]
and that
\[
    \tilde\bfA\tilde\bfW^{1-2/p}\tilde\bfA \preceq (1+\eps)\bfA^\top\bfW^p(\bfA)^{1-2/p}\bfA,
\]
\end{Theorem}
\begin{proof}
For simplicity of notation, let $\bfw = \bfw^{p,\OL}(\bfA)$ and $\bfW = \bfW^{p,\OL}(\bfA)$.

Our proof closely follows \cite{CMP2020}. Let
\[
    \bfG \coloneqq \bfA^\top \bfW^p(\bfA)^{1-2/p}\bfA
\]
where $\bfW^p(\bfA)$ is the diagonal matrix of the offline Lewis weights, and let
\[
    \bfu_i \coloneqq (\bfG^-)^{1/2}\bfa_i.
\]
Note that
\begin{equation}\label{eq:w-t-bound}
\begin{aligned}
    \bfu_i^\top\bfu_i &= \bfa_i^\top(\bfA^\top \bfW^p(\bfA)^{1-2/p}\bfA)^-\bfa_i = \bfw_i^p(\bfA)^{2/p}.
\end{aligned}
\end{equation}

We consider the matrix martingale $0 = \bfY_0, \bfY_1, \dots, \bfY_n\in\mathbb R^{d\times d}$ with difference sequence $\bfX_1, \dots, \bfX_n$. If $\norm*{\bfY_{i-1}}_2 \geq \eps$, then we set $\bfX_i \coloneqq 0$, otherwise
\[
    \bfX_i \coloneqq \begin{cases}
        \parens*{\frac{1}{\bfp_i}\tilde\bfw_i^{1-2/p} - \bfw_i^{1-2/p}}\bfu_i\bfu_i^\top & \text{if $\bfa_i$ is sampled in $\tilde\bfA$} \\
        -\bfw_i^{1-2/p}(\bfu_i\bfu_i^\top) & \text{otherwise}
    \end{cases}
\]
This gives
\[
    \bfY_{i-1} = (\bfG^-)^{1/2}\parens*{\tilde\bfA_{i-1}^\top\tilde\bfW_{i-1}^{1-2/p}\tilde\bfA_{i-1} - \bfA_{i-1}^\top\bfW_{i-1}^{1-2/p}\bfA_{i-1}}(\bfG^-)^{1/2}
\]

\paragraph{Bounds on the Difference Sequence.}

We now bound $\norm*{\bfX_j}_2$ and $\bfW_i \coloneqq \sum_{j=1}^i \E_{j-1}[\bfX_j^2]$ for use in the matrix Freedman inequality. These bounds are trivial when $\norm*{\bfY_{j-1}}_2 \geq \eps$, so suppose that $\norm*{\bfY_{j-1}}_2 < \eps$.

We will first show that $\bfu_i^\top\bfu_i \leq (1+\eps)\tilde\bfw_i^{2/p}$. Because $\tilde\bfA_{i-1}^\top\tilde\bfW_{i-1}^{1-2/p}\tilde\bfA_{i-1}$ and $\bfA_{i-1}^\top\bfW_{i-1}^{1-2/p}\bfA_{i-1}$ always have the same row space and are symmetric, we can write them as $\bfV\tilde\bfR\bfV^\top$ and $\bfV\bfR\bfV^\top$, respectively, where $\bfV\in\mathbb R^{d\times r}$ is an orthonormal basis of $\rowspan(\bfA_{i-1})$. If $\bfa_i\notin\rowspan(\bfA_{i-1})$, then $\bfX_i = 0$, so suppose that $\bfa_i\in\rowspan(\bfA_{i-1})$. Note then that $\bfa_i$ can be written as $\bfa_i = \bfV\bfb$ for some $\bfb\in\mathbb R^r$. Then,
\begin{equation}\label{eq:w-bound}
\begin{aligned}
    \tilde\bfw_i^{2/p} &= \bfa_i^\top(\tilde\bfA_{i-1}^\top\tilde\bfW_{i-1}^{1-2/p}\tilde\bfA_{i-1})^{-}\bfa_i \\
    &= \bfb^\top\bfV^\top(\bfV\tilde\bfR^{-1}\bfV^\top)\bfV\bfb && \text{Lemma \ref{lem:pinv-psd}} \\
    &= \bfb^\top \tilde\bfR^{-1}\bfb \\
    &= \bfb^\top (\bfR + (\tilde\bfR-\bfR))^{-1}\bfb.
\end{aligned}
\end{equation}
Now let $\bfP = \bfV\bfV^\top$ be the projection matrix onto $\rowspan(\bfA_{i-1})$. Let $\bfE\in\mathbb R^{r\times r}$ be such that $\bfP\bfG\bfP = \bfV\bfE\bfV^\top$. Then, we have that
\begin{align*}
    \bfV\bfE^{-1/2}(\tilde\bfR-\bfR)\bfE^{-1/2}\bfV^\top = \bfP\bfY_{i-1}\bfP \preceq \bfY_{i-1} \preceq \eps\bfI_d.
\end{align*}
so for every $\bfx\in\mathbb R^r$,
\[
    \bfx^\top \bfE^{-1/2}(\tilde\bfR-\bfR)\bfE^{-1/2} \bfx \leq \eps\bfx^\top\bfx.
\]
By a change of variable into $\bfy = \bfE^{-1/2}\bfx$, this gives
\[
    \bfy^\top(\tilde\bfR-\bfR)\bfy \leq \eps \bfy^\top \bfE\bfy
\]
for every $\bfy\in\mathbb R^r$, that is, that $\tilde\bfR - \bfR \preceq \eps\bfE$. Also note that by Lemma \ref{lem:online-lewis-bound} and the fact that $1-2/p < 0$,
\begin{align*}
    \bfW^p(\bfA) \preceq \bfW &\implies \bfW^p(\bfA)^{1-2/p} \succeq \bfW^{1-2/p} \\
    &\implies \bfG = \bfA^\top\bfW^p(\bfA)^{1-2/p}\bfA \succeq \bfA_{i-1}^\top\bfW^p(\bfA)_{i-1}^{1-2/p}\bfA_{i-1} \succeq \bfA_{i-1}^\top\bfW_{i-1}^{1-2/p}\bfA_{i-1} \\
    &\implies \bfP\bfG\bfP\succeq \bfP(\bfA_{i-1}^\top\bfW_{i-1}^{1-2/p}\bfA_{i-1})\bfP = \bfA_{i-1}^\top\bfW_{i-1}^{1-2/p}\bfA_{i-1} \\
    &\implies \bfE \succeq \bfR.
\end{align*}
Thus, $\bfR + (\tilde\bfR-\bfR)\preceq \bfE + \eps\bfE$. Then, continuing the calculation from \eqref{eq:w-bound}, we have
\begin{equation}\label{eq:w-bound-pt2}
\begin{aligned}
    \tilde\bfw_i^{2/p} &= \bfb^\top(\bfR + (\tilde\bfR-\bfR))^{-1}\bfb \\
    &\geq \bfb^\top(\bfE + \eps\bfE)^{-1}\bfb \\
    &= \frac1{1+\eps} \bfb^\top\bfE^{-1}\bfb \\
    &= \frac1{1+\eps} \bfu_i^\top\bfu_i \\
    &= \frac1{1+\eps} \bfw_i^p(\bfA)^{2/p} && \text{Equation \eqref{eq:w-t-bound}}
\end{aligned}
\end{equation}

Given our bound on $\bfu_i^\top\bfu_i$ in \eqref{eq:w-t-bound} and \eqref{eq:w-bound-pt2}, as well as Lemma \ref{lem:online-lewis-bound} we can now bound $\bfX_i$. We first have that
\begin{equation}\label{eq:spec-bound}
\begin{aligned}
    \norm*{\bfX_i}_2 &\leq \frac{\tilde\bfw_i^{1-2/p}}{\bfp_i} \norm*{\bfu_i\bfu_i^\top}_2 + \bfw_i^{1-2/p}\norm*{\bfu_i\bfu_i^\top}_2 \\
    &\leq \frac{(1+\eps)\tilde\bfw_i}{\bfp_i} + \bfw_i \\
    &\leq O(\alpha + \beta)
\end{aligned}
\end{equation}
almost surely. Furthermore, we have that
\begin{align*}
    \E_{i-1}[\bfX_i^2] &\preceq \bfp_i\cdot \parens*{\frac{\tilde\bfw_i^{1-2/p}}{\bfp_i} - \bfw_i^{1-2/p}}^2 (\bfu_i\bfu_i^\top)^2 + (1-\bfp_i)\bfw_i^{2(1-2/p)}(\bfu_i\bfu_i^\top)^2 \\
    &\preceq 2\frac{\tilde\bfw_i^{2(1-2/p)}}{\bfp_i}(\bfu_i\bfu_i^\top)^2 + 2\bfw_i^{2(1-2/p)}(\bfu_i\bfu_i^\top)^2 \\
    &\preceq 2(1+\eps)\frac{\tilde\bfw_i}{\bfp_i}\cdot\tilde\bfw_i^{1-2/p}\bfu_i\bfu_i^\top + 2\bfw_i\cdot \bfw_i^{1-2/p}\bfu_i\bfu_i^\top \\
    &\preceq O(\alpha + \beta)(\tilde\bfw_i^{1-2/p} + \bfw_i^{1-2/p})\bfu_i\bfu_i^\top \\
    &\preceq O(\alpha + \beta)\bfw_i^p(\bfA)^{1-2/p}\bfu_i\bfu_i^\top
\end{align*}
where the last inequality uses that $1-2/p < 0$. Then, for the predictable quadratic variation process $\bfW_i \coloneqq \sum_{k=1}^i \E_{k-1}[\bfX_k^2]$ of the martingale $\{\bfY_i\}$, we have
\begin{equation}\label{eq:quad-bound}
\begin{aligned}
    \norm*{\bfW_i}_2 &= O(\alpha+\beta)\norm*{(\bfG^-)^{1/2}\parens*{\sum_{k=1}^i \bfw^p_k(\bfA)^{1-2/p}\bfa_k\bfa_k^\top}(\bfG^-)^{1/2}}_2 \\
    &= O(\alpha+\beta)\norm*{((\bfA^\top \bfW^p(\bfA)^{1-2/p}\bfA)^-)^{1/2}\parens*{\bfA_k^\top \bfW^p(\bfA)^{1-2/p}_k\bfA_k}((\bfA^\top \bfW^p(\bfA)^{1-2/p}\bfA)^-)^{1/2}}_2 \\
    &\leq O(\alpha+\beta)\norm*{((\bfA^\top \bfW^p(\bfA)^{1-2/p}\bfA)^-)^{1/2}\parens*{\bfA^\top \bfW^p(\bfA)^{1-2/p}\bfA}((\bfA^\top \bfW^p(\bfA)^{1-2/p}\bfA)^-)^{1/2}}_2 \\
    &= O(\alpha+\beta)
\end{aligned}
\end{equation}
using Lemma \ref{lem:psd-flip}.

\paragraph{The Matrix Freedman's Inequality.}

We may now apply Theorem \ref{thm:matrix-freedman} along with the bounds of \eqref{eq:spec-bound} and \eqref{eq:quad-bound}, which gives that
\[
    \Pr\braces*{\norm*{\bfY_n}_2 \geq \eps} \leq d\cdot\exp\parens*{-\Omega(1)\cdot\frac{\eps^2}{\alpha+\beta}} \leq d\cdot \exp\parens*{-\log\frac{d}{\delta}} = \delta.
\]
Then under this event, \eqref{eq:w-bound-pt2} holds for every $i\in[n]$. Furthermore, we also have under this event that
\[
    \tilde\bfA\tilde\bfW^{1-2/p}\tilde\bfA \preceq \bfA\bfW^{1-2/p}\bfA + \eps\bfA\bfW^p(\bfA)^{1-2/p}\bfA \preceq (1+\eps)\bfA^\top\bfW^p(\bfA)^{1-2/p}\bfA,
\]
again using that $1-2/p < 0$.
\end{proof}

\subsubsection{Upper Bound on the Sum of Online Lewis Weight Estimates}

Unlike in the case for $\ell_2$, the proof in Theorem \ref{thm:estimate-lb} is not sufficient for upper bounding the estimated online Lewis weights for every $i\in[n]$. This is due to the fact that the spectral error in our Lewis quadratic is $\bfA\bfW^p(\bfA)^{1-2/p}\bfA$, which is in fact spectrally greater than $\bfA\bfW^{p,\OL}(\bfA)^{1-2/p}\bfA$ since $1-2/p < 0$. We thus appeal to the strategy of Lemma 3.4 of \cite{CMP2020}, which uses a different proof to bound the sum of online Lewis weights.

\begin{Theorem}\label{thm:lewis-estimate-ub}
With probability at least $1 - \delta$, we have that
\[
    \sum_{i=1}^n \tilde\bfw_i \leq O(d)\log(n\kappa^\OL) + O\parens*{\log\frac1\delta}
\]
\end{Theorem}
\begin{proof}
We closely follow \cite{CMP2020}. Let $\lambda = (\max_{i=1}^n \norm*{\bfA_i^-}_2)^{-1}/n^C$, where $C$ will be a sufficiently large constant exponent which can change from line to line. We then let
\[
    \Delta_i \coloneqq \log\det(\tilde\bfA_i^\top\tilde\bfW_i^{1-2/p}\tilde\bfA_i + \lambda\bfI_d) - \log\pdet(\tilde\bfA_{i-1}^\top\tilde\bfW_{i-1}^{1-2/p}\tilde\bfA_{i-1} + \lambda\bfI_d)
\]
We first show that $\E_{i-1}\bracks*{\exp(\tilde\bfw_i/8 - \Delta_i}$ is always at most $1$ whenever $\bfa_i\in\rowspan(\bfA_{i-1})$. Then by the pseudodeterminant lemma, we have that
\begin{align*}
    \E_{i-1}\bracks*{\exp\parens*{\frac{\tilde\bfw_i}{8} - \Delta_i}} &= \bfp_i\cdot \frac{\exp(\tilde\bfw_i/8)}{1 + (\tilde\bfw_i^{1/2-1/p}\bfa_i)^\top(\tilde\bfA_{i-1}^\top\tilde\bfW_{i-1}^{1-2/p}\tilde\bfA_{i-1} + \lambda\bfI_d)^{-1}(\tilde\bfw_i^{1/2-1/p}\bfa_i)/\bfp_i} \\
    &\hspace{5em}+ (1-\bfp_i)\exp(\tilde\bfw_i/8).
\end{align*}
Note that since $\tilde\bfw_i \leq 1$, we have $\exp(\tilde\bfw_i/8) \leq 1+\tilde\bfw_i/4$. Now if $\tilde\bfw_i/\alpha < 1$, then $\bfp_i = \tilde\bfw_i/\alpha$ and
\begin{align*}
    \tilde\bfw_i &= (\tilde\bfw_i^{1/2-1/p}\bfa_i)^\top(\tilde\bfA_{i-1}^\top\tilde\bfW_{i-1}^{1-2/p}\tilde\bfA_{i-1})^-(\tilde\bfw_i^{1/2-1/p}\bfa_i) \\
    &= (1\pm n^{-C})(\tilde\bfw_i^{1/2-1/p}\bfa_i)^\top(\tilde\bfA_{i-1}^\top\tilde\bfW_{i-1}^{1-2/p}\tilde\bfA_{i-1}+\lambda\bfI_d)^{-1}(\tilde\bfw_i^{1/2-1/p}\bfa_i)
\end{align*}
by rearranging and taking $p/2$th powers. Then,
\begin{align*}
    \E_{i-1}\bracks*{\exp\parens*{\frac{\tilde\bfw_i}{8} - \Delta_i}} &\leq \bfp_i\frac{1+\tilde\bfw_i/4}{1+(1-n^{-C})\alpha} + (1-\bfp_i)(1+\tilde\bfw_i/4) \\
    &= (1+\tilde\bfw_i/4)\parens*{1 - \bfp_i\frac{(1-n^{-C})\alpha}{1+(1-n^{-C})\alpha}} \\
    &= \parens*{1+\bfp_i \frac{\alpha}{4}}\parens*{1 - \bfp_i\frac{(1-n^{-C})\alpha}{1+(1-n^{-C})\alpha}} \leq 1
\end{align*}
for $\alpha$ sufficiently small. Otherwise, if $\bfp_i = 1$, then
\begin{align*}
    \E_{i-1}\bracks*{\exp\parens*{\frac{\tilde\bfw_i}{8} - \Delta_i}} &\leq \frac{1+\tilde\bfw_i/4}{1+(1-n^{-C})\tilde\bfw_i} \leq 1.
\end{align*}

Next, we analyze the expected product of $\exp(\tilde\bfw_i/8-\Delta_i)$ over the first $k$ steps. If $\bfa_k\notin\rowspan(\bfA_{k-1})$, we have that
\begin{align*}
    \E\bracks*{\exp\parens*{\sum_{i=1}^k \frac{\tilde\bfw_i}{8}-\Delta_i}} &= \E_{\text{first $k-1$ steps}}\bracks*{\exp\parens*{\sum_{i=1}^{k-1} \frac{\tilde\bfw_i}{8}-\Delta_i}\E_{k-1}\bracks*{\frac{\tilde\bfw_k}{8}-\Delta_k}} \leq \E\bracks*{\exp\parens*{\sum_{i=1}^{k-1} \frac{\tilde\bfw_i}{8}-\Delta_i}}.
\end{align*}
Inductively, we have that
\[
    \E\bracks*{\exp\parens*{\sum_{i=1}^n \frac{\tilde\bfw_i}{8}-\Delta_i}} \leq 1
\]
By Markov's inequality, we then have that
\[
    \Pr\braces*{\sum_{i=1}^n \tilde\bfw_i > 8\log\frac1\delta + 8\sum_{i=1}^n \Delta_i} \leq \delta.
\]
Note that
\begin{align*}
    \sum_{i=1}^n \Delta_i &= \log\det(\tilde\bfA^\top\tilde\bfW^{1-2/p}\tilde\bfA+\lambda\bfI_d) - \log\det(\lambda\bfI_d) \\
    &\leq \log\det((1+O(\eps))(\bfA^\top\bfW^p(\bfA)^{1-2/p}\bfA)) - \log\det(\lambda\bfI_d)
\end{align*}
by Theorem \ref{thm:estimate-lb}. Furthermore, for any unit vector $\bfx\in\mathbb R^d$, we have by properties of Lewis weights that
\[
    \norm*{\bfW^p(\bfA)^{1/2-1/p}\bfA\bfx}_2 = \poly(d) \norm*{\bfA\bfx}_p = \poly(n,d)\norm*{\bfA\bfx}_2,
\]
so the operator norm of $\bfA^\top\bfW^p(\bfA)^{1-2/p}\bfA$ is within a $\poly(n,d)$ factor of the operator norm of $\bfA^\top\bfA$. Thus,
\[
    \sum_{i=1}^n \tilde\bfw_i \leq O(d)\log\frac{\poly(n,d)\norm*{\bfA}_2}{\lambda} = O(d)\log(n\kappa^\OL) + O\parens*{\log\frac1\delta}.
\]
This yields the desired result.
\end{proof}

\subsection{\texorpdfstring{$\ell_p$}{lp} Subspace Embeddings via Online Lewis Weight Sampling}

Given our online Lewis weight estimates, we may now conclude by sampling proportionally to these weights. We use fresh randomness to sample proportionally to these weights, independently of the success of the random sampling process used to estimate the online Lewis weights. Thus, we can condition on the successes of Theorems \ref{thm:estimate-lb} and \ref{thm:lewis-estimate-ub} to obtain Lewis weight upper bounds with a small sum. We then get our main result:

\begin{Theorem}\label{thm:online-lewis-p<2}
Let $\bfA\in\mathbb R^{n\times d}$ and $p\in(0,2)$. Let $\delta\in(0,1)$ be a failure rate parameter and let $\eps\in(0,1)$ be an accuracy parameter. Then there is an online coreset algorithm $\mathcal A$ such that, with probability at least $1-\delta$, $\mathcal A$ outputs a weighted subset of $m$ rows with sampling matrix $\bfS$ such that
\[
    \norm*{\bfS_i\bfA_i\bfx}_p^p = (1\pm\eps)\norm*{\bfA_i\bfx}_p^p
\]
for all $\bfx\in\mathbb R^d$ and every $i\in[n]$, and
\[
    m = \begin{dcases}
        O\parens*{\frac{T}{\eps^2}\bracks*{(\log d)^2\log n + \log\frac1\delta}} & p \in (1,2) \\
        O\parens*{\frac{T}{\eps^2}\log\frac{n}{\delta}} & p = 1 \\
        O\parens*{\frac{T}{\eps^2}\bracks*{(\log d)^3 + \log\frac1\delta}} & p\in(0,1)
    \end{dcases}
\]
for $T = O(d)\log(n\kappa^\OL)$.
\end{Theorem}
\begin{proof}
We first use Lemma \ref{lem:flatten} to flatten the rows down to an online Lewis weight of $O(1/\log(d/\delta))$, which can be done in an online fashion, given knowledge of an upper bound on the online Lewis weight. We can then obtain upper bounds $\tilde\bfw_i$ on online Lewis weights using Algorithm \ref{alg:online-lewis-weight-estimate} and Theorem \ref{thm:estimate-lb} with $\eps = O(1)$ so that the upper bounds sum to at most
\[
    T = O(d)\log(n\kappa^\OL).
\]
This requires $O(T\log(d/\delta))$ samples. Using these Lewis weight estimates, we then sample using Theorem \ref{thm:lewis-weight-sampling-0<p<2}. The number of samples used here dominates the samples used to approximate the Lewis weights, and is as given in the theorem statement.
\end{proof}
\section{Applications: Online Coresets for Generalized Linear Models}\label{sec:glm}

As described in the introduction, our results for online Lewis weight sampling give new algorithms for optimizing generalized linear models in one pass in a stream. In this section, we show the first results showing the necessity of a linear dependence on the $\mu$-complexity parameter.

\subsection{Lower Bound for \texorpdfstring{$\mu$}{mu}-Complex Datasets}

We start with the following geometric lemma, which helps in constructing an input instance with bounded $\mu$-complexity.

\begin{figure}[ht]
	\centering
	\begin{tikzpicture}[scale=1]

    \draw[-] (-1.3,0) -- (1.3,0);
    \draw[-] (0,-1.3) -- (0,1.3);
	  
    \node[circle,fill,inner sep=1.5pt] (a) at (1,0) {};
    \node[circle,fill,inner sep=1.5pt] (a) at (-1,0) {};
    \node[circle,fill,inner sep=1.5pt] (a) at (0,1) {};
    \node[circle,fill,inner sep=1.5pt] (a) at (0,-1) {};
	  
    \draw[blue,line width=0.5mm] (-1.2,-0.8)--(1.2,0.8);
    \draw[gray, dashed] (24/52,1-36/52)--(0,1);
    \draw[gray, dashed] (-24/52,-1+36/52)--(0,-1);

    \draw[gray,dashed] (0,0) circle (0.353);
    \draw[red,line width=0.5mm] (-3/4,5/4)--(5/4,-3/4);

	\end{tikzpicture}
	\caption{A line cannot be close to all four points and the origin. The blue line is close to the origin, but far from $(0,\pm1)$. The red line is close to $(0,1)$ and $(1,0)$ but far from the origin.}
	\label{fig:four-pts}
\end{figure}
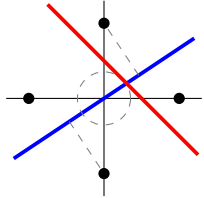

\begin{Lemma}\label{lem:four-pts}
Consider the four points $S = \braces*{(\pm1,0), (0,\pm1)}$ in $\mathbb R^2$. Let $L = \braces*{\bfx\in\mathbb R^2 : \angle*{\bfu,\bfx} = b}$ be an affine set in $\mathbb R^2$, for a unit vector $\bfu\in\mathbb R^2$ and offset $b$. Then, at least one of the following is true:
\begin{itemize}
    \item $\abs*{b}\geq \frac1{2\sqrt2}$
    \item there are $\bfs_\pm\in S$ such that $\angle*{\bfu,\bfs_+} \geq b + \frac1{2\sqrt2}$ and $\angle*{\bfu,\bfs_-} \leq b - \frac1{2\sqrt2}$
\end{itemize}
\end{Lemma}
\begin{proof}
Assume WLOG that $\bfu_1\geq\bfu_2\geq 0$. Then, $\angle*{\bfu,\bfs}$ are $\bfu_1,-\bfu_1,\bfu_2,-\bfu_2$ for $\bfs\in S$. Since $\bfu$ is a unit vector, we have that $\bfu_1 \geq 1/\sqrt2$, so
\[
    \bfu_1 = \angle*{\bfu,(1,0)} \geq \frac1{\sqrt2}, \qquad -\bfu_1 = \angle*{\bfu,(-1,0)} \leq -\frac1{\sqrt2}.
\]
Then if $\abs*{b} \leq \frac1{2\sqrt2}$, then
\[
    \bfu_1 = \angle*{\bfu,(1,0)} \geq b + \frac1{2\sqrt2}, \qquad -\bfu_1 = \angle*{\bfu,(-1,0)} \leq b-\frac1{2\sqrt2}.\qedhere
\]
\end{proof}

A similar conclusion continues to hold if we replace the standard basis vectors by approximately orthogonal points:
\begin{Corollary}\label{cor:four-pts}
Let $S\subseteq\mathbb R^2$ be a set of four points such that $\abs*{\angle*{\bfs,\bfs'}} \leq 1/100$ for each $\bfs\neq \bfs'\in S$. Then, at least one of the following is true:
\begin{itemize}
    \item $\abs*{b}\geq \frac1{3\sqrt2}$
    \item there are $\bfs_\pm\in S$ such that $\angle*{\bfu,\bfs_+} \geq b + \frac1{3\sqrt2}$ and $\angle*{\bfu,\bfs_-} \leq b - \frac1{3\sqrt2}$
\end{itemize}
\end{Corollary}

Using the above lemma, we will construct a variant of the lower bound instance of \cite{MSSW2018} for logistic regression with bounded $\mu$-complexity. Recall that the instance of \cite{MSSW2018} shows an $\Omega(n)$ lower bound for an instance with unbounded $\mu$-complexity by arranging $n$ points in a circle on the plane. Then, they reduce the \textsf{INDEX} problem to the problem of computing a coreset for logistic regression as follows. Alice's input point set is taken to be the points on the circle corresponding to $A\subseteq[n]$, where $A$ is Alice's input set, where all of the labels are $1$. Then, Alice computes a coreset for logistic regression and then sends the coreset to Bob. Bob then adds the circle point corresponding to his index $b\in[n]$ with label $-1$. If $b\notin A$, then there exists a hyperplane separating Alice's points and Bob's point, and thus the cost of logistic regression can be shown to be arbitrarily small, whereas if $b\in A$, then the cost is at least a constant. This shows an $\Omega(n)$ lower bound for any constant approximation.

In order to construct an input instance with bounded $\mu$-complexity, we first consider adding the four points as in the configuration of Figure \ref{fig:four-pts} and Lemma \ref{lem:four-pts}. We also add the origin twice, once with label $1$ and once with label $-1$. This will turn out to be enough to argue that the $\mu$-complexity will be bounded by $O(n)$. In order to prove our lower bound statement, we will in fact need a further modification and take a high-dimensional version of the circle instance, by using the following theorem from coding theory. This result has been used several times for related lower bounds \cite{LWW2021, MMWY2021, WY2022}.

\begin{Theorem}[Theorem 7, \cite{PTB2013}]\label{thm:ptb2013}
Let $m, D$ be integers. Then, there exists a set $\mathcal F\subseteq\{\pm1\}^d$ of binary vectors of size $\abs*{\mathcal F} = n$ for
\begin{align*}
    d &= 2^m - 1 \\
    n &= \begin{cases}
    \frac{2^{(D-\floor{D/4})m}-1}{2^m - 1} & \text{if $m$ is odd} \\
    2^{(D-\floor{D/4})m}-1 & \text{if $m$ is even}
    \end{cases}
\end{align*}
and for any $\bfx,\bfx'\in\mathcal F$ with $\bfx \neq \bfx'$,
\[
    \abs*{\angle*{\bfx,\bfx'}} = \abs*{\sum_{i=1}^n (-1)^{\mathbbm{1}(\bfx_i=\bfx_i')}} \leq 1 + 2(D-1)2^{m/2}.
\]
\end{Theorem}

\begin{Definition}[Hard $p$-Probit Instance]\label{def:hard-log-reg-inst}
We define a $p$-probit coreset instance as follows. Let $n\in\mathbb N$ and let $\Delta\in\mathbb N$. Let $A\subseteq[n]$ and let $b\in[n]$. Let $\mathcal F\subseteq\{\pm1\}^d$ be as given in Theorem \ref{thm:ptb2013} with parameters $n + 4\leq \abs*{\mathcal F} \leq O(n)$, $d = \Theta(\log^2 n)$, and $D = \Theta(\frac{\log n}{\log d}) = \Theta(\frac{\log n}{\log\log n})$. We associate each $i\in[n]$ with some $\bfx^{(i)}\in\mathcal F$. Note that we have four remaining points in $\mathcal F$ unused by those corresponding to $[n]$. We add these to the input dataset with label $1$, calling these the ``first four points''.

Now for each $a\in A$, we add $\Delta$ copies of the unit vector $\bfx'^{(a)}\coloneqq \bfx^{(a)} / \sqrt d$ for $\bfx^{(a)}\in\mathcal F$ with label $1$ to the input dataset. We also add $(0,0)$ with label $1$ and $(0,0)$ with label $-1$. Finally, we add $n$ copies of $\bfx'^{(b))} \coloneqq \bfx^{(b)} / \sqrt d$ for $\bfx^{(b)}\in\mathcal F$ with label $-1$. We define the associated matrix $\bfA\in\mathbb R^{m\times (d+1)}$ for $m = \Delta\abs*{A} + n + 4 + 2$, where each row $\bfa_i$ is $y_i \cdot (\bfb_i, 1)$ where $y_i$ is the label and $\bfb_i\in\mathbb R^d$ is the added point.
\end{Definition}

We first argue that this instance has bounded $\mu_p$-complexity.

\begin{Lemma}[Bounded $\mu$-Complexity]\label{lem:bdd-mu}
Consider the input instance of Definition \ref{def:hard-log-reg-inst}. Then, for any $A\subseteq[n]$ and $b\in[n]$,
\[
    \mu_p(\bfA) \leq O(\Delta n).
\]
\end{Lemma}
\begin{proof}
We consider any $\bfx\in\mathbb R^{d+1}$. Note that the definition of $\mu_p(\bfA)$ is scale invariant, so we can scale $\bfx$ so that its first $d$ coordinates form a unit vector. We refer to the first $d$ coordinates as $\bfu\in\mathbb R^d$ and the last coordinate as $b$. We now case on $b$. First, if $\abs*{b} \geq 2$, then for $\bfx\in\mathbb R^d$ with $\norm*{\bfx}_2 \leq 1$ and $y\in\{\pm1\}$,
\[
    \abs*{\angle*{\bfu,\bfx}} \leq 1 \leq \frac{\abs*{b}}{2}.
\]
Thus, the sign of $\angle*{\bfu,\bfx}+by$ is just the sign of $by$, and the magnitude is $\Theta(\abs*{b})$. By construction, the dataset contains at least one point with label $1$ and one point with label $-1$, so we have that
\[
    \frac{\norm*{(\bfA\bfx)^+}_p^p}{\norm*{(\bfA\bfx)^-}_p^p} \leq \frac{\Theta(\Delta n\abs*{b}^p)}{\Theta(\abs*{b}^p)} \leq O(\Delta n).
\]
Next, suppose that $\abs*{b} \in [\frac1{3\sqrt2}, 2]$. In this case, at least one of the points with $\bfu = (0,0)$ will have $by < 0$, so we have that $\norm*{(\bfA\bfx)^-}_p^p \geq \frac1{(3\sqrt2)^p}$. On the other hand, we have
\[
    \abs*{\angle*{\bfu,\bfx}+by} \leq 1 + 2 = 3
\]
for any $\norm*{\bfx}_2 \leq1$ and $y\in\{\pm1\}$, so $\norm*{(\bfA\bfx)^+}_p^p \leq 3^p\Delta n$. Thus, we again have that
\[
    \frac{\norm*{(\bfA\bfx)^+}_p^p}{\norm*{(\bfA\bfx)^-}_p^p} \leq O(\Delta n).
\]
Finally, suppose that $\abs*{b} < \frac1{3\sqrt2}$. Note that for any $\bfx,\bfx'\in\mathcal F$, we have 
\[
    \abs*{\angle*{\bfx,\bfx'}} \leq O\parens*{\frac{\frac{\log n}{\log\log n}\sqrt d}{d}} = O\parens*{\frac1{\log\log n}} \leq \frac1{100}.
\]
Then by Corollary \ref{cor:four-pts}, there is at least one point $\bfx$ of the ``first four points'' (see Definition \ref{def:hard-log-reg-inst}) such that
\[
    \angle*{\bfu,\bfx} \leq -b - \frac1{3\sqrt2} \implies \angle*{\bfu,\bfx} + b \leq -\frac1{3\sqrt2}
\]
Thus, we again have that $\norm*{(\bfA\bfx)^-}_p^p \geq \frac1{(3\sqrt2)^p}$ and $\norm*{(\bfA\bfx)^+}_p^p \leq (1+\frac1{3\sqrt2})^p\Delta n$ so we have
\[
    \frac{\norm*{(\bfA\bfx)^+}_p^p}{\norm*{(\bfA\bfx)^-}_p^p} \leq O(\Delta n)
\]
again. This covers all cases, so we conclude.
\end{proof}

To do cost calculations, we need several approximations on the $p$-probit cost function:

\begin{Lemma}\label{lem:p-probit-asymptotics}
Let $r \geq 1$. Then,
\[
    c\exp(-r^p/p) \leq \Phi_p(-r) \leq \exp(-r^p/p)
\]
for some sufficiently small constant $c>0$, and
\[
    \Phi_p(r) = \Theta(r^p)
\]
\end{Lemma}
\begin{proof}
The lower bound is given in \cite[Lemma 2.6]{MOP2022}. The upper bound is given by the following calculation:
\begin{align*}
    \int_{-\infty}^{-r} \exp(-\abs*{t}^p/p)~dt &= \int_{r}^\infty \exp(-t^p/p)~dt \\
    &\leq -\int_{r}^\infty -t^{p-1} \exp(-t^p/p)~dt \leq -\exp(-t^p/p)\vert_r^\infty = \exp(-r^p/p).
\end{align*}
The second item follows from \cite[Lemma C.3]{MOP2022}.
\end{proof}

Next, we lower bound the cost for any instance such that $b\in A$. 

\begin{Lemma}[Cost Lower Bound]\label{lem:logreg-cost-lb}
Suppose that $b\in A\subseteq[n]$. Then, the instance of Definition \ref{def:hard-log-reg-inst} has cost at least
\[
    \sum_{i=1}^n \psi_p([\bfA\bfx](i)) \geq \Omega\parens*{\Delta \log \frac{n}{\Delta}}
\]
for any $\bfx\in\mathbb R^{d+1}$. 
\end{Lemma}
\begin{proof}
If $b\in A$, then we show that the cost on the $n$ copies of Bob's points and the $\Delta$ copies of Alice's points already incurs a cost of at least $\Delta\log\frac{n}{\Delta}$. Note that the cost on these $n+\Delta$ points is at least
\begin{align*}
    \min_{x\geq 0} n\log(\Phi_p(-x)) + \Delta\log(\Phi_p(x)) &\geq \min_{x\geq0} c(n\exp(-x^p) + \Delta x^p) \\
    &= \min_{y\geq0} c(n\exp(-y) + \Delta y)
\end{align*}
for some sufficiently small constant $c$, by Lemma \ref{lem:p-probit-asymptotics}. This is a convex function with critical point $y$ which satisfies $n\exp(y) = \Delta$, or $y = \log(n/\Delta)$, which has a cost of
\[
    \Omega\parens*{\Delta\log\frac{n}{\Delta}}.\qedhere
\]
\end{proof}

On the other hand, if $b\notin A$, we show that we can upper bound the cost.

\begin{Lemma}[Cost Upper Bound]\label{lem:logreg-cost-ub}
Suppose that $b\notin A\subseteq[n]$. Then, there exists a $\bfx\in\mathbb R^{d+1}$ such that the instance of Definition \ref{def:hard-log-reg-inst} has cost at most
\[
    \sum_{i=1}^n \psi_p([\bfA\bfx](i)) \leq O(\log(\Delta n)).
\]
\end{Lemma}
\begin{proof}
Because $b\notin A$, if $\bfu = \bfx'^{(b)} = \bfx^{(b)} / \sqrt d$, then
\[
    \angle*{\bfu,\bfx'^{(b)}} = 1
\]
while for any other $\bfx^{(i)}\in\mathcal F$,
\[
    \abs*{\angle*{\bfu,\bfx'^{(i)}}} \leq \Theta\parens*{\frac{\frac{\log n}{\log\log n}\sqrt d}{d}} = \Theta\parens*{\frac{1}{\log\log n}}.
\]
Thus, we may set $b = \Theta(\log(\Delta n)^{1/p})$ and $\lambda = \Theta(\log(\Delta n)^{1/p})$ such that with the hyperplane in the scaled direction $\lambda \bfu$ and offset $b$, the cost from Bob's points is at most
\[
    n\log\parens*{1+\exp\parens*{-\parens*{\angle*{\lambda \bfu,\bfx'^{(b)}} + b}}} = n\log(\Theta(-\log(\Delta n))) = O(1),
\]
the cost of all of Alice's points and the first four points is at most
\[
    O(\Delta n)\log\parens*{1+\exp\parens*{-\parens*{\angle*{\lambda \bfu,\bfx'^{(b)}} + b}}} = O(\Delta n)\log(\Theta(-\log(\Delta n))) = O(1),
\]
and the cost from the two points at the origin is at most $O(\log(\Delta n))$. 
\end{proof}

By combining Lemmas \ref{lem:bdd-mu}, \ref{lem:logreg-cost-lb}, and \ref{lem:logreg-cost-lb}, we obtain the following hardness theorem:

\begin{Theorem}[Coreset Lower Bound for $p$-Probit Regression]\label{thm:mu-p-probit-lb}
There exists $\bfA\in\mathbb R^{m\times d}$ with $d = O(\log^2 m)$ and $\mu$-complexity at most $O(m)$ such that for any $1 \leq \Delta \leq O(m^{1/3})$, a mergeable coreset which approximates the optimal $p$-probit cost up to a $\Delta$ relative error must use $\Omega(m/\Delta)$ bits of space. In particular, a constant factor approximation to the optimal $p$-probit regression cost for a $\mu_p$-complex dataset requires $\Omega(\mu_p)$ bits of space.
\end{Theorem}
\begin{proof}
Consider the instance of Definition \ref{def:hard-log-reg-inst}. We set $m = O(\Delta n)$, so that the $\mu$-complexity is $O(m)$ and the dataset consists of $O(m)$ points in $d = O(\log^2 n) = O(\log^2 m)$ dimensions. Since $\Delta \leq O(m^{1/3})$, the lower bound on the cost when $b\in A$ is $\Delta \log (n/\Delta) = \Omega(\Delta \log m)$ while the upper bound on the cost when $b\notin A$ is $\log (n\Delta) = O(\log m)$. Thus, a $\Delta$-approximation can differentiate between these two instances. Thus, such a coreset can solve the \textsf{INDEX} problem on $n = O(m / \Delta)$ items, so the total number of bits used must be at least $\Omega(m/\Delta)$.
\end{proof}

Because the asymptotics of the logistic regression loss is the same as that of $p$-probit regression for $p = 1$, up to constant factors, so we get a similar statement for logistic regression:

\begin{Theorem}[Coreset Lower Bound for Logistic Regression]\label{thm:mu-logreg-lb}
There exists $\bfA\in\mathbb R^{m\times d}$ with $d = O(\log^2 m)$ and $\mu$-complexity at most $O(m)$ such that for any $1 \leq \Delta \leq O(m^{1/3})$, a mergeable coreset which approximates the optimal $p$-probit cost up to a $\Delta$ relative error must use at least $\Omega(m/\Delta)$ bits of space. In particular, a constant factor approximation to the optimal $p$-probit regression cost for a $\mu$-complex dataset requires $\tilde\Omega(\mu)$ bits of space.
\end{Theorem}

\section{Acknowledgements}

David P.\ Woodruff and Taisuke Yasuda were supported by ONR grant N00014-18-1-2562 and a Simons Investigator Award. We thank Samson Zhou for comments. We thank Praneeth Kacham for pointing out an error in an earlier version of the draft concerning adversarial streaming.

\bibliographystyle{alpha}
\bibliography{citations}

\newcommand{\etalchar}[1]{$^{#1}$}
\begin{thebibliography}{DMMW12}

\bibitem[BDM{\etalchar{+}}20]{BDMMUWZ2020}
Vladimir Braverman, Petros Drineas, Cameron Musco, Christopher Musco, Jalaj
  Upadhyay, David~P. Woodruff, and Samson Zhou.
\newblock Near optimal linear algebra in the online and sliding window models.
\newblock In {\em 61st {IEEE} Annual Symposium on Foundations of Computer
  Science, {FOCS} 2020, Durham, NC, USA, November 16-19, 2020}, pages 517--528.
  {IEEE}, 2020.

\bibitem[BDR21]{BDR2021}
Adam Block, Yuval Dagan, and Alexander Rakhlin.
\newblock Majorizing measures, sequential complexities, and online learning.
\newblock In Mikhail Belkin and Samory Kpotufe, editors, {\em Conference on
  Learning Theory, {COLT} 2021, 15-19 August 2021, Boulder, Colorado, {USA}},
  volume 134 of {\em Proceedings of Machine Learning Research}, pages 587--590.
  {PMLR}, 2021.

\bibitem[BLM89]{BLM1989}
J.~Bourgain, J.~Lindenstrauss, and V.~Milman.
\newblock Approximation of zonoids by zonotopes.
\newblock {\em Acta Math.}, 162(1-2):73--141, 1989.

\bibitem[BSS12]{BSS2012}
Joshua~D. Batson, Daniel~A. Spielman, and Nikhil Srivastava.
\newblock Twice-ramanujan sparsifiers.
\newblock {\em {SIAM} J. Comput.}, 41(6):1704--1721, 2012.

\bibitem[CD21]{CD2021}
Xue Chen and Michal Derezinski.
\newblock Query complexity of least absolute deviation regression via robust
  uniform convergence.
\newblock In Mikhail Belkin and Samory Kpotufe, editors, {\em Conference on
  Learning Theory, {COLT} 2021, 15-19 August 2021, Boulder, Colorado, {USA}},
  volume 134 of {\em Proceedings of Machine Learning Research}, pages
  1144--1179. {PMLR}, 2021.

\bibitem[CLM{\etalchar{+}}15]{CLMMPS2015}
Michael~B. Cohen, Yin~Tat Lee, Cameron Musco, Christopher Musco, Richard Peng,
  and Aaron Sidford.
\newblock Uniform sampling for matrix approximation.
\newblock In Tim Roughgarden, editor, {\em Proceedings of the 2015 Conference
  on Innovations in Theoretical Computer Science, {ITCS} 2015, Rehovot, Israel,
  January 11-13, 2015}, pages 181--190. {ACM}, 2015.

\bibitem[CLS22]{CLS2022}
Cheng Chen, Yi~Li, and Yiming Sun.
\newblock Online active regression.
\newblock In Kamalika Chaudhuri, Stefanie Jegelka, Le~Song, Csaba
  Szepesv{\'{a}}ri, Gang Niu, and Sivan Sabato, editors, {\em International
  Conference on Machine Learning, {ICML} 2022, 17-23 July 2022, Baltimore,
  Maryland, {USA}}, volume 162 of {\em Proceedings of Machine Learning
  Research}, pages 3320--3335. {PMLR}, 2022.

\bibitem[CMP20]{CMP2020}
Michael~B. Cohen, Cameron Musco, and Jakub Pachocki.
\newblock Online row sampling.
\newblock {\em Theory Comput.}, 16:1--25, 2020.

\bibitem[CP15]{CP2015}
Michael~B. Cohen and Richard Peng.
\newblock L\({}_{\mbox{p}}\) row sampling by lewis weights.
\newblock In Rocco~A. Servedio and Ronitt Rubinfeld, editors, {\em Proceedings
  of the Forty-Seventh Annual {ACM} on Symposium on Theory of Computing, {STOC}
  2015, Portland, OR, USA, June 14-17, 2015}, pages 183--192. {ACM}, 2015.

\bibitem[CWW19]{CWW2019}
Kenneth~L. Clarkson, Ruosong Wang, and David~P. Woodruff.
\newblock Dimensionality reduction for tukey regression.
\newblock In Kamalika Chaudhuri and Ruslan Salakhutdinov, editors, {\em
  Proceedings of the 36th International Conference on Machine Learning, {ICML}
  2019, 9-15 June 2019, Long Beach, California, {USA}}, volume~97 of {\em
  Proceedings of Machine Learning Research}, pages 1262--1271. {PMLR}, 2019.

\bibitem[DBPS18]{DBPS2018}
Alex Dytso, Ronit Bustin, H~Vincent Poor, and Shlomo Shamai.
\newblock Analytical properties of generalized gaussian distributions.
\newblock {\em Journal of Statistical Distributions and Applications},
  5(1):1--40, 2018.

\bibitem[DDH{\etalchar{+}}09]{DDHKM2009}
Anirban Dasgupta, Petros Drineas, Boulos Harb, Ravi Kumar, and Michael~W.
  Mahoney.
\newblock Sampling algorithms and coresets for $\ell_p$ regression.
\newblock {\em {SIAM} J. Comput.}, 38(5):2060--2078, 2009.

\bibitem[DMMW12]{DMMW2012}
Petros Drineas, Malik Magdon{-}Ismail, Michael~W. Mahoney, and David~P.
  Woodruff.
\newblock Fast approximation of matrix coherence and statistical leverage.
\newblock {\em J. Mach. Learn. Res.}, 13:3475--3506, 2012.

\bibitem[EMMZ22]{EMMZ2022}
Alessandro Epasto, Mohammad Mahdian, Vahab~S. Mirrokni, and Peilin Zhong.
\newblock Improved sliding window algorithms for clustering and coverage via
  bucketing-based sketches.
\newblock In Joseph~(Seffi) Naor and Niv Buchbinder, editors, {\em Proceedings
  of the 2022 {ACM-SIAM} Symposium on Discrete Algorithms, {SODA} 2022, Virtual
  Conference / Alexandria, VA, USA, January 9 - 12, 2022}, pages 3005--3042.
  {SIAM}, 2022.

\bibitem[FLPS22]{FLPS2022}
Maryam Fazel, Yin~Tat Lee, Swati Padmanabhan, and Aaron Sidford.
\newblock Computing lewis weights to high precision.
\newblock In Joseph~(Seffi) Naor and Niv Buchbinder, editors, {\em Proceedings
  of the 2022 {ACM-SIAM} Symposium on Discrete Algorithms, {SODA} 2022, Virtual
  Conference / Alexandria, VA, USA, January 9 - 12, 2022}, pages 2723--2742.
  {SIAM}, 2022.

\bibitem[FSS20]{FSS2020}
Dan Feldman, Melanie Schmidt, and Christian Sohler.
\newblock Turning big data into tiny data: Constant-size coresets for k-means,
  pca, and projective clustering.
\newblock {\em {SIAM} J. Comput.}, 49(3):601--657, 2020.

\bibitem[JLS22]{JLS2021}
Arun Jambulapati, Yang~P. Liu, and Aaron Sidford.
\newblock Improved iteration complexities for overconstrained \emph{p}-norm
  regression.
\newblock In Stefano Leonardi and Anupam Gupta, editors, {\em {STOC} '22: 54th
  Annual {ACM} {SIGACT} Symposium on Theory of Computing, Rome, Italy, June 20
  - 24, 2022}, pages 529--542. {ACM}, 2022.

\bibitem[Lee16]{Lee2016}
Yin~Tat Lee.
\newblock {\em Faster algorithms for convex and combinatorial optimization}.
\newblock PhD thesis, Massachusetts Institute of Technology, 2016.

\bibitem[Lew78]{Lew1978}
D.~R. Lewis.
\newblock Finite dimensional subspaces of ${L}_p$.
\newblock {\em Studia Mathematica}, 63(2):207--212, 1978.

\bibitem[LT91]{LT1991}
Michel Ledoux and Michel Talagrand.
\newblock {\em Probability in Banach Spaces: isoperimetry and processes},
  volume~23.
\newblock Springer Science \& Business Media, 1991.

\bibitem[LWW21]{LWW2021}
Yi~Li, Ruosong Wang, and David~P. Woodruff.
\newblock Tight bounds for the subspace sketch problem with applications.
\newblock {\em {SIAM} J. Comput.}, 50(4):1287--1335, 2021.

\bibitem[LWYZ20]{LWYZ2020}
Yi~Li, Ruosong Wang, Lin Yang, and Hanrui Zhang.
\newblock Nearly linear row sampling algorithm for quantile regression.
\newblock In {\em Proceedings of the 37th International Conference on Machine
  Learning, {ICML} 2020, 13-18 July 2020, Virtual Event}, volume 119 of {\em
  Proceedings of Machine Learning Research}, pages 5979--5989. {PMLR}, 2020.

\bibitem[MM13]{MM2013}
Xiangrui Meng and Michael~W. Mahoney.
\newblock Low-distortion subspace embeddings in input-sparsity time and
  applications to robust linear regression.
\newblock In Dan Boneh, Tim Roughgarden, and Joan Feigenbaum, editors, {\em
  Symposium on Theory of Computing Conference, STOC'13, Palo Alto, CA, USA,
  June 1-4, 2013}, pages 91--100. {ACM}, 2013.

\bibitem[MMM{\etalchar{+}}22]{MMMWZ2022}
Raphael~A. Meyer, Cameron~N. Musco, Christopher~P. Musco, David~P. Woodruff,
  and Samson Zhou.
\newblock Fast regression for structured inputs.
\newblock In {\em International Conference on Learning Representations}, 2022.

\bibitem[MMWY21]{MMWY2021}
Cameron Musco, Christopher Musco, David~P. Woodruff, and Taisuke Yasuda.
\newblock Active sampling for linear regression beyond the $\ell_2$ norm.
\newblock {\em CoRR}, abs/2111.04888, 2021.

\bibitem[MOP22]{MOP2022}
Alexander Munteanu, Simon Omlor, and Christian Peters.
\newblock $p$-generalized probit regression and scalable maximum likelihood
  estimation via sketching and coresets.
\newblock {\em CoRR}, 2022.

\bibitem[MOW21]{MOW2021}
Alexander Munteanu, Simon Omlor, and David~P. Woodruff.
\newblock Oblivious sketching for logistic regression.
\newblock In Marina Meila and Tong Zhang, editors, {\em Proceedings of the 38th
  International Conference on Machine Learning, {ICML} 2021, 18-24 July 2021,
  Virtual Event}, volume 139 of {\em Proceedings of Machine Learning Research},
  pages 7861--7871. {PMLR}, 2021.

\bibitem[MRM21]{MRM2021}
Tung Mai, Anup~B. Rao, and Cameron Musco.
\newblock Coresets for classification - simplified and strengthened.
\newblock In {\em Advances in Neural Information Processing Systems 34: Annual
  Conference on Neural Information Processing Systems 2021, NeurIPS 2021,
  December 6-14, 2021, virtual}, 2021.

\bibitem[MSSW18]{MSSW2018}
Alexander Munteanu, Chris Schwiegelshohn, Christian Sohler, and David~P.
  Woodruff.
\newblock On coresets for logistic regression.
\newblock In Samy Bengio, Hanna~M. Wallach, Hugo Larochelle, Kristen Grauman,
  Nicol{\`{o}} Cesa{-}Bianchi, and Roman Garnett, editors, {\em Advances in
  Neural Information Processing Systems 31: Annual Conference on Neural
  Information Processing Systems 2018, NeurIPS 2018, December 3-8, 2018,
  Montr{\'{e}}al, Canada}, pages 6562--6571, 2018.

\bibitem[Pan03]{Pan2003}
Dmitry Panchenko.
\newblock Symmetrization approach to concentration inequalities for empirical
  processes.
\newblock {\em Ann. Probab.}, 31(4):2068--2081, 2003.

\bibitem[PPP21]{PPP2021}
Aditya Parulekar, Advait Parulekar, and Eric Price.
\newblock {L1} regression with lewis weights subsampling.
\newblock In Mary Wootters and Laura Sanit{\`{a}}, editors, {\em Approximation,
  Randomization, and Combinatorial Optimization. Algorithms and Techniques,
  {APPROX/RANDOM} 2021, August 16-18, 2021, University of Washington, Seattle,
  Washington, {USA} (Virtual Conference)}, volume 207 of {\em LIPIcs}, pages
  49:1--49:21. Schloss Dagstuhl - Leibniz-Zentrum f{\"{u}}r Informatik, 2021.

\bibitem[PTB13]{PTB2013}
Udaya Parampalli, Xiaohu Tang, and Serdar Boztas.
\newblock On the construction of binary sequence families with low correlation
  and large sizes.
\newblock {\em {IEEE} Trans. Inf. Theory}, 59(2):1082--1089, 2013.

\bibitem[RST10]{RST2010}
Alexander Rakhlin, Karthik Sridharan, and Ambuj Tewari.
\newblock Online learning: Random averages, combinatorial parameters, and
  learnability.
\newblock In John~D. Lafferty, Christopher K.~I. Williams, John Shawe{-}Taylor,
  Richard~S. Zemel, and Aron Culotta, editors, {\em Advances in Neural
  Information Processing Systems 23: 24th Annual Conference on Neural
  Information Processing Systems 2010. Proceedings of a meeting held 6-9
  December 2010, Vancouver, British Columbia, Canada}, pages 1984--1992. Curran
  Associates, Inc., 2010.

\bibitem[Sar06]{Sar2006}
Tam{\'{a}}s Sarl{\'{o}}s.
\newblock Improved approximation algorithms for large matrices via random
  projections.
\newblock In {\em 47th Annual {IEEE} Symposium on Foundations of Computer
  Science {(FOCS} 2006), 21-24 October 2006, Berkeley, California, USA,
  Proceedings}, pages 143--152. {IEEE} Computer Society, 2006.

\bibitem[Sch87]{Sch1987}
Gideon Schechtman.
\newblock More on embedding subspaces of {$L_p$} in {$l^n_r$}.
\newblock {\em Compositio Math.}, 61(2):159--169, 1987.

\bibitem[Sch11]{Sch2011}
Gideon Schechtman.
\newblock Tight embedding of subspaces of {$L_p$} in {$\ell_p^n$} for even
  {$p$}.
\newblock {\em Proc. Amer. Math. Soc.}, 139(12):4419--4421, 2011.

\bibitem[SS11]{SS2011}
Daniel~A. Spielman and Nikhil Srivastava.
\newblock Graph sparsification by effective resistances.
\newblock {\em {SIAM} J. Comput.}, 40(6):1913--1926, 2011.

\bibitem[SW11]{SW2011}
Christian Sohler and David~P. Woodruff.
\newblock Subspace embeddings for the $l_1$-norm with applications.
\newblock In Lance Fortnow and Salil~P. Vadhan, editors, {\em Proceedings of
  the 43rd {ACM} Symposium on Theory of Computing, {STOC} 2011, San Jose, CA,
  USA, 6-8 June 2011}, pages 755--764. {ACM}, 2011.

\bibitem[SZ01]{SZ2001}
Gideon Schechtman and Artem Zvavitch.
\newblock Embedding subspaces of $l_p$ into $l_p^n$, $0< p< 1$.
\newblock {\em Mathematische Nachrichten}, 227(1):133--142, 2001.

\bibitem[Tal90]{Tal1990}
Michel Talagrand.
\newblock Embedding subspaces of {$L_1$} into {$l^N_1$}.
\newblock {\em Proc. Amer. Math. Soc.}, 108(2):363--369, 1990.

\bibitem[Tal95]{Tal1995}
Michel Talagrand.
\newblock Embedding subspaces of {$L_p$} in {$l^N_p$}.
\newblock In {\em Geometric aspects of functional analysis ({I}srael,
  1992--1994)}, volume~77 of {\em Oper. Theory Adv. Appl.}, pages 311--325.
  Birkh\"{a}user, Basel, 1995.

\bibitem[Tro11]{Tro2011}
Joel~A. Tropp.
\newblock Freedman's inequality for matrix martingales.
\newblock {\em Electron. Commun. Probab.}, 16:262--270, 2011.

\bibitem[UU21]{UU2021}
Jalaj Upadhyay and Sarvagya Upadhyay.
\newblock A framework for private matrix analysis in sliding window model.
\newblock In Marina Meila and Tong Zhang, editors, {\em Proceedings of the 38th
  International Conference on Machine Learning, {ICML} 2021, 18-24 July 2021,
  Virtual Event}, volume 139 of {\em Proceedings of Machine Learning Research},
  pages 10465--10475. {PMLR}, 2021.

\bibitem[Ver18]{Ver2018}
Roman Vershynin.
\newblock {\em High-dimensional probability}, volume~47 of {\em Cambridge
  Series in Statistical and Probabilistic Mathematics}.
\newblock Cambridge University Press, Cambridge, 2018.

\bibitem[VH14]{Van2014}
Ramon Van~Handel.
\newblock Probability in high dimension.
\newblock Technical report, Princeton University, 2014.

\bibitem[WW19]{WW2019}
Ruosong Wang and David~P. Woodruff.
\newblock Tight bounds for $\ell_p$ oblivious subspace embeddings.
\newblock In Timothy~M. Chan, editor, {\em Proceedings of the Thirtieth Annual
  {ACM-SIAM} Symposium on Discrete Algorithms, {SODA} 2019, San Diego,
  California, USA, January 6-9, 2019}, pages 1825--1843. {SIAM}, 2019.

\bibitem[WY22]{WY2022}
David~P. Woodruff and Taisuke Yasuda.
\newblock High-dimensional geometric streaming in polynomial space.
\newblock {\em CoRR}, 2022.

\bibitem[WZ13]{WZ2013}
David~P. Woodruff and Qin Zhang.
\newblock Subspace embeddings and $\ell_p$-regression using exponential random
  variables.
\newblock In Shai Shalev{-}Shwartz and Ingo Steinwart, editors, {\em {COLT}
  2013 - The 26th Annual Conference on Learning Theory, June 12-14, 2013,
  Princeton University, NJ, {USA}}, volume~30 of {\em {JMLR} Workshop and
  Conference Proceedings}, pages 546--567. JMLR.org, 2013.

\end{thebibliography}

\appendix

\section{High Probability \texorpdfstring{$\ell_p$}{lp} Lewis Weight Sampling, \texorpdfstring{$0<p<2$}{0 < p < 2}}\label{sec:high-prob-lws}

Note that the Lewis weight sampling algorithm of \cite{CP2015} samples from Lewis weights with replacement, while we sample each row once with probability proportional to the Lewis weight estimate. We thus carry out a similar analysis for this sampling process in the following discussion. A similar analysis has been carried out by \cite[Lemma 3.2]{CD2021} for the case of $\ell_1$ Lewis weights, in the context of active $\ell_1$ linear regression. In order to obtain a $\log\frac1\delta$ dependence on the failure rate $\delta$, we provide generalizations of the analysis conducted in \cite{LT1991, SZ2001} by analyzing higher moments.

We first show the following moment bounds, which are a modification of results by \cite{LT1991} in a similar way as we did for the case of $p > 2$ in Theorem \ref{thm:one-sided-lt}.

\begin{Theorem}[Rademacher moment bounds]\label{thm:one-sided-lt-0<p<2}
    Let $\bfA\in\mathbb R^{n\times d}$ and $p\in(0,2)$. Let $\bfw$ be $\ell_p$ Lewis weights for $\bfA$. Suppose that
    \[
        \frac{\bfw_i}{d} \leq \beta
    \]
    for all $i\in[n]$, for some $\beta>0$. Define the quantity
    \[
        \Lambda \coloneqq \sup_{\norm*{\bfA\bfx}_p = 1}\abs*{\sum_{i=1}^n \bfsigma_i \abs*{\angle*{\bfa_i,\bfx}}^p}
    \]
    Let $l\geq 1$ and let $\bfsigma = \{\bfsigma_i\}_{i=1}^n$ be independent Rademacher variables. Then,
    \[
        \E_{\bfsigma}[\Lambda^l] \leq \begin{dcases}
        \bracks*{C(p)\beta\cdot d[(\log d)^2(\log n)+l]}^{l/2} & p\in(0,1) \\
        n\bracks*{C(p)\beta\cdot d\cdot l}^{l/2} & p = 1 \\
        \bracks*{C(p)\beta\cdot d[(\log d)^3+l]}^{l/2} & p\in(1,2)
        \end{dcases}
    \]
    where $C(p)$ is a constant depending only on $p$.   
\end{Theorem}
\begin{proof}
For $p = 1$, this is Lemma 8.4 of \cite{CP2015} with a slight change in the normalization. For $p\in(0,1)\cup(1,2)$, we only briefly sketch this result, since a similar result is worked out in detail in Section \ref{sec:one-sided-lewis-moment-bound} for $p>2$. As in this result, we use Lemma \ref{lem:panchenko} to instead bound the same Gaussian process considered in \cite{LT1991}. The only difference is then that we apply a tail version of Dudley's entropy integral, which requires a diameter bound, which is easily seen to be $O(\sqrt d)$ from \cite{LT1991} for $1 < p < 2$ and \cite{SZ2001} for $0 < p < 1$. Then by integrating this tail bound, we attain the claimed moment bounds.
\end{proof}

We use these results to bound the distortion of the subspace embedding, as done in \cite{CP2015, CD2021}. 

\begin{Theorem}[High probability one-shot Lewis weight sampling]\label{thm:lewis-weight-sampling-0<p<2}
Let $\bfA\in\mathbb R^{n\times d}$ and $0 < p < 2$. Let $\delta\in(0,1)$ be a failure rate parameter and let $\eps\in(0,1)$ be an accuracy parameter. Let $\bfw\in\mathbb R^n$ be the $\ell_p$ Lewis weights. Suppose that we set $\bfs_i = 1/\bfp_i^{1/p}$ with probability $\bfp_i$, where $\bfp_i \geq \min\{\bfw_i/(d\beta), 1\}$, for
\[
    \beta = \begin{dcases}
    \frac{\eps^2}{d[(\log d)^2(\log n)+\log\frac1\delta]} & p \in (1, 2) \\
    \frac{\eps^2}{d\log\frac{n}{\delta}} & p = 1 \\
    \frac{\eps^2}{d[(\log d)^3+\log\frac1\delta]} & p \in (0, 1)
    \end{dcases}
\]
Then, with probability at least $1-\delta$,
\[
    \norm*{\bfS\bfA\bfx}_p^p = (1\pm O(\eps))\norm*{\bfA\bfx}_p^p
\]
for all $\bfx\in\mathbb R^d$.
\end{Theorem}
\begin{proof}
In what follows $C$ will be a constant which can change from line to line. Consider the $l$th moment of the error, i.e.,
\[
    \E_\bfs \bracks*{\sup_{\norm*{\bfA\bfx}_p = 1}\abs*{\parens*{\sum_{i=1}^n \abs*{\angle*{\bfs_i\cdot \bfa_i, \bfx}}^p} - 1}^l}.
\]
We will use $l = O(\log\frac1\delta))$ for $p\in(0,2)\setminus\{1\}$ and $l = O(\log\frac{n}{\delta})$ for $p = 1$. By a standard symmetrization argument \cite{CP2015, CD2021}, this is bounded above by
\[
    2^l\E_\bfs \bracks*{\sup_{\norm*{\bfA\bfx}_p = 1}\abs*{\sum_{i=1}^n \sigma_i \abs*{\angle*{\bfs_i\cdot \bfa_i, \bfx}}^p}^l},
\]
where $\sigma = \{\sigma_i\}_{i=1}^n$ are independent Rademacher variables.

Let $\beta$ be the Lewis weight upper bound required by the statement of the current theorem. Then let $\bfA'\in\mathbb R^{r\times d}$ be a matrix with Lewis weights uniformly bounded by $\beta$ such that
\[
    \bfA'^\top \bfW^p(\bfA')^{1-2/p}\bfA' \succeq \bfA^\top\bfW^p(\bfA)^{1-2/p}\bfA
\]
and $\norm*{\bfA'\bfx}_p = O(\norm*{\bfA\bfx}_p)$ for all $\bfx\in\mathbb R^d$. Such a $\bfA'$ exists with $r = \tilde O(d/\beta)$ rows by Lemma B.1 of \cite{CP2015} (see also \cite[Lemma B.1]{CD2021}) for $p\in(1,2)$, while we can take $r = O(n)$ for $p\in(0,1)$ by splitting rows of $\bfA$, as suggested in \cite{CP2015} for results that are independent of $n$. We then define the vertical concatenation
\[
    \bfA'' \coloneqq \begin{bmatrix}\bfS\bfA \\ \bfA'\end{bmatrix}
\]
where $\bfS = \diag(\bfs)$. By the monotonicity of Lewis weights for $p\in(0,2)$ \cite[Lemma 5.5]{CP2015}, the Lewis weights of a row in $\bfA''$ are at most the Lewis weights of the row in the original matrix (either $\bfS\bfA$ or $\bfA'$). We have that
\[
    \bfA''^\top\bfW^p(\bfA'')^{1-2/p}\bfA'' \succeq \bfA'^\top\bfW^p(\bfA')^{1-2/p}\bfA'\succeq \bfA^\top\bfW^p(\bfA)^{1-2/p}\bfA
\]
by construction of $\bfA'$. Then for any row $i$ in $\bfA''$ corresponding to $\bfS\bfA$,
\begin{align*}
    \bfw_i^p(\bfA'')^{2/p} &= (\bfs_i\bfa_i)^\top(\bfA''^\top\bfW^p(\bfA'')^{1-2/p}\bfA'')^-(\bfs_i\bfa_i) \\
    &\leq (\bfs_i\bfa_i)^\top(\bfA^\top\bfW^p(\bfA)^{1-2/p}\bfA)^-(\bfs_i\bfa_i) \\
    &\leq \frac{\bfw_i^p(\bfA)^{2/p}}{\bfp_i^{2/p}} \\
    &\leq \beta^{2/p}
\end{align*}
by Lemma \ref{lem:psd-flip}, so $\bfw_i^p(\bfA'') \leq \beta$. For any row $i\in\bfA''$ corresponding to $\bfA'$, we immediately have that $\bfw_i^p(\bfA'') \leq \bfw_i^p(\bfA') \leq \beta$ by monotonicity of Lewis weights for $p\in(0,2)$. Thus, $\bfA''$ has Lewis weights uniformly bounded by $\beta$. Applying Theorem \ref{thm:one-sided-lt-0<p<2}, we find that
\[
    \E_\bfs \bracks*{\sup_{\norm*{\bfA\bfx}_p = 1}\abs*{\parens*{\sum_{i=1}^n \abs*{\angle*{\bfs_i\cdot \bfa_i, \bfx}}^p} - 1}^l} \leq O(\eps)^l
\]
for $p\in(0,2)\setminus\{1\}$ and
\[
    \E_\bfs \bracks*{\sup_{\norm*{\bfA\bfx}_1 = 1}\abs*{\parens*{\sum_{i=1}^n \abs*{\angle*{\bfs_i\cdot \bfa_i, \bfx}}} - 1}^l} \leq n O(\eps)^l
\]
for $p = 1$, by our choice of $l$ and $\beta$. By Markov's inequality, we then have that
\[
    \Pr\braces*{\sup_{\norm*{\bfA\bfx}_p=1}\abs*{\norm*{\bfS\bfA}_p^p-1}^l \geq \frac1\delta(C\cdot \eps)^l} \leq \delta
\]
for $p\in(0,2)\setminus\{1\}$ and
\[
    \Pr\braces*{\sup_{\norm*{\bfA\bfx}_1=1}\abs*{\norm*{\bfS\bfA}_1-1}^l \geq \frac{n}\delta(C\cdot \eps)^l} \leq \delta.
\]
Then taking $l$th roots for $l = O(\log\frac1\delta)$ for $p\in(0,2)\setminus\{1\}$ and $l = O(\log\frac{n}{\delta})$ for $p = 1$, we have that
\[
    \Pr\braces*{\sup_{\norm*{\bfA\bfx}_p=1}\abs*{\norm*{\bfS\bfA}_p^p-1} \geq C\cdot \eps} \leq \delta.\qedhere
\]
\end{proof}
\section{Preliminaries from Probability in Banach Spaces}

We introduce generalizations from the theory of probability in Banach spaces that we will need for our purposes.

\subsection{Entropy Bounds}

\begin{Definition}[Covering numbers]
Let $B_1,B_2\subseteq\mathbb R^d$ be two sets. Define $E(B_1, B_2)$ to be the minimum number of translates of $B_2$ required to cover $B_1$. For a metric $d$ and radius $t$, define $E(B_1, d_X, t)$ to be the minimum number $d_X$-balls of radius $t$ required to cover $B_1$.
\end{Definition}

\begin{Definition}[Levy mean]
The Levy mean is defined as
\[
    M_X = \int_{\mathbb S^{d-1}} \norm*{\bfx}~d\sigma(\bfx) = \E_{\bfx\sim\mathbb S^{d-1}}\norm*{\bfx}.
\]
\end{Definition}

\begin{Remark}
By noting that $\bfx\sim\mathbb S^{d-1}$ is the same as drawing a Gaussian vector and normalizing, this is
\[
    M_X = \E_{\bfg\sim\mathcal N(0, \bfI_d)}\norm*{\frac{\bfg}{\norm*{\bfg}_2}} = \frac{\E\norm*{\bfg}_2}{\E\norm*{\bfg}_2}\E\norm*{\frac{\bfg}{\norm*{\bfg}_2}} = \frac1{\E\norm*{\bfg}_2}\E\norm*{\bfg}
\]
since the norm of the Gaussian is independent of its direction.
\end{Remark}

\begin{Lemma}[Dual Sudakov minoration (Proposition 4.2, \cite{BLM1989})]\label{lem:dual-sudakov}
Let $(X, \norm*{\cdot})$ be Banach space on $\mathbb R^d$ and let be the Levy mean of $\norm*{\cdot}$. Then, for some constant $C >0$, we have that
\[
    \log E(B_2, t\cdot B_X) \leq C\cdot d\parens*{\frac{M_X}{t}}^2
\]
where $B_2 = \braces*{\bfx : \norm*{\bfx}_2 \leq 1}$ and $B_X = \braces*{\bfx : \norm*{\bfx} \leq 1}$. 
\end{Lemma}

\subsection{Subspaces of \texorpdfstring{$\ell_p$}{lp} in Generalized Lewis' Position}

\begin{Definition}[One-sided $\bfU$-weights]\label{def:orthonormal-weights}
Let $\bfU\in\mathbb R^{n\times d}$ be an orthonormal matrix and let $\gamma\in(0,1]$. We then define $\gamma$-one-sided $\bfU$-weights as any weights $\bfv\in\mathbb R^n$ satisfying
\[
    \bfv_i \geq \gamma\frac{\norm*{\bfe_i^\top\bfU}_2^2}{d}.
\]
We define the normalized one-sided $\bfU$-weights to be $\bar\bfv_i \coloneqq \bfv_i/T$ for $T \coloneqq \sum_{i=1}^n \bfv_i$. 
\end{Definition}

\begin{Definition}\label{def:orthonormal-lp-norms}
Let $\bfU\in\mathbb R^{n\times d}$ be an orthonormal matrix and let $\bar\bfv$ be normalized $\gamma$-one-sided $\bfU$-weights (Definition \ref{def:orthonormal-weights}). Let $0 < q < \infty$. We then define the following (quasi-)norm on $\mathbb R^d$:
\[
    \norm*{\bfx}_{\bar\bfv,q} \coloneqq \bracks*{\sum_{i=1}^n \bar\bfv_i \abs*{[\bar\bfV^{-1/2}\bfU\bfx](i)}^q}^{1/q} = \norm*{\bar\bfV^{1/q-1/2}\bfU\bfx}_q
\]
where $\bar\bfV \coloneqq \diag(\bar\bfv)$. For $q = \infty$, define
\[
    \norm*{\bfx}_{\bar\bfv,q} \coloneqq \norm*{\bar\bfV^{-1/2}\bfU\bfx}_\infty.
\]
\end{Definition}

We note the following equivalence bounds of these norms:

\begin{Lemma}\label{lem:orthonormal-lp-norms-monotonicity}
    Let $T = \sum_{i=1}^n \bfv_i$. The following hold for all $\bfx\in\mathbb R^d$:
\begin{itemize}
    \item For $0 < p < q \leq \infty$, $\norm*{\bfx}_{\bar\bfv,p} \leq \norm*{\bfx}_{\bar\bfv,q}$
    \item For $0 < p < 2$, $\norm*{\bfx}_{\bar\bfv,2} \leq (Td/\gamma)^{1/p-1/2}\norm*{\bfx}_{\bar\bfv,p}$
    \item For $2 < p \leq \infty$, $\norm*{\bfx}_{\bar\bfv,p} \leq (Td/\gamma)^{1/2-1/p}\norm*{\bfx}_{\bar\bfv,2}$
    \item For $0 < p < 2 < q \leq \infty$, $\norm*{\bfx}_{\bar\bfv,p} \leq (Td/\gamma)^{1/p - 1/q}\norm*{\bfx}_{\bar\bfv,q}$
\end{itemize}
\end{Lemma}
\begin{proof}
For $0 < q < \infty$, we have by Jensen's inequality that
\[
    \bracks*{\sum_{i=1}^n \bar\bfv_i \abs*{[\bar\bfV^{-1/2}\bfU\bfx](i)}^p}^{q/p} \leq \sum_{i=1}^n \bar\bfv_i \abs*{[\bar\bfV^{-1/2}\bfU\bfx](i)}^q
\]
and taking $q$th roots on both sides gives the first inequality. For $q = \infty$, we have that
\[
    \sum_{i=1}^n \bar\bfv_i \abs*{[\bar\bfV^{-1/2}\bfU\bfx](i)}^p \leq \norm*{\bar\bfV^{-1/2}\bfU\bfx}_\infty^p \sum_{i=1}^n \bar\bfv_i = \norm*{\bar\bfV^{-1/2}\bfU\bfx}_\infty^p
\]
and taking $p$th roots on both sides gives the result.

We then have that
\begin{align*}
    \norm*{\bar\bfV^{-1/2}\bfU\bfx}_\infty &= \max_{i=1}^n \abs*{\bfe_i^\top\bar\bfV^{-1/2}\bfU\bfx} \\
    &= \sqrt{T}\max_{i=1}^n \abs*{\bfe_i^\top\bfV^{-1/2}\bfU\bfx} \\
    &\leq \sqrt{T}\max_{i=1}^n \frac{\sqrt d}{\sqrt\gamma\norm*{\bfe_i^\top\bfU}_2}\abs*{\bfe_i^\top\bfU\bfx}  && \text{$\gamma$-one-sidedness} \\
    &\leq \sqrt{T}\max_{i=1}^n \frac{\sqrt d}{\sqrt{\gamma}\norm*{\bfe_i^\top\bfU}_2}\norm*{\bfe_i^\top\bfU}_2\norm*{\bfx}_2 && \text{Cauchy--Schwarz} \\
    &= \sqrt{Td/\gamma}\norm*{\bfx}_2 = \sqrt{Td/\gamma}\norm*{\bfU\bfx}_2 = \sqrt{Td/\gamma}\norm*{\bfx}_{\bar\bfv,2}
\end{align*}
which gives the bound on $\norm*{\bfx}_{\bar\bfv,\infty}$. Finally, for $0 < p < 2$,
\[
    \norm*{\bfx}_{\bar\bfv,2}^2 \leq \norm*{\bfx}_{\bar\bfv,\infty}^{2-p}\norm*{\bfx}_{\bar\bfv,p}^p \leq (Td/\gamma)^{(2-p)/2}\norm*{\bfx}_{\bar\bfv,2}^{2-p}\norm*{\bfx}_{\bar\bfv,p}^p \implies \norm*{\bfx}_{\bar\bfv,2} \leq (Td/\gamma)^{1/p-1/2}\norm*{\bfx}_{\bar\bfv,p}
\]
and for $p \geq 2$,
\[
    \norm*{\bfx}_{\bar\bfv,p}^p \leq \norm*{\bfx}_{\bar\bfv,\infty}^{p-2}\norm*{\bfx}_{\bar\bfv,2}^2 \leq (Td/\gamma)^{(p-2)/2}\norm*{\bfx}_{\bar\bfv,2}^{p-2}\norm*{\bfx}_{\bar\bfv,2}^2 \implies \norm*{\bfx}_{\bar\bfv,p} \leq (Td/\gamma)^{1/2-1/p}\norm*{\bfx}_{\bar\bfv,2}
\]

The last inequality follows by combining the previous two inequalities.
\end{proof}

We now bound the Levy mean for $\ell_p$ norms induced by orthonormal matrices $\bfU$.

\begin{Lemma}\label{lem:levy-mean-bound}
Let $\bfU\in\mathbb R^{n\times d}$ be an orthonormal matrix and let $\bfv$ be $\gamma$-one-sided $\bfU$-weights, let $T$ be their sum, and let $\bar\bfv$ be their normalization. Let $1\leq q<\infty$. The Levy mean for the (quasi-)norm $\norm*{\cdot}_{\bar\bfv,q}$ (Definition \ref{def:orthonormal-lp-norms}) is at most
\[
    M_X \leq O((Tq)^{1/2}).
\]
\end{Lemma}
\begin{proof}
We have that
\begin{align*}
    M_X &= \frac{\E\norm*{\bfg}_{\bar\bfv,q}}{\E\norm*{\bfg}_2} \\
    &= \frac1{\E\norm*{\bfg}_2} \E\bracks*{\sum_{i=1}^n \abs*{\bar\bfv_i^{-1/2}\cdot\bfe_i^\top\bfU\bfg}^q \bar\bfv_i}^{1/q} \\
    &= \frac1{\E\norm*{\bfg}_2} \bracks*{\sum_{i=1}^n \E\abs*{\bar\bfv_i^{-1/2}\cdot\bfe_i^\top\bfU\bfg}^q \bar\bfv_i}^{1/q} && \text{Jensen} \\
    &\leq \frac1{\E\norm*{\bfg}_2} O(q^{1/2})\bracks*{\sum_{i=1}^n \norm*{\bar\bfv_i^{-1/2}\cdot\bfe_i^\top\bfU}_2^q \bar\bfv_i}^{1/q} && \text{Gaussian moments} \\
    &= \frac1{\E\norm*{\bfg}_2} O((Tq)^{1/2})\bracks*{\sum_{i=1}^n \norm*{\bfv_i^{-1/2}\cdot\bfe_i^\top\bfU}_2^q \bar\bfv_i}^{1/q} \\
    &\leq \frac{\sqrt d}{\E\norm*{\bfg}_2} O((Tq/\gamma)^{1/2})\bracks*{\sum_{i=1}^n \norm*{\frac{\bfe_i^\top\bfU}{\norm*{\bfe_i^\top\bfU}_2}}_2^q \bar\bfv_i}^{1/q} && \text{$\gamma$-one-sidedness} \\
    &= \frac{\sqrt d}{\E\norm*{\bfg}_2} O((Tq/\gamma)^{1/2})\bracks*{\sum_{i=1}^n \bar\bfv_i}^{1/q} \\
    &\leq O((Tq/\gamma)^{1/2}).
\end{align*}
\end{proof}

The above bound translates into entropy bounds in $\mathbb R^d$ by dual Sudakov minoration (Lemma \ref{lem:dual-sudakov}), which in turn implies entropy bounds in the subspace spanned by $\bfV^{-1/2}\bfU$:

\begin{Corollary}\label{cor:entropy-bound}
Let $\bfU\in\mathbb R^{n\times d}$ be an orthonormal matrix and let $1 \leq q < \infty$. Let $\bfv\in\mathbb R^n$ be $\gamma$-one-sided $\bfU$-weights, let $T$ be their sum, and let $\bar\bfv$ be their normalization. Let $\bar\bfV = \diag(\bar\bfv)$. Let $E\subseteq\mathbb R^n$ be the subspace spanned by $\bar\bfV^{-1/2}\bfU$ and for $0 < p < \infty$, and define the norm
\[
    \norm*{\bfy}_{\bar\bfv,p} \coloneqq \bracks*{\sum_{i=1}^n \bar\bfv_i\abs*{\bfy(i)}^p}^{1/p}.
\]
on $E$. Denote by $B_{\bar\bfv,p}(E)$ the unit balls of $\norm*{\cdot}_{\bar\bfv,p}$ in the subspace $E$. Then, for some constant $C>0$, we have that
\[
    \log E(B_{\bar\bfv,2}(E), t\cdot B_{\bar\bfv,q}(E)) \leq C\cdot \frac{Tqd}{\gamma t^2}.
\]
\end{Corollary}
\begin{proof}
We have by Lemmas \ref{lem:dual-sudakov} and \ref{lem:levy-mean-bound} that
\[
    \log E(B_2(\mathbb R^d), t\cdot B_{\bar\bfv,q}(\mathbb R^d)) \leq C\cdot \frac{Tqd}{\gamma t^2}
\]
where $B_2(\mathbb R^d)\subseteq\mathbb R^d$ is the unit $\ell_2$ ball in $d$ dimensions, and $B_{\bar\bfv,q}(\mathbb R^d)\subseteq \mathbb R^d$ is the unit $\norm*{\cdot}_{\bar\bfv,q}$ ball in $d$ dimensions. Note that $E$ equipped with the norm $\norm*{\cdot}_{\bar\bfv,2}$ is isometric with $\mathbb R^d$ equipped with the usual $\ell_2$ norm, since
\[
    \norm*{\bar\bfV^{-1/2}\bfU\bfx}_{\bfv,2} = \bracks*{\sum_{i=1}^n \bar\bfv_i\abs*{[\bar\bfV^{-1/2}\bfU\bfx](i)}^2}^{1/2} = \norm*{\bfU\bfx}_2 = \norm*{\bfx}_2.
\]
Furthermore, $E$ equipped with the norm $\norm*{\cdot}_{\bar\bfv,q}$ is isometric with $\mathbb R^d$ equipped with the $\norm*{\cdot}_{\bar\bfv,q}$ norm. It follows that the covering numbers must then be the same.
\end{proof}

\section{One-Sided Lewis Weight Sampling, \texorpdfstring{$2<p<\infty$}{2 < p < inf}}\label{sec:one-sided-lewis-moment-bound}

We now work out the changes necessary to \cite{LT1991} to make the $\gamma$-one-sided weights sampling work. A similar proof for usual Lewis weights is worked out in detail in \cite{MMWY2021}.

\LedouxTalagrand*

We first change the position of the subspace to the generalized Lewis's position associated with the orthonormal matrix $\bfU = \bfW^{1/2-1/p}\bfA\bfR$ (see Corollary \ref{cor:entropy-bound}) and $\gamma$-one-sided $\bfU$-weights $\bfv = \bfw/d$ (note that $\bfw$ is a factor of $d$ off from the normalization of Definition \ref{def:orthonormal-weights}). That is, we write
\[
    \sup_{\norm*{\bfA\bfx}_p = 1}\abs*{\sum_{i=1}^n \bfsigma_i \abs*{\angle*{\bfa_i,\bfx}}^p}^l = \sup_{\norm*{\bfA\bfR\bfx}_p = 1}\abs*{\sum_{i=1}^n \bfw_i \bfsigma_i \abs*{[\bfW^{-1/p}\bfA\bfR\bfx](i)}^p}^l
\]
so that the subspace $E$ of Corollary \ref{cor:entropy-bound} is the same as the one spanned by $\bfW^{-1/2}\cdot\bfW^{1/2-1/p}\bfA\bfR = \bfW^{-1/p}\bfA\bfR$.

We now partition $[n]$ into two sets of coordinates, the coordinates $J$ such that $\bfw_i \geq 1/\poly(n)$, and its complement. For coordinates $i\notin J$, note that
\[
    \sup_{\norm*{\bfA\bfR\bfx}_p = 1}\abs*{\sum_{i\notin J} \bfw_i \bfsigma_i \abs*{[\bfW^{-1/p}\bfA\bfR\bfx](i)}^p} \leq \frac1{\poly(n)}\sup_{\norm*{\bfA\bfR\bfx}_p = 1}\sum_{i\notin J} \abs*{[\bfW^{-1/p}\bfA\bfR\bfx](i)}^p \leq \frac1{\poly(n)}
\]
by Lemma \ref{lem:oslw-sensitivity}, for any $\bfsigma$, so it suffices to consider the coordinates in $J$. 

\paragraph{Bounding by a Gaussian process.}

We first show that it suffices to bound a certain Gaussian process. To make this comparison, we will use the following lemma of Panchenko (see also Lemma 7.6 of \cite{Van2014}):

\begin{Lemma}[Lemma 1, \cite{Pan2003}]\label{lem:panchenko}
Let $X, Y$ be random variables such that
\[
    \E[\Phi(X)] \leq \E[\Phi(Y)]
\]
for every increasing convex function $\Phi$. If
\[
    \Pr\braces*{Y\geq t} \leq c_1 \exp(-c_2 t^\alpha)\qquad\mbox{for all $t\geq 0$},
\]
for some $c_1,\alpha\geq 1$ and $c_2>0$, then
\[
    \Pr\braces*{X\geq t} \leq c_1 \exp(1-c_2 t^\alpha)\qquad\mbox{for all $t\geq 0$}.
\]
\end{Lemma}

Let $\Phi$ be a convex increasing function. We first note that since $\bfw_i \leq d\beta$, $\bfw_i \leq \parens{d\beta\bfw_i}^{1/2}$, so by the Rademacher contraction principle \cite[Theorem 4.12]{LT1991}, we have that
\begin{align*}
    &\E_{\bfsigma}\Phi\parens*{\sup_{\norm*{\bfA\bfR\bfx}_p = 1}\abs*{\sum_{i\in J} \bfw_i \bfsigma_i \abs*{[\bfW^{-1/p}\bfA\bfR\bfx](i)}^p}} \leq \\ &\hspace{5em}\E_{\bfsigma}\Phi\parens*{2\beta^{1/2}\sup_{\norm*{\bfA\bfR\bfx}_p = 1}\abs*{\sum_{i\in J} (d\bfw_i)^{1/2} \bfsigma_i \abs*{[\bfW^{-1/p}\bfA\bfR\bfx](i)}^p}}.
\end{align*}
Then by a comparison theorem between Rademacher and Gaussian averages (see Equation 4.8 of \cite{LT1991}), we have that
\begin{align*}
    &\E_{\bfsigma}\Phi\parens*{2\beta^{1/2}\sup_{\norm*{\bfA\bfR\bfx}_p = 1}\abs*{\sum_{i\in J} (d\bfw_i)^{1/2} \bfsigma_i \abs*{[\bfW^{-1/p}\bfA\bfR\bfx](i)}^p}}\leq \\ &\hspace{5em}\E_{g}\Phi\parens*{\sqrt{2\pi}\beta^{1/2}\sup_{\norm*{\bfA\bfR\bfx}_p = 1}\abs*{\sum_{i\in J} (d\bfw_i)^{1/2} g_i \abs*{[\bfW^{-1/p}\bfA\bfR\bfx](i)}^p}}.
\end{align*}
for independent standard Gaussians $g_i$. Thus, by Lemma \ref{lem:panchenko}, it suffices to obtain tail bounds on
\begin{equation}\label{eq:gp}
    \sqrt{2\pi}\beta^{1/2}\sup_{\norm*{\bfA\bfR\bfx}_p = 1}\abs*{\sum_{i\in J} (d\bfw_i)^{1/2} g_i \abs*{[\bfW^{-1/p}\bfA\bfR\bfx](i)}^p}.
\end{equation}
Note that in the original proof of \cite{LT1991}, the contraction principle and comparison theorems are directly used to bound the expected supremum, while we pass through Lemma \ref{lem:panchenko} to obtain tail bounds.

\paragraph{Bounding the Gaussian process.}

We now bound the Gaussian process of \eqref{eq:gp}. We will obtain tail bounds via the following tail bound version of Dudley's inequality:
\begin{Theorem}[Theorem 8.1.6, \cite{Ver2018}]\label{thm:dudley-tail}
Let $(X_t)_{t\in T}$ be a Gaussian process with pseudo-metric $d_X(s,t)\coloneqq \norm*{X_s - X_t}_2$. Let $E(T, d_X, u)$ denote the minimal number of $d_X$-balls of radius $u$ required to cover $T$. Then, for every $z\geq 0$, we have that
\[
	\Pr\braces*{\sup_{t\in T}X_t \geq C\bracks*{\int_0^\infty \sqrt{\log E(T, d_X, u)}~du + z\cdot \diam(T)}} \leq 2\exp(-z^2)
\]
\end{Theorem}

We thus need to bound the metric
\[
    d_X(\bfy,\bfy') \coloneqq \bracks*{\sum_{i\in J} d\bfw_i\parens*{\abs*{\bfW^{-1/p}\bfy(i)}^p - \abs*{\bfW^{-1/p}\bfy'(i)}^p}^2}^{1/2}
\]
in order to obtain entropy diameter bounds. This is the exact same metric bounded by \cite{LT1991}, with Lewis weights replaced by one-sided Lewis weights. Modifying their bound straightforwardly leads to the following:
\begin{Lemma}\label{lem:dx-bound}
    Define $\norm*{\bfy}_J \coloneqq \max_{i\in J}\abs*{\bfy(i)}$. Then,
\[
    d_X(\bfy,\bfy') \leq 2p\sqrt d(\norm*{\bfw}_1^{1/2-1/p})^{p/2-1}\norm*{\bfW^{-1/p}\bfy-\bfW^{-1/p}\bfy'}_J
\]
\end{Lemma}
\begin{proof}
We closely follow the proof of (15.18) of \cite{LT1991}. We first handle $p>2$. For $a,b\geq0$, we have the elementary
\[
    a^p-b^p \leq p(a^{p-1}+b^{p-1})\abs{a-b}.
\]
This gives that
\begin{align*}
    d_X(\bfy,\bfy')^2 &\leq 2p^2\sum_{i\in J}d\bfw_i\max\braces*{\abs*{\bfW^{-1/p}\bfy(i)}^{2p-2},\abs*{\bfW^{-1/p}\bfy'(i)}^{2p-2}}\abs*{\bfW^{-1/p}\bfy(i)-\bfW^{-1/p}\bfy'(i)}^2 \\
    &\leq 2p^2\norm*{\bfW^{-1/p}\bfy-\bfW^{-1/p}\bfy'}_J^2\sum_{i\in J}d\bfw_i\max\braces*{\abs*{\bfW^{-1/p}\bfy(i)}^{2p-2},\abs*{\bfW^{-1/p}\bfy'(i)}^{2p-2}} \\
    &\leq 2p^2 (\norm*{\bfw}_1^{1/2-1/p})^{p-2}\norm*{\bfW^{-1/p}\bfy-\bfW^{-1/p}\bfy'}_J^2\sum_{i\in J}d\bfu_i\max\braces*{\abs*{\bfW^{-1/p}\bfy(i)}^{p},\abs*{\bfW^{-1/p}\bfy'(i)}^{p}} \\
    &\leq 2p^2 d(\norm*{\bfw}_1^{1/2-1/p})^{p-2}\norm*{\bfW^{-1/p}\bfy-\bfW^{-1/p}\bfy'}_J^2(\norm*{\bfy}_p^p + \norm*{\bfy'}_p^p) \\
    &\leq 4p^2 d(\norm*{\bfw}_1^{1/2-1/p})^{p-2}\norm*{\bfW^{-1/p}\bfy-\bfW^{-1/p}\bfy'}_J^2,
\end{align*}
where we have used Lemma \ref{lem:oslw-sensitivity} to bound $\abs*{\bfW^{-1/p}\bfy(i)}$. Taking square roots yields the claim.
\end{proof}

As a consequence, we get diameter bounds.

\begin{Lemma}\label{lem:diameter-p>2}
The $d_X$-diameter of $B_p$ is at most $4p(\norm*{\bfw}_1^{p/2-1}d)^{1/2} \leq 4p(\norm*{\bfw}_1^{p/2})^{1/2}$.
\end{Lemma}
\begin{proof}
Let $\bfy,\bfy'\in B_p$. Then, by Lemma \ref{lem:oslw-sensitivity},
\begin{align*}
    \norm*{\bfW^{-1/p}(\bfy-\bfy')}_J^p &\leq \norm*{\bfW^{-1/p}(\bfy-\bfy')}_\infty^p \\
    &\leq \max_{i=1}^n \bfw_i^{-1}\cdot \norm*{\bfw}_1^{p/2-1}\bfw_i\norm*{\bfy-\bfy'}_p^p \\
    &\leq 2^p\norm*{\bfw}_1^{p/2-1}.
\end{align*}
Then by Lemma \ref{lem:dx-bound},
\begin{align*}
    \sup_{\bfy,\bfy'\in B_p}d_X(\bfy,\bfy') &\leq \sup_{\bfy,\bfy'\in B_p} 2p\sqrt d(\norm*{\bfw}_1^{1/2-1/p})^{p/2-1}\norm*{\bfW^{-1/p}\bfy-\bfW^{-1/p}\bfy'}_J \\
    &\leq \sup_{\bfy,\bfy'\in B_p} 4p\sqrt d(\norm*{\bfw}_1^{1/2-1/p})^{p/2-1}\norm*{\bfw}_1^{1/2-1/p} \\
    &\leq 4p\norm*{\bfw}_1^{p/4-1/2}\sqrt d.\qedhere
\end{align*}
\end{proof}

Let $T_\bfw$ denote the sum of the one-sided Lewis weights. Then,
\begin{align*}
    \norm*{\bfA\bfR\bfx}_p^p &= \sum_{i=1}^n \abs*{[\bfA\bfR\bfx](i)}^p = \sum_{i=1}^n \bfw_i \abs*{[\bfW^{-1/p}\bfA\bfR\bfx](i)}^p \\
    &= d\sum_{i=1}^n \bfv_i \abs*{[\bfW^{-1/p}\bfA\bfR\bfx](i)}^p = T_\bfw\sum_{i=1}^n \bar\bfv_i \abs*{[\bfW^{-1/p}\bfA\bfR\bfx](i)}^p = T_\bfw\norm*{\bfW^{-1/p}\bfA\bfR\bfx}_{\bar\bfv,p}^p
\end{align*}
so the unit $\ell_p$ ball
\[
    B_p \coloneqq \braces*{\bfA\bfx : \norm*{\bfA\bfx}_p \leq 1}
\]
is isometric to $T_\bfw^{-1/p}\cdot B_{\bar\bfv,q}(E)$, which is the unit ball in the subspace $E = \colspan(\bfW^{-1/p}\bfA\bfR)$ from Corollary \ref{cor:entropy-bound}, scaled down by $T_\bfw^{1/p}$.

Let
\[
    \Delta = 2p\sqrt{d}(T_\bfw^{1/2-1/p})^{p/2-1}.
\]
Then for $q = O(\log n)$, note that
\[
    d_X(\bfy,\bfy') \leq \Delta\norm*{\bfW^{-1/p}\bfy-\bfW^{-1/p}\bfy'}_J \leq O(\Delta)\norm*{\bfW^{-1/p}\bfy-\bfW^{-1/p}\bfy'}_{\bar\bfv,q}
\]
since $\bfw_i \geq 1/\poly(n)$ for $i\in J$. We may then bound the covering number of $B_p$ by $d_X$ (the metric associated with the Gaussian process) as
\begin{align*}
    \log E(B_p,d_X,t) &\leq \log E(T_\bfw^{-1/p}\cdot B_{\bar\bfv,q}(E),\norm*{\cdot}_{\bar\bfv,q},\Theta(t/\Delta)) \\
    &= \log E(B_{\bar\bfv,q}(E),\norm*{\cdot}_{\bar\bfv,q},\Theta(T_\bfw^{1/p}t/\Delta)) \\
    &\leq \log E\parens*{B_{\bar\bfv,q}(E),\norm*{\cdot}_{\bar\bfv,q},\Theta(1)\frac{t}{T_\bfw^{p/4-1/2}}}.
\end{align*}
We now compute Dudley's entropy integral. We have that
\begin{align*}
    &\int_0^\infty\sqrt{\log E(B_p,d_X,t)}~dt \\
    \leq~&O(1)T_\bfw^{p/4-1/2}\int_0^\infty\sqrt{\log E(B_{\bar\bfv,p},\norm*{\cdot}_{\bar\bfv,q},t)}~dt \\
    \leq~&O(1)T_\bfw^{p/4-1/2}\bracks*{\int_0^1\sqrt{\log E(B_{\bar\bfv,p},\norm*{\cdot}_{\bar\bfv,q},t)}~dt + \int_1^\infty\sqrt{\log E(B_{\bar\bfv,p},\norm*{\cdot}_{\bar\bfv,q},t)}~dt} \\
    \leq~&O(1)T_\bfw^{p/4-1/2}\bracks*{\int_0^1\sqrt{d\log\frac{d}{t}}~dt + \int_1^{\poly(d)}\frac{\sqrt{T_\bfw\log n}}{\sqrt\gamma t}~dt} \\
    \leq~&O(1)\frac{T_\bfw^{p/4}}{\sqrt\gamma}(\log d)\sqrt{\log n}
\end{align*}
where we have used a standard volume argument for the entropy bound for $t\in(0,1)$ and Corollary \ref{cor:entropy-bound} for the entropy bound for $t\in(1,\infty)$. Then by combining the above calculation with Theorem \ref{thm:dudley-tail} and the diameter calculation of Lemma \ref{lem:diameter-p>2} that for some constant $C>0$,
\[
    \Pr\braces*{\Lambda' \geq C\cdot T_\bfw^{p/4}\bracks*{\gamma^{-1/2}(\log d)\sqrt{\log n} + z}} \leq 2\exp(-z^2)
\]
for
\[
    \Lambda' = \sup_{\norm*{\bfA\bfR\bfx}_p = 1}\abs*{\sum_{i\in J} (d\bfw_i)^{1/2} g_i \abs*{[\bfW^{-1/p}\bfA\bfR\bfx](i)}^p}.
\]

\paragraph{Moment Bounds.}

We now piece together our work above to obtain moment bounds on $\Lambda$. We start by getting moment bounds for $\Lambda'$. We have that
\begin{align*}
    \E[\Lambda'^l] &= (C\cdot T_\bfw^{p/4})^{l} \E\bracks*{\parens*{\frac{\Lambda'}{C\cdot T_\bfw^{p/4}}}^l} \\
    &= l(C\cdot T_\bfw^{p/4})^{l}\int_0^\infty z^l \cdot \Pr\braces*{\frac{\Lambda'}{C\cdot T_\bfw^{p/4}}\geq z}~dz \\
    &= l(C\cdot T_\bfw^{p/4})^{l}\left[\int_0^{4\gamma^{-1/2}(\log d)\sqrt{\log n}} z^l \cdot \Pr\braces*{\frac{\Lambda'}{C\cdot T_\bfw^{p/4}}\geq z}~dz + \right. \\
    &\hspace{8em}\left.\int_{4\gamma^{-1/2}(\log d)\sqrt{\log n}}^\infty z^l \cdot \Pr\braces*{\frac{\Lambda'}{C\cdot T_\bfw^{p/4}}\geq z}~dz \right] \\
    &\leq l(C\cdot T_\bfw^{p/4})^{l}\left[(4\gamma^{-1/2}(\log d)\sqrt{\log n})^{l+1}  + \int_{4\gamma^{-1/2}(\log d)\sqrt{\log n}}^\infty z^l \cdot \Pr\braces*{{\Lambda'} \geq {C\cdot T_\bfw^{p/4}} z}~dz\right] \\
    &\leq l(C\cdot T_\bfw^{p/4})^{l}\left[(4\gamma^{-1/2}(\log d)\sqrt{\log n})^{l+1} +\right. \\
    & \hspace{8em}\left. \int_{4\gamma^{-1/2}(\log d)\sqrt{\log n}}^\infty z^l \cdot \Pr\braces*{{\Lambda'} \geq {C\cdot T_\bfw^{p/4}} [\gamma^{-1/2}(\log d)\sqrt{\log n} + 3z/4]}~dz\right] \\
    &\leq l(C\cdot T_\bfw^{p/4})^{l}\bracks*{(4\gamma^{-1/2}(\log d)\sqrt{\log n})^{l+1} + 2\int_{4\gamma^{-1/2}(\log d)\sqrt{\log n}}^\infty z^l \exp(-z^2/2)~dz} \\
    &\leq l(C\cdot T_\bfw^{p/4})^{l}\bracks*{(4\gamma^{-1/2}(\log d)\sqrt{\log n})^{l+1} + 2\frac{l!}{2^{l/2}(l/2)!}} \\
    &\leq l(C\cdot T_\bfw^{p/4})^{l}\bracks*{(4\gamma^{-1/2}(\log d)\sqrt{\log n})^{l+1} + 2(l/2)^{l/2}} \\
    &= \bracks*{O(1)T_\bfw^{p/2}([\gamma^{-1}(\log d)^2\log n]^{1+1/l} + l)}^{l/2}
\end{align*}
Thus,
\begin{align*}
    \E[\Lambda^l] &\leq 2^l\parens*{(1/\poly(n))^l + \E[(\sqrt{2\pi}\beta^{1/2}\Lambda')^l]} \\
    &\leq \bracks*{O(1)\beta \cdot T_\bfw^{p/2}([\gamma^{-1}(\log d)^2\log n]^{1+1/l} + l)}^{l/2}
\end{align*}
as desired.

\end{document}